\documentclass[11pt,a4paper]{article}
\pdfoutput=1 
             
\usepackage{graphicx}
             
\usepackage[utf8x]{inputenc}	
\usepackage[english]{babel}
\usepackage{paralist}
\usepackage{adjustbox}

\usepackage{amsmath}
\numberwithin{equation}{section}
\usepackage{amssymb}
\usepackage{amsthm}
\usepackage{MnSymbol}
\usepackage{mathtools}		
\usepackage{dsfont}		
\usepackage{stmaryrd}		
\usepackage{cite}		
\usepackage[svgnames]{xcolor}
\usepackage{enumerate}
\usepackage{setspace}
\usepackage{booktabs}
\usepackage{tikz}
\usetikzlibrary{shapes}
\usetikzlibrary{positioning}
\usetikzlibrary{chains}
\usetikzlibrary{arrows,fit,decorations.pathreplacing, shapes.misc}
\usetikzlibrary{decorations.markings}
\tikzstyle{every picture}+=[remember picture]
\tikzstyle{na} = [baseline=-.5ex]
\usepackage{todonotes}
\usepackage{tikz-3dplot}
\usepackage{pgfplots}
\usepackage[font=small,labelfont=bf]{caption}
\usepackage[font=small,labelfont=bf]{subcaption}
\usepackage[pdftex,breaklinks,colorlinks=true,urlcolor=RoyalBlue,linkcolor=blue,
citecolor=blue]{hyperref}

\usepackage[ddmmyyyy,hhmmss]{datetime}

\def\beq{\begin{equation}}
\def\eeq{\end{equation}}
\def\beqn{\begin{eqnarray}}
\def\eeqn{\end{eqnarray}}

\def\Tr{{\rm Tr}}

\def\={\!&=&}
\def\+{\!&+&}
\def\-{\!&-&}

\newcommand{\diag}{\mathrm{diag}}

\newcommand{\Ncal}{\mathcal{N}}

\newcommand{\Fcal}{\mathcal{F}}

\newcommand{\Z}{\mathbb{Z}}
\newcommand{\FF}{\mathbb{F}}

\newcommand{\surm}{\mathrm{SU}}
\newcommand{\urm}{\mathrm{U}}

\newcommand{\sorm}{\mathrm{SO}}
\newcommand{\sprm}{\mathrm{Sp}}

\newcommand{\surmL}{\mathfrak{su}}
\newcommand{\sprmL}{\mathfrak{sp}}
\newcommand{\sormL}{\mathfrak{so}}

\newcommand{\vol}{\mathrm{Vol}}

\newcommand{\geom}{\mathrm{geom}}
\newcommand{\trun}{\mathrm{trun}}
\newcommand{\eff}{\mathrm{eff}}

\def\P{\mathbb{P}}

\begin{document}
\begin{titlepage}
\setcounter{page}{0}

\begin{center}

{\Large\bf 
Fibre-base duality of 5d KK theories
}

\vspace{15mm}

{\large Andreas P.\ Braun${}^{1}$},\ 
{\large Jin Chen${}^{2}$},\ 
{\large Babak Haghighat${}^{2}$},\ 
{\large Marcus Sperling${}^{2}$},\  and \ 
{\large Shuhang Yang${}^{2}$}
\\[5mm]
\noindent ${}^{1}${\em Department of Mathematical Sciences, Durham University,}\\
{\em Lower Mountjoy, Stockton Rd, Durham DH1 3LE, UK}\\
{Email: {\tt andreas.braun@durham.ac.uk}} 
\\[5mm]
\noindent ${}^{2}${\em Yau Mathematical Sciences Center, Tsinghua University}\\
{\em Haidian District, Beijing, 100084, China}\\
{Email: {\tt jinchen@mail.tsinghua.edu.cn},} \\  
{{\tt babakhaghighat@tsinghua.edu.cn},} \\  
{{\tt msperling@mail.tsinghua.edu.cn }}\\  
{{\tt yangsh018@mail.tsinghua.edu.cn }}
\\[5mm]
\vspace{15mm}

\begin{abstract}
We study circle compactifications of 6d superconformal field theories giving rise to 5d rank $1$ and rank $2$ Kaluza-Klein theories. We realise the resulting theories as M-theory compactifications on local Calabi-Yau 3-folds and match the prepotentials from geometry and field theory. One novelty in our approach is that we include explicit dependence on bare gauge couplings and mass parameters in the description which in turn leads to an accurate parametrisation of the prepotential including all parameters of the field theory. We find that the resulting geometries admit ``fibre-base" duality which relates their six-dimensional origin with the purely five-dimensional quantum field theory interpretation. The fibre-base duality is realised simply by swapping base and fibre curves of compact surfaces in the local Calabi-Yau which can be viewed as the total space of the anti-canonical bundle over such surfaces. Our results show that such swappings precisely occur for surfaces with a zero self-intersection of the base curve and result in an exchange of the 6d and 5d pictures.
\end{abstract}

\end{center}

\end{titlepage}
{\baselineskip=12pt
{\footnotesize
\tableofcontents
}
}

\section{Introduction}
Five-dimensional $\mathcal{N}=1$ supersymmetric gauge theories play an important role in our understanding of supersymmetric gauge theories in general. 
Initially, these theories have been studied from various aspects: field theory \cite{Seiberg:1996bd,Morrison:1996xf,Intriligator:1997pq}, brane constructions \cite{Aharony:1997ju,Aharony:1997bh,DeWolfe:1999hj}, and geometry via M-theory backgrounds with Calabi-Yau singularities \cite{Douglas:1996xp}.
Recently renewed interest has culminated in significant progress: notably from field theory \cite{Kim:2012gu,Bergman:2013ala,Bergman:2013aca,Zafrir:2014ywa,Tachikawa:2015mha,Zafrir:2015uaa,Yonekura:2015ksa,Cremonesi:2015lsa,Jefferson:2017ahm,Ferlito:2017xdq}, using brane systems \cite{Bao:2011rc,Bergman:2014kza,Kim:2015jba,Hayashi:2015fsa,Gaiotto:2015una,Bergman:2015dpa,Zafrir:2015rga,Hayashi:2015zka,Ohmori:2015tka,Zafrir:2015ftn,Hayashi:2015vhy,Zafrir:2016jpu,Hayashi:2016abm,Hayashi:2017btw,Hayashi:2018bkd,Hayashi:2018lyv,Cabrera:2018jxt,Hayashi:2019yxj,Hayashi:2019jvx,Bourget:2020gzi,vanBeest:2020civ,vanBeest:2020kou,Akhond:2020vhc,Hayashi:2021pcj}, and from geometry \cite{DelZotto:2017pti,Xie:2017pfl,Esole:2017rgz,Esole:2017qeh,Esole:2017hlw,Jefferson:2018irk,Esole:2018csl,Esole:2018mqb,Bhardwaj:2018yhy,Bhardwaj:2018vuu,Apruzzi:2018nre,Banerjee:2018syt,Closset:2018bjz,Esole:2019hgr,Esole:2019asj,Apruzzi:2019vpe,Apruzzi:2019opn,Apruzzi:2019enx,Bhardwaj:2019jtr,Bhardwaj:2019fzv,Bhardwaj:2019ngx,Saxena:2019wuy,Bhardwaj:2019xeg,Apruzzi:2019kgb,Closset:2019juk,Kashani-Poor:2019jyo,Closset:2020scj,Closset:2020afy,Duan:2020imo,Morrison:2020ool,Bhardwaj:2020phs,Albertini:2020mdx}. 
On the one hand, compactifications of such theories give rise to lower dimensional field theories and moreover, what has recently emerged as a major research line, constructing 4d domain walls within them is a stepping stone for finding UV Lagrangians for a wide class of 4d $\mathcal{N}=1$ superconformal theories \cite{Gaiotto:2015una,Razamat:2016dpl,Bah:2017gph,Kim:2017toz,Kim:2018bpg,Kim:2018lfo,Razamat:2018gro,Chen:2019njf,Razamat:2019mdt,Pasquetti:2019hxf}. On the other hand, it is believed that those 5d field theories that are conformal (SCFTs) arise from circle compactification of 6d SCFTs \cite{Jefferson:2017ahm,Jefferson:2018irk}. Thus a thorough study of these theories will shed light on the landscape of their six-dimensional parents and possibly unveil new dualities \cite{Bhardwaj:2018yhy,Bhardwaj:2018vuu,Apruzzi:2019opn,Apruzzi:2019enx,Apruzzi:2019vpe}. 

One instance where the above 6d/5d relation becomes very transparent and, thus, more tractable is the case of 5d Kaluza-Klein (KK) theories arising from twisted circle compactifications \cite{Kim:2015jba,Hayashi:2015fsa,Zafrir:2015rga,Hayashi:2015zka,Hayashi:2015vhy,Jefferson:2018irk,Bhardwaj:2018yhy,Bhardwaj:2018vuu,Hayashi:2019yxj,Bhardwaj:2019fzv}. To be more explicit, in these cases the six-dimensional parent theory emerges as a UV fixed point of a 5d supersymmetric gauge theory by taking the gauge couplings to infinity or alternatively sending $1/g_{\textrm{YM}}^2 \rightarrow 0$ which translates to decompactifying the 6d circle. It has recently been conjectured \cite{Jefferson:2018irk} that all 5d SCFTs are connected to 5d KK theories via RG flows and that these flows arise by integrating out BPS particles from the 5d KK theory. From a geometric engineering perspective arising from an M-theory compactification, such BPS particles correspond to M2-branes wrapping holomorphic curves inside the Calabi-Yau with their masses being identified with the volume of the curves as measured by the K\"ahler form. Integrating out such states means performing a flop transition in the geometry. In fact, the resulting 5d SCFTs at the end of such RG flows have long been known to arise from M-theory compactifications on non-compact Calabi-Yau three-fold singularities obtained from collapsing surfaces \cite{Morrison:1996xf,Douglas:1996xp,Intriligator:1997pq,Jefferson:2018irk,Closset:2018bjz}. The starting point of such RG flows is then an M-theory compactification which lifts to an F-theory compactification as the Calabi-Yau becomes elliptically fibred \cite{Bhardwaj:2019fzv}. This story is well-known in the case of the E-string theory where the starting point of the RG flow corresponds to the total space of the anti-canonical bundle over del Pezzo $9$, which is sometimes also called the half-K3 surface. This is a rational elliptic surface obtained from blowing up $\mathbb{P}^2$ nine times at different points. Mass-deformations of this theory then result in the chain of 5d SCFTs corresponding to $\mathcal{N}=1$ $\surm(2)$ gauge theories with $N_f < 8$ as first described by Seiberg \cite{Seiberg:1996bd}.

Starting from a 5d SCFT, there is yet another mass-deformation one can turn on, corresponding to a non-zero $1/g_{\textrm{YM}}^2$, which results in a weakly coupled gauge theory description in the IR. Geometrically, this corresponds to the phase where one moves away from the boundary of the K\"ahler cone such that the volume of the collapsing surface is restored to a non-zero value. This phase is important for studying magnetic monopole strings \cite{Haghighat:2011xx,Haghighat:2012bm,Beaujard:2020sgs} and their BPS spectra as well as for connecting to four-dimensional gauge theories \cite{Katz:1996fh}. Typically, in such phases the Calabi-Yau is the anti-canonical bundle over a bouquet of surfaces which are Hirzebruch surfaces or their blowups. Such surfaces admit a description as a $\mathbb{P}^1$ fibration over $\mathbb{P}^1$. For fibred surfaces the self-intersection number of the fibre vanishes, so that we can distinguish the base $\mathbb{P}^1$ (identified with sections of the fibration) by its non-trivial self-intersection. We henceforth denote the class of the fibre by $f$, and the class of the section with negative self-intersection number\footnote{The other section then has a positive self-intersection number.} by $e$. Whenever the self-intersection of $e$ is zero, which happens in the case of a Hirzebruch $0$ surface ($\mathbb{F}_0$), the identification of base and fibre is ambiguous and their roles can be swapped. This leads to fibre-base duality among the resulting gauge theory descriptions as first noticed in \cite{Katz:1997eq} and from a brane-web perspective in \cite{Aharony:1997bh}. This duality becomes particularly interesting when one is dealing with the class of 5d KK theories. Whereas in one frame the theory can be naturally viewed as resulting from a circle-compactification of a six-dimensional theory, as is the viewpoint adopted in \cite{Bhardwaj:2019fzv}, there is another frame where the theory admits a description in terms of a purely five-dimensional supersymmetric QFT. These two frames are precisely related through fibre-base duality. The more familiar frame is the six-dimensional frame where the ``fibre" is identified with the elliptic fibre of the Calabi-Yau which decomposes into intersecting $\mathbb{P}^1$s over the base $-n$-curves. Here, the volume of the fibral curves can be naturally identified with (combinations of) gauge fugacities of the six-dimensional theory. But there is a dual frame where the same fibral curves can be viewed as gauge nodes in a five-dimensional QFT and it is more natural to identify their volume with the inverse of the square of the gauge coupling. In this frame the volume of the base curves gets identified with gauge fugacities of the five-dimensional theory and it is more natural to swap the roles of base and fibre.

Speaking from a more practical/technical point of view, in the prepotentials of 5d KK theories presented in \cite{Bhardwaj:2019fzv} the dependence on flavour fugacities corresponding to gauge couplings and masses has been omitted. This is partly due to the fact that, as discussed above, the superconformal fixed points of KK theories and their deformations arise in the limits where these parameters are turned off. However, in order to arrive at a complete picture of fibre-base duality and various other applications, it is beneficial to include the dependence on flavour parameters explicitly in the prepotential. From the geometric viewpoint, this means to associate non-compact divisors of the Calabi-Yau to them and parametrise the K\"ahler form accordingly. Among the main applications of such a parametrisation would be the computation of 5d BPS spectra using the 6d blowup equations of \cite{Gu:2018gmy,Gu:2019dan,Gu:2019pqj,Gu:2020fem} which has also been advocated in \cite{Kim:2020hhh}. In this paper, we employ a parametrisation of the K\"ahler form which makes both the 6d and 5d frames manifest and restores the dependence on gauge couplings and mass parameters in the 5d prepotential. To this end, we introduce two sets of divisors where for ease of exposition we restrict to the case of a single 5d gauge node. The first set, denoted by $f_i$, restricts to the fibres of compact surfaces $S_i$ which are at the same time the irreducible components of the elliptic fibre of the Calabi-Yau. The second set, denoted by $F$, restricts to the base curve $e_i$ whenever the surface $S_i$ is a blowup of $\mathbb{F}_0$ and otherwise to $f_i$. In the 6d frame the $f_i$ play the role of the fibres of the geometry, while in the 5d frame the $F$ correspond to the fibres. The $F$ is precisely introduced for measuring the gauge coupling strengths $1/g^2$ in the 5d frame and is dual to the $f_i$ in the sense that they measure the volume of the curves $f_i$ subject to the constraint that the total volume of the elliptic fibre is expressed as $1/g^2$. This ensures that the infinite coupling limit corresponds to the F-theory limit where the theory becomes six-dimensional. 

In this paper we identify the 6d and 5d duality frames for the class of rank $1$ and rank $2$ KK theories, see the Table \ref{tab:KK_theories}. These frames are connected by exchanging $e$ and $f$ whenever possible as can be explicitly seen in the examples we study. For instance, in Section \ref{sec:SU3_on_3_Z2_twist}, equation \eqref{eq:-3Z26d} shows the 6d frame where the $f_i$ correspond to the fibres of the geometry and give rise to the roots of the 6d gauge group. On the other hand, equation \eqref{eq:-3Z25d} shows the 5d frame where $f_1$ and $e_1$ have been exchanged and the resulting intersection matrix is the Cartan matrix of the 5d gauge group. Another example can be found in Section \ref{sec:SO8_on_4_Z3_twist} where the 6d frame can be seen in equation \eqref{eq:-4Z36d} corresponding to the F-theory geometry over a $-4$-curve where the $\sorm(8)$ gauge group has been modded out by $\mathbb{Z}_3$ in a twisted compactification on $S^1$. The 5d dual frame is obtained by swapping $f_2$ and $e_2$ and gives rise to the Cartan matrix \eqref{eq:-4Z35d} of the 5d gauge group, namely $\surm(4)$. Similar examples can be found in Section \ref{sec:SU2_1Adj}: 6d fibre \eqref{eq:fibre_SU2+Adj} vs 5d fibre \eqref{eq:root_via_volume_SU2+Adj};  Section \ref{sec:SU3_on_2_Z2_twist}: 6d fibre \eqref{eq:6d_fibre_SU3_on_2_Z2_twist} vs 5d fibre \eqref{eq:Sp2_3Lambda_roots}; Section \ref{sec:SU3_on_1_Z2_twist}: 6d fibre \eqref{eq:6d_fibre_SU3_on_1_Z2_twist} vs 5d fibre \eqref{eq:G2_6F_roots}; Section \ref{sec:Sp1_on_1_no_twist}: 6d fibre \eqref{eq:6d_fibre_Sp1_on_1_no_twist} vs 5d fibre \eqref{eq:roots_via_volume_Sp2_10F}. In all cases we study in this paper, the geometric prepotential we obtain by taking the cube of the K\"ahler form agrees precisely with the expectation from the corresponding 5d $\mathcal{N}=1$ QFTs. When discussing fibre-base dual theories, we only present results for one specific 5d duality frame. Other 5d frames can be obtained from that one by various geometric operations, but are not just related by 6d-5d fibre-base duality. 

The organisation of the present paper is as follows. In Section \ref{sec:geometry} we start by reviewing known results about five-dimensional gauge theories including how to obtain their prepotentials from the 5d and 6d frames, respectively, as well as geometrically in terms of triple intersections of compact surfaces. We then proceed to describe the geometry in more detail and explain how to define non-compact divisors in order to include the contributions of mass parameters and gauge couplings in the prepotential. This is done in Section \ref{sec:non-compact}. In Section \ref{sec:instructive} we present some instructive examples consisting of the E-string geometry, the geometry of $\surm(n)$ gauge groups over a $-2$-curve, as well as twisted compactifications of the non-Higgsable $\surm(3)$ and $\sorm(8)$ theories over $-3$ and $-4$ curves, respectively. 
In Section \ref{sec:fibre-base_6d-5d}, we proceed to more exotic cases in the class of rank $1$ and $2$ KK theories, which admit a fibre-base duality between the 6d frame and one 5d frame. 
Thereafter, we exemplify some fibre-base like dualities between dual 5d theories realised on geometries that are related via an exchange of $e\leftrightarrow f$ in an $\FF_0$.
Finally, we present our conclusions in Section \ref{sec:conclusions}. The Appendices \ref{app:background} and \ref{app:rulingsofdp9} contain further details about our conventions on Lie algebras, mathematical details on the geometry of Hirzebruch surfaces, and a self-contained exposition about the del Pezzo $9$ surface giving rise to the E-string theory.

\begin{table} 
\begin{adjustbox}{width=\textwidth}
\renewcommand{\arraystretch}{1.25}
\centering
\begin{tabular}{cc|lcc}
\toprule
\multicolumn{2}{c|}{6d theory }
 & 5d KK theory & \multicolumn{2}{c}{5d geometry}  \\ \midrule
%
 \raisebox{-.5\height}{
 \begin{tikzpicture}
\node (a1) at (0,0) {$\mathbf{1}$};
  \node at (0,0.35) {$\scriptstyle{ \sprmL(0)^{(1)}}$};
   \end{tikzpicture}
 }
 & \scriptsize{6d E-string}
 & $\surm(2)$ +$8$F
 & $dP_9$ &
  \\
%
 \raisebox{-.5\height}{
 \begin{tikzpicture}
\node (a1) at (0,0) {$\mathbf{2}$};
  \node at (0,0.35) {$\scriptstyle{ \surmL(1)^{(1)}}$};
   \end{tikzpicture}
 }
 &$\substack{\text{6d $\Ncal=(2,0)$ $A_1$ theory,}\\\text{rank-1 M-string} }$
 & $\surm(2)_0$ +$1$Adj
 & \raisebox{-.5\height}{ 
 \begin{tikzpicture}
  \node (v0) at (0,0) {$\FF_{0}^{1+1}$};  
\draw (v0) to [out=140-100,in=220+100,looseness=3] (v0);
    \node at (1.25,0.3) {$\scriptscriptstyle{e-x}$};
        \node at (1.25,-0.3) {$\scriptscriptstyle{e-y}$};
 \end{tikzpicture}
 }
 &
 \\ 
%
 \raisebox{-.5\height}{
 \begin{tikzpicture}
\node (a1) at (0,0) {$\mathbf{2}$};
  \node at (0,0.35) {$\scriptstyle{ \surmL(1)^{(1)}}$};
  \draw (a1) to [out=-45,in=-135,looseness=3] (a1);
   \end{tikzpicture}
 }
 & $\substack{\text{6d $\mathcal{N}=(2,0)$ $A_2$  theory ,} \\ \text{w/  permutation twist}  }$
 & $\surm(2)_{\pi}$ +$1$Adj
 & \raisebox{-.5\height}{
\begin{tikzpicture}
  \node (v0) at (0,0) {$\FF_{1}^{1+1}$};  
\draw (v0) to [out=140-100,in=220+100,looseness=3] (v0);
    \node at (1,0.3) {$\scriptscriptstyle{x}$};
        \node at (1,-0.3) {$\scriptscriptstyle{y}$};
 \end{tikzpicture}}
 & $\star$
 \\ \midrule
%
 \raisebox{-.5\height}{
 \begin{tikzpicture}
\node (a1) at (0,0) {$\mathbf{3}$};
  \node at (0,0.35) {$\scriptstyle{ \surmL(3)^{(2)}}$};
   \end{tikzpicture}
 }
 &  $\substack{\text{6d min $\surm(3)$ SCFT,} \\ \text{w/  automorphism twist}  }$
 & $\surm(3)_9$
 & \raisebox{-.5\height}{
 \begin{tikzpicture}
  \node (v0) at (0,0) {$\FF_{10}$};  
  \node (v1) at (3,0){$\FF_{0}$};  
  \draw  (v0) edge (v1);
  \node at (0.5,0.2) {$\scriptscriptstyle{e}$};
  \node at (2.25,0.2) {$\scriptscriptstyle{4e+f}$};
 \end{tikzpicture}
 } &
 \\
%
 \raisebox{-.5\height}{
 \begin{tikzpicture}
\node (a1) at (0,0) {$\mathbf{2}$};
  \node at (0,0.35) {$\scriptstyle{ \surmL(3)^{(2)}}$};
   \end{tikzpicture}
 }
 & $\substack{\text{6d $\surm(3)$ +6F theory,} \\ \text{w/ automorphism twist}  }$
 & $\sprm(2)$ + $3\ \Lambda^2$
 & \raisebox{-.5\height}{
 \begin{tikzpicture}
  \node (v0) at (0,0) {$\FF_{6}$};  
  \node (v1) at (3,0){$\FF_{0}^3$};  
  \draw  (v0) edge (v1);
  \node at (0.5,0.2) {$\scriptscriptstyle{e}$};
  \node at (1.75,0.2) {$\scriptscriptstyle{4e+2f-2\sum_i x_i}$};
 \end{tikzpicture}
 } &
  \\
%
 \raisebox{-.5\height}{
 \begin{tikzpicture}
\node (a1) at (0,0) {$\mathbf{1}$};
  \node at (0,0.35) {$\scriptstyle{ \surmL(3)^{(2)}}$};
   \end{tikzpicture}
 }
 & $\substack{\text{6d $\surm(3)$ +12F theory,} \\ \text{w/ automorphism twist}  }$
 & 
\raisebox{-.25\height}{
$\begin{cases}
  \surm(3)_4  + 6\ \mathrm{F}\\
 G_2 + 6\ \mathrm{F} \\
 \sprm(2) + 2\ \Lambda^2 + 4\ \mathrm{F}
 \end{cases}
$}
 & \raisebox{-.5\height}{
 \begin{tikzpicture}
  \node (v0) at (0,0) {$\FF_{2}$};  
  \node (v1) at (3,0){$\FF_{0}^6$};  
  \draw  (v0) edge (v1);
  \node at (0.5,0.2) {$\scriptscriptstyle{e}$};
  \node at (2,0.2) {$\scriptscriptstyle{3e+4f-2\sum_i x_i}$};
 \end{tikzpicture}
 } &
 \\
%
 \raisebox{-.5\height}{
 \begin{tikzpicture}
  \node (a1) at (0,0) {$\mathbf{1}$};
  \node at (0,0.35) {$\scriptstyle{ \sprmL(1)^{(1)}}$};
   \end{tikzpicture}
 }
 & \scriptsize{6d $\sprm(1)$ +10F theory}
 & 
\raisebox{-.25\height}{
$ \begin{cases}
   \surm(3)_0  + 10\ \mathrm{F}\\
   \sprm(2)  + 10\ \mathrm{F}
  \end{cases}
$}
 &  \raisebox{-.5\height}{
 \begin{tikzpicture}
  \node (v0) at (0,0) {$\FF_{0}$ };  
  \node (v1) at (3,0) {$\FF_{1}^{10}$};  
  \draw  (v0) edge (v1);
%
  \node at (0.75,0.25) {$\scriptscriptstyle{2e{+}f}$};
  \node at (2,0.25) {$\scriptscriptstyle{2h{-}\sum_i x_i }$};
 \end{tikzpicture}
 } &
  \\
%
 \raisebox{-.5\height}{
 \begin{tikzpicture}
  \node (a1) at (0,0) {$\mathbf{1}$};
  \node at (0,0.35) {$\scriptstyle{ \sprmL(0)^{(1)}}$};
  \node  (a2) at (1.5,0) {$ \mathbf{2}$};
  \node at (1.5,0.35) {$\scriptstyle{ \surmL(1)^{(1)}}$};
  \draw  (a1) edge (a2);
   \end{tikzpicture}
 }
 & \footnotesize{6d rank 2 E-string}
 & 
\raisebox{-.25\height}{
$\begin{cases}
 \surm(3)_{\frac{3}{2}}  + 9\ \mathrm{F} \\
 \sprm(2) + 1\ \Lambda^2 + 8\ \mathrm{F}
\end{cases} $
}
 & \raisebox{-.5\height}{
 \begin{tikzpicture}
  \node (v0) at (0,0) {$\FF_{0}^{1{+}1}$};  
  \node (v1) at (3,0){$\FF_{1}^{8}$};  
  \draw  (v0) edge (v1);
    \draw (v0) to [out=140,in=220,looseness=2] (v0);
      \node at (-1,0.4) {$\scriptscriptstyle{e{-}w}$};
        \node at (-1,-0.4) {$\scriptscriptstyle{e{-}z}$};
%
  \node at (0.5,0.25) {$\scriptscriptstyle{f}$};
  \node at (2.0,0.25) {$\scriptscriptstyle{2e+3f-\sum_i x_i}$};
 \end{tikzpicture}
 } &
 \\
%
 \raisebox{-.5\height}{
 \begin{tikzpicture}
  \node (a1) at (0,0) {$\mathbf{2}$};
  \node at (0,0.35) {$\scriptstyle{ \surmL(2)^{(1)}}$};
   \end{tikzpicture}
 }
 & \scriptsize{6d $\surm(2)$ +4F theory} 
 & 
  $\surm(2){\times }\surm(2)$ +$2$bi-F
 & 
\raisebox{-.5\height}{
 \begin{tikzpicture}
  \node (v0) at (0,0) {$\FF_{0}^4$};  
  \node (v1) at (3,0){$\FF_{2}$};  
  \node (gh) at (1.5,0){$\scriptscriptstyle{2}$}; 
  \draw  (v0) edge (gh);
    \draw  (gh) edge (v1);
  \node at (0.75,0.25) {$\scriptscriptstyle{\substack{e\\ e-\sum_i x_i}}$};
  \node at (2.25,0.25) {$\scriptscriptstyle{\substack{e\\ h}}$};
 \end{tikzpicture}
 } &
 \\
%
 \raisebox{-.5\height}{
 \begin{tikzpicture}
  \node (a1) at (0,0) {$\mathbf{2}$};
  \node at (0,0.35) {$\scriptstyle{ \surmL(1)^{(1)}}$};
  \node  (a2) at (1.5,0) {$ \mathbf{2}$};
  \node at (1.5,0.35) {$\scriptstyle{ \surmL(1)^{(1)}}$};
  \draw  (a1) edge (a2);
   \end{tikzpicture}
 }
 & \scriptsize{ 6d $\mathcal{N}=(2,0)$ $A_2$ theory}
 & 
  $\surm(3)_{0}$  +$1$Adj
 & 
\raisebox{-.5\height}{
 \begin{tikzpicture}
  \node (v0) at (0,0) {$\FF_{0}^{1{+}1}$};  
  \node (v1) at (3,0){$\FF_{0}^{1{+}1}$};  
  \node (gh) at (1.5,0){$\scriptscriptstyle{2}$}; 
  \draw  (v0) edge (gh);
    \draw  (gh) edge (v1);
    \draw (v0) to [out=140,in=220,looseness=2] (v0);
      \node at (-1,0.4) {$\scriptscriptstyle{e{-}x}$};
        \node at (-1,-0.4) {$\scriptscriptstyle{e{-}y}$};
    \draw (v1) to [out=140-100,in=220+100,looseness=2] (v1);
    \node at (3+1,0.4) {$\scriptscriptstyle{e{-}z}$};
        \node at (3+1,-0.4) {$\scriptscriptstyle{e{-}w}$};
  \node at (0.75,0.3) {$\scriptscriptstyle{\substack{f-x\\ x}}$};
  \node at (2.25,0.3) {$\scriptscriptstyle{\substack{f-z\\ z}}$};
 \end{tikzpicture}
 } &
 \\
%
 \raisebox{-.5\height}{
 \begin{tikzpicture}
  \node (a1) at (0,0) {$\mathbf{2}$};
  \node at (0,0.35) {$\scriptstyle{ \surmL(1)^{(1)}}$};
  \node  (a2) at (1.5,0) {$ \mathbf{2}$};
  \node at (1.5,0.35) {$\scriptstyle{ \surmL(1)^{(1)}}$};
  \node (gh) at (0.75,0) {$\scriptscriptstyle{2} $};
  \draw  (a1) edge (gh);
  \draw[->] (gh) edge (a2);
   \end{tikzpicture}
 }
 &  $\substack{\text{6d $\Ncal{=}(2,0)$ $A_3$ theory,} \\ \text{w/ permutation twist}  }$ 
 & 
 $\sprm(2)_{0}$  +$1$Adj
 & \raisebox{-.5\height}{
 \begin{tikzpicture}
  \node (v0) at (0,0) {$\FF_{0}^{1{+}1}$};  
  \node (v1) at (3,0){$\FF_{0}^{1{+}1}$};  
  \node (gh) at (1.5,0){$\scriptscriptstyle{2}$}; 
  \draw  (v0) edge (gh);
    \draw  (gh) edge (v1);
    \draw (v0) to [out=140,in=220,looseness=2] (v0);
      \node at (-1,0.4) {$\scriptscriptstyle{e{-}x}$};
        \node at (-1,-0.4) {$\scriptscriptstyle{e{-}y}$};
    \draw (v1) to [out=140-100,in=220+100,looseness=2] (v1);
    \node at (3+1,0.4) {$\scriptscriptstyle{e{-}z}$};
        \node at (3+1,-0.4) {$\scriptscriptstyle{e{-}w}$};
  \node at (0.75,0.3) {$\scriptscriptstyle{\substack{f{-}x\\ x}}$};
  \node at (2.25,0.3) {$\scriptscriptstyle{\substack{2f{-}z\\ z}}$};
 \end{tikzpicture}
 } & $\star$
  \\
%
 \raisebox{-.5\height}{
 \begin{tikzpicture}
  \node (a1) at (0,0) {$\mathbf{2}$};
  \node at (0,0.35) {$\scriptstyle{ \surmL(1)^{(1)}}$};
  \node  (a2) at (1.5,0) {$ \mathbf{2}$};
  \node at (1.5,0.35) {$\scriptstyle{ \surmL(1)^{(1)}}$};
  \draw  (a1) edge (a2);
  \draw (a1) to [out=-45,in=-135,looseness=3] (a1);
   \end{tikzpicture}
 }
 &  $\substack{\text{6d $\Ncal{=}(2,0)$ $A_4$ theory,} \\ \text{w/ permutation twist}  }$  
 & 
 \raisebox{-.25\height}{
$\begin{cases}
 \surm(3)_{\frac{3}{2}}  + 1\ \mathrm{Sym}\\
  \sprm(2)_{\pi} + 1\ \mathrm{Adj} 
\end{cases} $
}
 &\raisebox{-.5\height}{
 \begin{tikzpicture}
  \node (v0) at (0,0) {$\FF_{6}^{1{+}1}$};  
  \node (v1) at (3,0){$\FF_{0}$};  
  \draw  (v0) edge (v1);
    \draw (v0) to [out=140,in=220,looseness=2] (v0);
      \node at (-1,0.4) {$\scriptstyle{x}$};
        \node at (-1,-0.4) {$\scriptstyle{y}$};
%
  \node at (0.75,0.25) {$\scriptstyle{e}$};
  \node at (2.25,0.25) {$\scriptstyle{2e{+}f}$};
 \end{tikzpicture}
 } & $\star$
 \\
%
 \raisebox{-.5\height}{
 \begin{tikzpicture}
\node (a1) at (0,0) {$\mathbf{2}$};
  \node at (0,0.35) {$\scriptstyle{ \surmL(2)^{(1)}}$};
  \draw (a1) to [out=-45,in=-135,looseness=3] (a1);
   \end{tikzpicture}
 }
 &  $\substack{\text{6d rank-2 $(A_1,A_1)$ theory,} \\ \text{w/ permutation twist}  }$
 & 
  $\surm(3)_0$  +$1$Sym +$1$F
 &
 \raisebox{-.5\height}{
 \begin{tikzpicture}
  \node (v0) at (0,0) {$\FF_{1}^{2}$};  
  \node (v1) at (3,0){$\FF_{0}^{1{+}1}$};  
  \node (gh) at (1.5,0){$\scriptscriptstyle{2}$}; 
  \draw  (v0) edge (gh);
    \draw  (gh) edge (v1);
    \draw (v1) to [out=140-100,in=220+100,looseness=3] (v1);
    \node at (3+1,0.4) {$\scriptscriptstyle{x}$};
        \node at (3+1,-0.4) {$\scriptscriptstyle{y}$};
  \node at (0.8,0.3) {$\scriptscriptstyle{\substack{h\\ h{-}\sum_i x_i}}$};
  \node at (2.2,0.3) {$\scriptscriptstyle{\substack{e+f{-}x{-}2y\\ e{-}x}}$};
 \end{tikzpicture}
 }
 & $\star$
  \\ 
%
 \raisebox{-.5\height}{
 \begin{tikzpicture}
  \node (a1) at (0,0) {$\mathbf{2}$};
  \node at (0,0.35) {$\scriptstyle{ \surmL(1)^{(1)}}$};
  \node  (a2) at (1.5,0) {$ \mathbf{2}$};
  \node at (1.5,0.35) {$\scriptstyle{ \surmL(1)^{(1)}}$};
  \node (gh) at (0.75,0) {$\scriptscriptstyle{3} $};
  \draw  (a1) edge (gh);
  \draw[->] (gh) edge (a2);
   \end{tikzpicture}
 }
 &  $\substack{\text{6d $\Ncal=(2,0)$ $D_4$ theory,} \\ \text{w/ permutation twist}  }$ 
 & 
 \raisebox{-.25\height}{
$\begin{cases}
 G_2  + 1\ \mathrm{Adj}\\
  \surm(3)_{\frac{15}{2}} + 1\ \mathrm{F} 
\end{cases} $
}
 & 
 \raisebox{-.5\height}{
 \begin{tikzpicture}
  \node (v0) at (0,0) {$\FF_{0}^{1{+}1}$};  
  \node (v1) at (3,0){$\FF_{0}^{1{+}1}$};
  \node (gh) at (1.5,0){$\scriptscriptstyle{2}$};  
  \draw  (v0) edge (gh);
  \draw  (gh) edge (v1);
  \node at (0.6,0.2) {$\scriptscriptstyle{f{-}x}$};
  \node at (0.6,-0.2) {$\scriptscriptstyle{x}$};
  \node at (2.25,0.2) {$\scriptscriptstyle{3f{-}x}$};
  \node at (2.25,-0.2) {$\scriptscriptstyle{x}$};
  \draw (v0) to [out=135,in=225,looseness=2] (v0);
  \draw (v1) to [out=45,in=-45,looseness=2] (v1);
  \node at (-1,0.3) {$\scriptscriptstyle{e{-}x}$};
  \node at (-1,-0.3) {$\scriptscriptstyle{e{-}y}$};
  \node at (4,0.3) {$\scriptscriptstyle{e{-}z}$};
  \node at (4,-0.3) {$\scriptscriptstyle{e{-}w}$};
\end{tikzpicture}
 }
 &
  \\ \bottomrule
\end{tabular}
\end{adjustbox}
\caption{Rank 1 and rank 2 KK theories \cite{Jefferson:2017ahm,Jefferson:2018irk,Bhardwaj:2019fzv}. Theories labelled with $\star$ are referred to as \emph{non-geometric}, in the sense that these are not conventional geometric descriptions. These theories are understood as algebraic proposals that mimic many features of local threefolds, but may not satisfy the consistency conditions reviewed in Section \ref{sec:geometry}.
}
\label{tab:KK_theories}
\end{table}

\section{5d gauge theories from geometry}
\label{sec:geometry}

An $\mathcal{N}=1$ 5d effective gauge theory is characterised by a gauge group $G$ and hypermultiplets in a representation $\mathbf{R}= \oplus \mathbf{R}_j$ of $G$. Furthermore, there can be topological data $k$ corresponding to classical Chern-Simons level, as for example in the case of $G = \surm(N\geq 3)$, or discrete $\theta$-angle as in the cases $G = \sprm(N)$. At a generic point in the Coulomb branch, the gauge symmetry $G$ is broken to the maximal torus $\urm(1)^r$. The Coulomb branch is parametrised by the expectation values of scalar fields $\phi$ in the vector multiplets. Here, the scalar field $\phi$ takes values in the Cartan subalgebra of the gauge group $G$. The low energy abelian action is determined by a prepotential $\mathcal{F}_{5d}$. The prepotential is 1-loop exact and the full quantum result is a cubic polynomial of the vector multiplet scalar $\phi$ and mass parameters $m_f$, given by:
\begin{equation} \label{eq:F}
	\Fcal_{5d} = \frac{1}{2g^2}h_{ij} \phi_i \phi_j + \frac{\kappa}{6}d_{ijk} \phi_i \phi_j \phi_k + \frac{1}{12}\left(\sum_{\boldsymbol{\alpha} \in \textrm{root}} |\boldsymbol{\alpha} \cdot \phi|^3 - \sum_f \sum_{\mathbf{w} \in \mathbf{R}_f} |\mathbf{w}\cdot \phi + m_f|^3\right),
\end{equation} 
where $h_{ij} = \textrm{Tr}(t_i t_j)$, and $d_{ijk} = \frac{1}{2} \textrm{Tr}_{\mathbf{F}}(t_i\{t_j,t_k\})$ with $\mathbf{F}$ in the fundamental representation, and we refer to Appendix \ref{sec:liealg} for our Lie algebra conventions. 

The 1-loop correction to the prepotential renormalises the gauge coupling. The effective coupling in the Coulomb branch is simply given by a second derivative of the quantum prepotential which fixes the exact metric on the Coulomb branch:
\begin{equation}
	(\tau_{\eff})_{ij} = (g_{\eff}^{-2})_{ij} = \partial_i \partial_j \Fcal_{5d}, \quad ds^2 = (\tau_{\eff})_{ij} d\phi_i d\phi_j\;.
	\label{eq:effective_coupling}
\end{equation}
In addition, the exact spectrum of magnetic monopoles on the Coulomb branch can be easily obtained from the quantum prepotential and their tension is given by
\begin{equation}
	\phi_{D_i} = \partial_i \Fcal_{5d}, \quad i =1 ,\ldots, r.
\end{equation}
Moreover, one can compute Chern-Simons couplings:
\begin{equation}
	\kappa_{ijk} = \partial_i \partial_j \partial_k \Fcal_{5d}.
\end{equation}
In the following paragraphs we see how to obtain the prepotential \eqref{eq:F} from various viewpoints ranging from a parent 6d SCFT to a geometric realisation in terms of intersecting complex rational surfaces.

\subsection{Prepotentials from 6d SCFTs}
\label{sec:6d5d}

In this section we review how 5d prepotentials are obtained from a given 6d SCFT upon twisted circle compactification following \cite{Bhardwaj:2019fzv}.

6d SCFTs can be described on their tensor branch in terms of tensor multiplets $B_i$ together with gauge multiplets associated to a gauge algebra $\mathfrak{g}_i$. It is known that the $\mathfrak{g}_i$ are either simple or trivial algebras. There are, moreover, fundamental BPS string excitations $s^j$ charged under the $B_j$ such that their charge is given by the Kronecker delta $\delta^i_j$. Let $\Omega^{ij}$ denote the Dirac pairing between $s^i$ and $s^j$. Then there is a   Green-Schwarz term in the Lagrangian on the tensor branch of the following form
\begin{equation}
	\Omega^{ij} B_i \wedge \textrm{tr}(F_j^2) 
\end{equation}
where $F_j$ is the field strength for the $j$-th gauge group if $\mathfrak{g}_j$ is simple and $F_j = 0$ if $\mathfrak{g}_j$ is trivial. In the classification of \cite{Heckman:2013pva}, the $\Omega^{ij}$ matrix is the intersection matrix of $-n$ curves in the base of the F-theory construction. As such it is symmetric, positive definite and all of its entries are integers. 

Upon compactification on a circle, there is the possibility to twist by an element of a discrete global symmetry group $\Gamma$. In order to generate the most general discrete symmetry of a 6d SCFT, one has to combine two kinds of basic discrete symmetries. The first type acts as an outer automorphisms of the gauge algebras $\mathfrak{g}_i$. 
The outer automorphism acting on the roots of $\mathfrak{g}$ induces an action on the irreducible representations; more precisely, the automorphism twist acts on the Dynkin coefficients of the corresponding weights. 
Therefore, we see that such an action is an action on representations of the gauge algebra. The second type of automorphism arises from permuting tensor multiplets $i \mapsto \sigma(i)$ such that
\begin{align}
\begin{aligned}
	\mathfrak{g}_{\sigma(i)} & = \mathfrak{g}_i  \\
	\Omega^{\sigma(i) \sigma(j)} & = \Omega^{ij} 
	\end{aligned}
\end{align}
for all $i, j$. 
Modding out by a given permutation $\sigma$ of a 6d SCFT, generates a different pairing matrix $\Omega^{\alpha \beta}_{\sigma}$; here the indices $\alpha$, $\beta$ label orbits of nodes $i$, which arise under the repeated action of the permutation $\sigma$. 
In more detail, the element $\Omega^{\alpha \beta}_{\sigma}$ can be computed by selecting a node $i$, which lies inside the orbit $\alpha$, and summing over all nodes $j$ that constitute the other orbit $\beta$, i.e.\
\begin{equation}
	\Omega^{\alpha \beta}_{\sigma} = \sum_{j \in \beta} \Omega^{ij}  \,.
\end{equation}
This matrix may not be symmetric, but must be positive definite. One can then combine the action of the two types of twists discussed and denote them by $\sigma, \{q_{\alpha}\}$ where $q_{\alpha}$ is the degree of the action of the outer automorphism on the gauge algebra corresponding to node $\alpha$. 
Naturally, one is led to consider the $\mathcal{O}^{(q_{\alpha})}$ invariant subalgebra $\mathfrak{h}_{\alpha} = \mathfrak{g}_{\alpha}/\mathcal{O}^{(q_{\alpha})}$ of $\mathfrak{g}_{\alpha}$. 
From the viewpoint of the compactified theory, every node $\alpha$ gives rise to the such defined low energy gauge algebra $\mathfrak{h}_{\alpha}$.
For a given 6d theory, the authors of  \cite{Bhardwaj:2019fzv} then define a graph associated to the twisted compactification, such that the value of each node $\alpha$ is $\mathfrak{g}_i^{(q_{\alpha})}\atop \mathbf{\Omega^{\alpha \alpha}}$ and the number of edges between two nodes $\alpha$ and $\beta$ is given by $|\Omega^{\alpha \beta}|$. In the first column of Table \ref{tab:KK_theories} we summarise all rank 1 and 2 5d theories so obtained using the notation of \cite{Bhardwaj:2019fzv}.

With these data, the authors of \cite{Bhardwaj:2019fzv} derive the following prepotential for the twisted circle compactification of the 6d theory
\begin{equation} \label{eq:6dprepotential}
	6\mathcal{F}_{6d} = \sum_{\alpha,\beta} 3 \Omega_{\sigma}^{\alpha \beta} \phi_{0,\alpha} \left(K^{ab}_{\beta} \phi_{a,\beta} \phi_{b,\beta}\right) + \frac{1}{2} \left(\sum_{\boldsymbol{\alpha}} |\boldsymbol{\alpha} \cdot \phi|^3 - \sum_f \sum_{\mathbf{w}(\mathcal{R}_f)} |\mathbf{w}(\mathcal{R}_f) \cdot \phi + m_f|^3\right),
\end{equation}
where $\phi_{0,\alpha}$ is the scalar living in the vector multiplet corresponding to $\urm(1)_{\alpha}$, i.e.\ the affine node.  
The $\phi_{a,\beta}$ are scalar components of the $\urm(1)_{a,\beta}$ vector multiplets that parametrise the Cartan of $\mathfrak{h}_{\beta}$ coming from the degenerate fibre above the curve $\beta$. 
Here $K^{ab}_{\beta}$ is the Killing form on $\mathfrak{h}_{\beta}$.

Moreover, the hypermultiplets, which are charged under the low energy gauge algebra $\mathfrak{h}$, furnish a representation $\mathcal{R}$ which can be decomposed as $\mathcal{R} = \oplus_f \mathcal{R}_f$ into irreducible representations of $\mathfrak{h}$.
Note that contrary to the prepotential given in \eqref{eq:F}, the so obtained prepotential \eqref{eq:6dprepotential} from 6d does not depend on gauge coupling parameters $\frac{1}{g^2}$. However, such terms should be there and have been neglected in the initial derivation given in \cite{Bhardwaj:2019fzv}. One central goal of the current paper is to remedy this gap.

\subsection{Prepotential from geometry}
\label{sec:Fgeom}

As advocated in \cite{Bhardwaj:2019fzv}, the 5d gauge theory resulting from twisted circle compactification can be equivalently described in terms of an M-theory compactification on a Calabi-Yau $X_S$. 
This Calabi-Yau can be described as a local neighbourhood of an arrangement of surfaces -- which each need to be irreducible, compact, and holomorphic -- such that these surfaces intersect each other at most pairwise transversely.
Because these surfaces can be arranged into families, it is convenient to assign a corresponding label $\alpha$.
The irreducible surfaces in each family $\alpha$ are denoted as $S_{a,\alpha}$ where $0 \leq a \leq r_{\alpha}$ and $r_{\alpha}$ denotes the rank of $h_{\alpha}$. 
The K\"ahler parameter associated to $S_{a,\alpha}$ are identified with the parameter $\phi_{a,\alpha}$ of the Coulomb branch in the compactified KK theory discussed above. 
If the invariant algebra $h_{\alpha}$ for the node $\alpha$ is trivial then one associates just one surface $S_{0,\alpha}$.

A key role is played by the shifted prepotential $\Fcal_{\trun}$ which corresponds to the truncated prepotential of a purely five-dimensional QFT where all mass and gauge coupling deformations have been turned off. $\Fcal_{\trun}$ is obtained from $\Fcal_{6d}$ by performing the following replacement:
\begin{equation}
	\phi_{b,\alpha} \rightarrow \phi_{b,\alpha} - d^{\vee}_b \phi_{0,\alpha},
\end{equation}
for all $1 \leq b \leq r_{\alpha}$ and for all $\alpha$. 
The $d^{\vee}_a$ are the dual Coxeter labels of the affine twisted algebra associated to the 6d gauge algebra.
The prepotential $6 \Fcal_{\trun}$ is composed of cubic terms of the form
$\kappa_{a\alpha,b\beta,c\gamma} \ \phi_{a,\alpha} \phi_{b,\beta}\phi_{c,\gamma}$ and the appearing coefficients are geometrically realised by the following intersection numbers of compact surfaces inside the Calabi-Yau $X_S$:
\begin{subequations}
\begin{eqnarray}
	\kappa_{a\alpha,a\alpha,a\alpha} & = & S_{a,\alpha} \cap S_{a,\alpha} \cap S_{a,\alpha}\,, \\
	\kappa_{a\alpha,a\alpha,b\beta} & = & 3 S_{a,\alpha} \cap S_{a,\alpha} \cap S_{b,\beta} \,,\\
	\kappa_{a\alpha,b\beta,c\gamma} & = & 6 S_{a,\alpha} \cap S_{b,\beta} \cap S_{c,\gamma}\,,
\end{eqnarray}
\end{subequations}
and the $(a,\alpha)$, $(b,\beta)$, $(c,\gamma)$ indices are assumed to be distinct and non-equal.
For evaluating the triple intersection product of any three such surfaces one has the freedom to evaluate the intersection number inside any one of the three.
In more detail, the concept of ``gluing curves" needs to be introduced. Denote the intersection locus between two distinct surfaces $S_{a,\alpha}$ and $S_{b,\beta}$ in $X_S$ by $\mathcal{L}_{a\alpha,b\beta}$. In general, $\mathcal{L}_{a\alpha,b\beta}$ can be reducible and splits into geometrically irreducible components given by the sum $\sum_i \mathcal{L}^i_{a\alpha,b\beta}$. Each $\mathcal{L}^i_{a\alpha,b\beta}$ can then be associated to an irreducible curve $C^i_{a,\alpha;b,\beta}$ in $S_{a,\alpha}$ as well as an irreducible curve $C^i_{b,\beta;a,\alpha}$ in $S_{b,\beta}$. Put differently, the intersection of $S_{a,\alpha}$ and $S_{b,\beta}$ can be realised by identifying the curves
\begin{equation}
\label{eq:gluingcurves_two_equiv_notation}
	C^i_{a,\alpha;b,\beta} \sim C^i_{b,\beta;a,\alpha}
\end{equation}
with each other for all $i$. 
One may think of the process that identifies curves as in \eqref{eq:gluingcurves_two_equiv_notation} as an operation in which two surfaces are "gluing together" along such curves.
The above description of the local Calabi-Yau $X_S$ can be compactly summarised in terms of connected graphs as shown for the case of rank 1 and rank 2 5d theories in the last column of Table \ref{tab:KK_theories}. Each node of the graph corresponds to one of the surfaces $S_{a,\alpha}$ which in general are Hirzebruch surfaces and blowups thereof. An edge between two nodes indicates that the corresponding surfaces share one or more curves. If the number of gluing curves is greater than $1$, it is indicated on the edge. The gluing curves are written next to each surface.

Neglecting gauge coupling and mass terms for now, the truncated prepotential $\Fcal_{\trun}$ can be expressed purely geometrically using the above data by defining a K\"ahler form
\begin{equation}
	J_\phi = \sum_{a,\alpha} \phi_{a,\alpha} S_{a,\alpha},
\end{equation}
by the following identity
\begin{eqnarray}
    \mathcal{F}_{\trun} = \textrm{vol}(X_S) = \frac{1}{3!} \int_{X_S} J_\phi \wedge J_\phi \wedge J_\phi.
\end{eqnarray}
The K\"ahler form can also be used to measure the volume of a curve $C$ via
\begin{equation}
	\textrm{vol}(C) = - J_{\phi} \cdot C.
\end{equation}
where we have adopted the conventions of \cite{Bhardwaj:2019fzv}.
The identification \eqref{eq:gluingcurves_two_equiv_notation} implies that
\begin{equation}
	J_{\phi} \cdot C^i_{a,\alpha;b,\beta} = J_{\phi} \cdot C^i_{b,\beta;a,\alpha},
	\label{eq:consistency}
\end{equation}
and the gluing curves also have to satisfy the `Calabi-Yau condition' spelled out in \eqref{eq:CYcondition_gluing_curves}. 

\subsection{\texorpdfstring{The geometry of $X_S$}{The geometry of XS}}
\label{sec:non-compact}
So far, we have only described the compact divisors $S_{a,\alpha}$ which are related to the Coulomb branch parameters $\phi_{a,\alpha}$. In the Calabi-Yau threefold $X_S$ there can be further non-compact divisors related to the mass parameters and the gauge couplings. In order to describe such divisors, we discuss the geometry of $X_S$ in some more detail.

As every single one of the compact surfaces $S_{a,\alpha}$ sits inside a non-compact Calabi-Yau threefold $X_S$, their normal bundle must be equal to their 
canonical bundle, $K_{S_{a,\alpha}}$. For a single surface $S_{a,\alpha}$, this means that $X_S$ is simply the total space of the bundle $K_{S_{a,\alpha}}$. 
Gluing several surface $S_{a,\alpha}$ along $\mathcal{L}_{a\alpha,b\beta} = S_{a,\alpha} \cap S_{b,\beta}$ implies that we should think of $X_S$ as being glued from the total spaces of 
the bundles $K_{S_{a,\alpha}}$. The surfaces $S_{a,\alpha}$ are the zero loci of sections of the line bundles $K_{S_{a,\alpha}}$ on $X_S$. 

We can check that this gluing can be done consistently by examining the normal bundle of $\mathcal{L}_{a\alpha,b\beta}$ inside $X_S$.
This normal bundle is a sum of line bundles which can be described as 
\begin{equation}\label{eq:N_glue_curve_ab}
N_{\mathcal{L}_{a\alpha,b\beta} \backslash X_S} = \left. K_{S_{a,\alpha}}\right|_{\mathcal{L}_{a\alpha,b\beta}} \oplus N_{\mathcal{L}_{a\alpha,b\beta}\backslash S_{a,\alpha}} 
\end{equation}
by taking $\mathcal{L}_{a\alpha,b\beta}$ as an algebraic subvariety of $S_{a,\alpha}$, and as 
\begin{equation}\label{eq:N_glue_curve_ba}
N_{\mathcal{L}_{a\alpha,b\beta}\backslash X_S} = \left.K_{S_{b,\beta}}\right|_{\mathcal{L}_{a\alpha,b\beta}} \oplus N_{\mathcal{L}_{a\alpha,b\beta}\backslash S_{b,\beta}} 
\end{equation}
by taking $\mathcal{L}_{a\alpha,b\beta}$ as an algebraic subvariety of $S_{b,\beta}$. Under a consistent gluing, these two expression need to be identical. 
As $\mathcal{L}_{a\alpha,b\beta}$ is given by the intersection of $S_{a,\alpha}$ and $S_{b,\beta}$, adjunction tells us that 
$N_{\mathcal{L}_{a\alpha,b\beta}\backslash S_{a,\alpha}}$ is the restriction of $K_{S_{\alpha,b}}$ to $\mathcal{L}_{a\alpha,b\beta}$. Hence,
we find
\begin{equation}
\left. K_{S_{a,\alpha}}\right|_{\mathcal{L}_{a\alpha,b\beta}} = N_{\mathcal{L}_{a\alpha,b\beta}\backslash S_{b,\beta}}  \, ,
\end{equation}
which shows that the gluing identifies the direction transverse to $S_{a\alpha}$ in $X$ with the normal direction of $\mathcal{L}_{a\alpha,b\beta}$ in 
$S_{b,\beta}$, so that \eqref{eq:N_glue_curve_ab} agrees with \eqref{eq:N_glue_curve_ba}.

Using the realisation of $\mathcal{L}_{a\alpha,b\beta}\backslash S_{b,\beta}$ as a complete intersection, it is straightforward to work out using adjunction that
\begin{equation}\label{eq:CYcondition_gluing_curves}
\begin{aligned}
\chi(\mathcal{L}_{a\alpha,b\beta}) & =  \int_{\mathcal{L}_{a\alpha,b\beta}} c_1(\mathcal{L}_{a\alpha,b\beta}) = 
\int_{\mathcal{L}_{a\alpha,b\beta}} -[K_{S_{a,\alpha}}]-[K_{S_{b,\beta}}]\\
& = - \int_{X_S} [K_{S_{a,\alpha}}]\cdot [K_{S_{b,\beta}}] \cdot ([K_{S_{a,\alpha}}]+[K_{S_{b,\beta}}]) \\
& = - \int_{S_{a,\alpha}} (\mathcal{L}_{a\alpha,b\beta})^2  -  \int_{S_{b,\beta}} (\mathcal{L}_{b\beta, a\alpha})^2  \,.
\end{aligned}
\end{equation}
When $\mathcal{L}_{a\alpha,b\beta}$ is an irreducible curve of genus $g$, the above expression is equal to $2-2g$, which gives the `Calabi-Yau condition' of the gluing.

The description of $X_S$ we have given makes it clear that given a collection of curves $\{C_{a,\alpha}\}$ on the surfaces $S_{a,\alpha}$, we can construct a 
divisor in $X_S$ that restricts to these curves by simply taking the inverse image $\pi^{-1}_{a,\alpha}(C_{a,\alpha})$ under the projection 
$\pi_{a,\alpha}: K_{S_{a,\alpha}}\rightarrow S_{a,\alpha}$ for each $C_{a,\alpha}$, and then gluing these to a (reducible) divisor. As the total spaces 
$K_{S_{a,\alpha}}$ glue together consistently, the only extra consistency condition that arises is when one (or some) of the $\{C_{a,\alpha}\}$ is 
a gluing curve, i.e. a component of $\mathcal{L}_{a\alpha,b\beta}$. In this case, the divisor $\pi^{-1}(C_{a,\alpha})$ will restrict to non-trivial 
curves on both $S_{a,\alpha}$ and $S_{b,\beta}$. 

Having discussed the inclusion of non-compact divisors into our geometry, we are now ready to employ a parametrisation of the K\"ahler form which includes dependence on bare gauge couplings and mass parameters. To this end, we introduce divisors $F_i$ with the index $i$ running over $i=0,\ldots,n_0$. Here $n_0$ is the total number of surfaces $S_{a,\alpha}$ which can be written as blowups of $\mathbb{F}_0$, which is equal to the total number of bare gauge couplings. 
The $F_i$ are then given by the inverse images $\pi^{-1}_{a,\alpha}(e_{a,\alpha})\bigcup \pi^{-1}_{b,\beta}(l f_{b,\beta})$ where $e_{a,\alpha}$ is a base of $\mathbb{F}_0$ (or blowups of it) and $l f_{b,\beta}$ denotes the fibre (taken with an integer multiplicity $l$) of a surface with non-zero self-intersection number of base curve (i.e. which does not arise from $\mathbb{F}_0$). In cases where all $S_{a,\alpha}$ admit a description in terms of blowups of $\mathbb{F}_0$, the second factor in the inverse image is missing and the $F_i$ can be purely written in terms of inverse images of base curves $e_i$. The $F_i$ are associated with gauge coupling parameters. 

In case that there is only one 5d gauge algebra $\mathfrak{g}$, this can be summarised as follows: suppose there are $n=\mathrm{rk}(\mathfrak{g})$ compact surfaces $S_i$ and one is able to identify inside each $S_i$ a complex curve $\tilde{f}_i$ of zero self-intersection such that 
\begin{align}
    -\tilde{f}_i \cdot S_j = C_{ij}^{\mathfrak{g}}
    \label{eq:identify_gauge_algebra}
\end{align}
which implies that the $\tilde{f}_i$ act as simple roots $\alpha_i$ of $\mathfrak{g}$. The non-compact divisor $F$, associated to the gauge coupling $g$, has to intersect to compact surfaces as
\begin{align}
    - F \cdot S_i \cdot S_j = h_{ij}^{\mathfrak{g}} \,.
    \label{eq:identify_F_divisor}
\end{align}
Recalling the relation \eqref{eq:Cartan_vs_metric_tensor} between $C_{ij}^{\mathfrak{g}}$ and $ h_{ij}^{\mathfrak{g}}$, one finds 
\begin{align}
    F|_{S_i} = D_i^{-1} \tilde{f}_i = 
    \frac{2}{\langle \alpha_i,\alpha_i \rangle}  \tilde{f}_i \;,
    \label{eq:determine_F_divisor}
\end{align}
which is verified in all the examples considered in this paper.

We can find similar sets of non-compact divisors, denoted by $N_f$, which are associated to bare mass parameters $m_f$. The divisors $N_f$ can restrict to blowup curves $x_i$ or fibres $f_i$, but never to base curves $e_i$. Now we can parametrise the K\"ahler form as follows
\begin{eqnarray}
    J = - \sum_i \frac{1}{g_i^2} F_i + \sum_{a,\alpha} \phi_{a,\alpha} S_{a,\alpha} + \sum_f m_f N_f.
\end{eqnarray}
which results in the geometric prepotential
\begin{eqnarray}
    \Fcal_{\geom} = \frac{1}{3!} \int_{X_s} J \wedge J \wedge J.
\end{eqnarray}
Henceforth, we will omit the indices $\{a,\alpha\}$ and parametrise all non-compact as well as compact divisors in the geometry purely in terms of the Latin indices $i,j$ and $f$. Using the above parametrisation of the K\"ahler form, we can then express effective gauge couplings as volumes of various 2-cycles as follows:
\begin{eqnarray}
    \tau_{ij} = \partial_i \partial_j \Fcal_{\geom} = \textrm{vol}(S_i \bigcap S_j) = \int_{X_S} J \wedge S_i \wedge S_j.
    \label{eq:effectice_coupling_geom}
\end{eqnarray}
The parametrisation of the K\"ahler form is fixed in such a way that $\tau_{ij}$ matches the effective gauge coupling $\partial_i \partial_j \Fcal_{5d}$ of a five-dimensional supersymmetric QFT.

\section{Instructive examples}
\label{sec:instructive}
In this section, inclusion of gauge coupling and mass parameters into the geometric description is explored in a set of selected examples. Among these are untwisted compactification, like the the well-studied E-string theory or the 5d affine $A$-type quiver theory. Twisted compactifications are explored for the 6d minimal $\surm(3)$ and $\sorm(8)$ SCFTs with outer automorphism twist.

\subsection{\texorpdfstring{$-1$ curve: E-string theory on a circle}{-1 curve: E-string theory on a circle}}
\label{sec:E-string}
Let us start with 6d $\mathcal{N}=(1,0)$ E-string theory corresponding to one 
M5 brane probing an M9 wall. Compactifying this theory on a circle (with a 
possible twist) gives a 5d $\sprm(1)$ gauge theory with $N_f = n \leq 8$. Using 
the fact that $d_{ijk} = 0$ for $G=\sprm(1)=\surm(2)$, we obtain from \eqref{eq:F} 
\begin{equation}
	\Fcal_{5d} = \frac{1}{g^2} \phi^2 + \frac{1}{12} \left(2 |2\phi|^3 - 
\sum_{j=1}^8 |\phi + m_j|^3- 
\sum_{j=1}^8 |\phi - m_j|^3\right).
\end{equation}
For $n=8$, setting all masses to zero, i.e.\ $m_j = 0$ for $j=1, \ldots, 8$, we 
get 
\begin{equation}
	\Fcal_{5d} = \frac{1}{g^2}\phi^2.
\end{equation}  
In particular, we see that at infinite coupling, $g=\infty$, the 
monopole tension $\phi_D = \partial \mathcal{F}/\partial \phi$ vanishes 
regardless of the value of $\phi$. This is a remnant of the fact, that the UV 
completion of this 5d theory is in fact a 6d theory, namely the E-string theory.

Let us now assume that $\phi > m_j$ for all $j$. We can then write
\begin{equation}\label{eq:prepE-stringcircle}
\Fcal_{5d} = \frac{1}{g^2} \phi^2 + \frac{1}{6}\left((8-n)\phi^3 - 3 \phi 
\sum_j  m_j^2 \right) \;.
\end{equation}
The effective gauge coupling is given by 
\begin{equation}\label{eq:effgaugecouplingfromprep}
\tau_{\eff} = (\partial_\phi)^2 \Fcal_{5d} =  \frac{2}{g^2} + (8-n) \phi
\end{equation}
which matches the result of \cite{Seiberg:1996bd}. 

\paragraph{Geometric realisation.}
The E-string theory in 6d is engineered geometrically in F-theory by collapsing 
a rational elliptic surface (also called $dP_9$ surface) $S$ inside a non-compact elliptically fibred 
Calabi-Yau threefold. If we consider M-Theory on the same geometry, we find the 
5d theory that was discussed as a compactification of the 6d $\Ncal=(1,0)$ SCFT on 
a circle. Following the general discussion of geometric engineering of 5d gauge 
theories, we need to exploit a ruling on $S$. Such a ruling is discussed 
in \cite{Douglas:1996xp}, see also Appendix \ref{app:rulingsofdp9}, where it is 
shown that it originates from $S$ being a blowup of $\mathbb{F}_0$ at 8 points. 
We can generalise this to surfaces $S= dP_{n+1}$ which are blowups of 
$\mathbb{F}_0$ at $n$ points, and which in turn give rise to a $\sprm(1)$ gauge 
theory with $n$ flavours. We can model the relevant Calabi-Yau geometry $X_S$ as
\begin{equation}
X_S = \mathcal{O}_{dP_{n+1}}(K_{dP_{n+1}}) . 
\end{equation}
For every curve in $dP_{n+1}$, there is hence an associated non-compact divisor 
in $X_S$ by simply taking the preimage under the projection $\pi: X_S \rightarrow dP_{n+1}$.

Let us denote the fibre and base classes of the ruling by $f$ and $e$, and the 
$n$ exceptional divisors of the blowups by $x_i$. In terms of the $\P^1$ fibration on 
$S$, we can think of the $x_i$ as irreducible fibre components, i.e. over $n$ points of 
the base, the fibre splits into the two components $x_i$ and $f-x_i$. 

The intersections between the curves on $S$ can be summarised as (see 
Appendix \ref{app:rulingsofdp9} for details)
\begin{equation}
e \cdot f = 1\,\, , \hspace{1cm}
x_i \cdot x_j = - \delta_{ij}
\end{equation}
with all others vanishing. Note that this implies that
\begin{equation}
(f-x_i) \cdot (f-x_j) = - \delta_{ij} \,\,,\hspace{1cm} e \cdot (f-x_i) = 1 \, .
\end{equation}
Let us parametrise the K\"ahler form as 
\begin{equation}
J = a f + b [K] + \sum_i c_i x_i 
\end{equation}
where $[K]$ is the class of the $dP_{n+1}$ inside $X_S$,  
\begin{equation}
[K]|_S = -c_1(S) = -\left(2e +2f -\sum_i x_i \right)|_S \, .
\end{equation}
The flavour masses are given by integrating the K\"ahler form over the curves $x_i$ and $f-x_i$, they are 
given by $\phi \pm m_i$ \cite{Seiberg:1996bd}, where $\phi$ is the Coulomb branch parameter 
and $m_i$ are the mass parameters. We find 
\begin{equation}
-J \cdot x_i = b + c_i   \hspace{1cm}
-J \cdot (f - x_i) = b - c_i  
\end{equation}
so that we are led to associate $b = \phi$ and $c_i = m_i$. Note that we are in the phase where $\phi > m_i$. 

We can work out triple intersection numbers involving at least one compact 
divisor by using that $[K]$ is the Poincare dual to
$S$ inside $X$. The geometric prepotential is then
\begin{equation}
\begin{aligned}
\Fcal_{\geom} = \frac{1}{6} J^3 &= \frac{1}{6}\left( [K]^3 \phi^3 + 3 
([K]^2 \cdot f )\,  \phi^2  a  + 
\sum_{i=1}^{n} 3 ([K]^2 \cdot x_i) \, \phi^2 m_i 
+ 3  ([K] \cdot x_i^2) \, \phi m_i^2  
\right) \\
&= \frac{1}{6}\left( (8-n) \phi^3 - 6 \phi^2  a  -
\sum_{i=1}^{n} 3 \phi^2 m_i 
- 3  \, \phi m_i^2  
\right) 
\end{aligned}
\end{equation}
Comparing to \eqref{eq:prepE-stringcircle} we can fix $a = -\frac{1}{g^2} - \tfrac12 \sum_i m_i$.

We can now work out the volume of the elliptic curve defining F-theory on 
$X_S$. The class of the elliptic curve inside $S$ is simply $c_1(S)$. 
Its volume is hence given by
\begin{equation}
\vol(f_{\mathrm{ell}}) = - J \cdot f_\mathrm{ell} =  - J \cdot c_1(S) =  \frac{2}{g^2} + (8-n)
\label{eq:volume_of_ellipitic_fibre_of_Estring}
\end{equation}
The SCFT fixed point is reached by setting all $m_i$ to zero and letting $g 
\rightarrow \infty$, which implies for $n=8$ that $\vol(f_\mathrm{ell}) \rightarrow 0$. This 
is the F-theory limit and we recover that the marginal ($n=8$) theory has a 
SCFT fixed point in 6d. 
%
%
\subsection{\texorpdfstring{$-2$ curve: $\surm(n)$ without twist}{-2 curve: SU(n) without twist}}
\label{sec:affine_quiver}
Consider a $-2$ curve which supports an $\surmL(n)$ gauge algebra. The 5d KK theory is known to be the affine $\widehat{A}_{n-1}$ quiver gauge theory 
\begin{align}
\raisebox{-.5\height}{
 	\begin{tikzpicture}
	\tikzstyle{gauge} = [circle, draw,inner sep=3pt];
	\tikzstyle{flavour} = [regular polygon,regular polygon sides=4,inner 
sep=3pt, draw];
	\node (g1) [gauge,label=below:{$\scriptstyle{\surm(2)}$}] {};
	\node (g2) [gauge, right of=g1,label=below:{$\scriptstyle{\surm(2)}$}] {};
	\node (g3) [right of=g2] {$\cdots$};
	\node (g4) [gauge, right of=g3,label=below:{$\scriptstyle{\surm(2)}$}] {};
	\node (g5) [gauge, right of=g4,label=below:{$\scriptstyle{\surm(2)}$}] {};
	\node (g0) [gauge, above of= g3, label=above:{$\scriptstyle{\surm(2)}$}] {};
	\draw (g1)--(g2) (g2)--(g3) (g3)--(g4) (g4)--(g5) (g5)--(g0) 
(g0)--(g1);
		\draw[decoration={brace,mirror,raise=20pt},decorate,thick](-0.2,0) -- node[below=20pt] {\footnotesize{$n-1$ node} } (4.2,0);
	\end{tikzpicture}
	}
    \label{eq:5d_affine_A-type_quiver}
\end{align}
with $n$ $\surm(2)$ gauge groups and $\surm(2)_i\times\surm(2)_{i+1}$ bifundamental hypermultiplets. There are $n$ gauge couplings $g_i$, one for each gauge group, and $n$ mass parameters $m_{i,i+1}$, one for each bifundamental hypermultiplet.

\paragraph{5d gauge theory description.}
The prepotential of \eqref{eq:5d_affine_A-type_quiver} is given by
\begin{align}
\label{eq:prepot_affine_A}
\begin{aligned}
 6 \Fcal_{5d} &= \sum_{i=0}^{n-1} \frac{6}{g_i^2} \phi_i^2 +  \sum_{i=0}^{n-1} 8 \phi_i^3 
 - \frac{1}{2} \sum_{i=0}^{n-1} \left( |\phi_i +\phi_{i+1} \pm m_{i,i+1}|^3 +|-\phi_i +\phi_{i+1} \pm m_{i,i+1}|^3   \right) \,,
 \end{aligned}
\end{align}
with the index identification $i+1=n\sim 0$.
One restricts to the dominant Weyl chamber $\phi_i \geq 0$ for each $i$ and chooses a suitable Coulomb branch phase
\begin{align}
\label{eq:phase_choice_affine_A}
    \begin{cases}
     \phi_{i} + \phi_{i+1} \pm m_{i,i+1} &\geq 0 \,, \\
     -\phi_i +\phi_{i+1} +m_{i,i+1} &\geq 0\,, \\
     \phi_i -\phi_{i+1} +m_{i,i+1} &\geq 0\,,
    \end{cases}
    \qquad \text{for } i=0,1,\ldots,n-1 \,.
\end{align}
The prepotential \eqref{eq:prepot_affine_A} in this phase becomes:
\begin{align}
 6 \Fcal_{5d} =  
 \sum_{i=0}^{n-1} 
 \left( 
 \frac{6}{g_i^2} \phi_i^2 
 +  6 \phi_i^3   
 - 3 \phi_i \phi_{i+1}^2 
 - 3 \phi_{i+1} \phi_{i}^2
 -3 m_{i,i+1} (\phi_i^2 - 2\phi_i \phi_{i+1}+\phi_{i+1}^2 )
 - 3 m_{i,i+1}^2 (\phi_i +\phi_{i+1}) 
 \right) 
 \label{eq:prepot_affine_A_phase}
\end{align}
up to constant terms $\mathcal{O}(m_i^3)$.
\paragraph{5-brane web and toric geometry.}
The theory \eqref{eq:5d_affine_A-type_quiver} can be realised as world-volume theory of Type IIB 5-brane web. In the simplest case of an $\widehat{A}_1$ quiver, the periodic 5-brane web is
\begin{align}
\raisebox{-.5\height}{
    \begin{tikzpicture}
     \draw (0,0)--(1,0)--(2,-1)--(3,-1)--(4,0)--(5,0);
     \draw (2,-1)--(2,-2);
     \draw (3,-1)--(3,-2);
     \draw (0,2)--(1,1)--(4,1)--(5,2);
     \draw (1,1)--(1,0);
     \draw (4,1)--(4,0);
     \draw[dashed] (5,2)--(6,2)--(7,1);
     \draw[dashed] (5,0)--(7,0);
     \draw[dashed] (5,2)--(5,3);
     \draw[dashed] (6,2)--(6,3);
     \draw[dashed] (-1,2)--(0,2)--(0,3);
     \draw[dashed] (-1,0)--(0,0);
     \node at (-0.5,2) {$/$};
     \node at (-0.5,0) {$//$};
     \node at (5.5,2) {$/$};
     \node at (5.5,0) {$//$};
     \draw[<->,dashed,red,very thick] (2.25,-1)--(2.25,1);
     \node[red] at (2.75,0) {$2\phi_1$};
     \draw[<->,dashed,red,very thick] (-0.75,0)--(-0.75,2);
     \node[red] at (-0.25,1) {$2\phi_0$};
     \draw[<->,dashed,blue,very thick] (5,2)--(5,1);
     \node[blue] at (6.1,1.5) {$\phi_0-\phi_1+m$};
     \draw[<->,dashed,blue,very thick] (4.5,1)--(4.5,0);
     \node[blue] at (5.6,0.5) {$\phi_0+\phi_1-m$};
     \draw[dotted,blue,thick] (4,1)--(5,1);
     \draw[<->,dashed,blue,very thick] (4,0)--(4,-1);
     \node[blue] at (5.1,-0.5) {$\phi_1-\phi_0+m$};
     \draw[dotted,blue,thick] (3,-1)--(4,-1);
    \end{tikzpicture}
    }
\end{align}
which describes the phase \eqref{eq:phase_choice_affine_A}. Focusing on the part of the 5-brane web that describes $\surm(2) $ with 2 fundamental flavours, one can equivalently consider the dual graph. This yields the fan of $\FF_0$ blown up at two points.
\begin{align}
    \raisebox{-.5\height}{
    \begin{tikzpicture}
      \node[draw,circle,inner sep=0.8pt,fill,black] (origin) at (0,0) {};
      \node[draw,circle,inner sep=0.8pt,fill,black] (ff) at (0,1) {};
      \node[draw,circle,inner sep=0.8pt,fill,black] (f) at (0,-1) {};
      \node[draw,circle,inner sep=0.8pt,fill,black] (h) at (1,0) {};
      \node[draw,circle,inner sep=0.8pt,fill,black] (e) at (-1,0) {};
      \node[draw,circle,inner sep=0.8pt,fill,black] (x1) at (1,-1) {};
      \node[draw,circle,inner sep=0.8pt,fill,black] (x2) at (-1,-1) {};
      \draw[dotted] (ff)--(h)--(x1)--(f)--(x2)--(e)--(ff);
      \draw (origin)--(ff) (origin)--(f) (origin)--(e) (origin)--(x1) (origin)--(x2) (origin)--(h);
     \node at (0,1.35) {${f}$}; 
     \node at (-1.25,0.25) {${e-x_1}$};
     \node at (1.25,0.25) {${h-x_2}$};
     \node at (-1.35,-1.05) {$x_1$};
     \node at (1.35,-1.05) {$x_2$};
     \node at (0,-1.35) {${f-x_1-x_2}$};
    \end{tikzpicture}
    }
\end{align}
We have labelled the rays of the fan by the associated divisor (or curve) classes as follows: $e$ and $h$ are sections, and $f$ the fibre of the ruling. The classes of the exceptional divisors are denoted by $x_i$. As we see below, we can geometrically realise the $\surm(n)$ theories on a $-2$ curve without twist by combining these surfaces $S_i$ as building blocks along a circle. The gluing curves connecting every such surface to the one on the left/right are always given by the toric divisors that point to the left/right, $e-x_1$ and $e-x_2$ in the above notation. This can already be anticipated from the 5-brane web. 
\subsubsection{SU(2)}
\label{sec:SU2_on_2}
To begin with, consider the untwisted circle compactification of a $-2$ curve with an $\surmL(2)$ algebra. In the following, and throughout the paper, we employ the following notation for compact surfaces $S_i$ in the Calabi-Yau geometry. Namely, a surface $S_i$ obtained from blowing up $\mathbb{F}_n$ $k$ times is denoted by $\mathbf{i}^k_n$.
\paragraph{Geometry.}
Following \cite{Bhardwaj:2019fzv}, the geometry for the circle compactified theory reads
\begin{align}
\raisebox{-.5\height}{
 \begin{tikzpicture}
  \node (v0) at (0,0) {$\mathbf{0}_0^{4}$ };  
  \node (v1) at (6,0) {$\mathbf{1}_{2}$};  
  \node (vaux) at (3,0) {$\scriptscriptstyle{2}$};
  \draw  (v0) edge (vaux);
  \draw  (vaux) edge (v1);
%
  \node at (1.5,0.30) {$\substack{e_0 \\e_0 -\sum_{i=1}^4 x_i}$};
  \node at (5,0.30) {$\substack{e_1 \\ h_1  }$};
 \end{tikzpicture}
 }
  \label{eq:SU2_on_2}
\end{align} 
First, since the $e_0 -x_i$ curves have self-intersection $(e_0 -x_i)^2=-1$, due to $\FF_0$, one can flip any $e_0-x_i$. So, one might choose 
\begin{align}
\raisebox{-.5\height}{
 \begin{tikzpicture}
  \node (v0) at (0,0) {$\mathbf{0}_0^{2}$ };  
  \node (v1) at (6,0) {$\mathbf{1}_{2}^{2}$};  
  \node (vaux) at (3,0) {$\scriptscriptstyle{2}$};
  \draw  (v0) edge (vaux);
  \draw  (vaux) edge (v1);
%
  \node at (1.5,0.30) {$\substack{e_0 \\e_0 -\sum_{i=3}^4 x_i}$};
  \node at (5,0.30) {$\substack{e_1 \\ h_1  -\sum_{i=1}^2 x_i }$};
 \end{tikzpicture}
 }
\end{align} 
where $x_{1,2}$ have been flopped.
Now, by virtue of inverse of the isomorphism \eqref{eq:iso_Fn_Fn+1} for $x_1$ one finds
\begin{align}
\raisebox{-.5\height}{
 \begin{tikzpicture}
  \node (v0) at (0,0) {$\mathbf{0}_0^{2}$ };  
  \node (v1) at (6,0) {$\mathbf{1}_{1}^{2}$};  
  \node (vaux) at (3,0) {$\scriptscriptstyle{2}$};
  \draw  (v0) edge (vaux);
  \draw  (vaux) edge (v1);
%
  \node at (1.5,0.30) {$\substack{e_0 \\e_0 -\sum_{i=3}^4 x_i}$};
  \node at (5,0.30) {$\substack{e_1-x_1 \\ h_1  - x_2 }$};
 \end{tikzpicture}
 }
\end{align} 
where $h_1$ is inside $\FF_1^2$. Repeating the isomorphism \eqref{eq:iso_Fn_Fn+1} for $x_2$ yields
\begin{align}
\raisebox{-.5\height}{
 \begin{tikzpicture}
  \node (v0) at (0,0) {$\mathbf{0}_0^{2}$ };  
  \node (v1) at (6,0) {$\mathbf{1}_{0}^{2}$};  
  \node (vaux) at (3,0) {$\scriptscriptstyle{2}$};
  \draw  (v0) edge (vaux);
  \draw  (vaux) edge (v1);
%
  \node at (1.5,0.30) {$\substack{e_0 \\e_0 -\sum_{i=3}^4 x_i}$};
  \node at (5,0.30) {$\substack{e_1-\sum_{i=1}^2 x_i \\ e_1   }$};
 \end{tikzpicture}
 }
\end{align} 
Finally, one splits the blowups symmetrically 
\begin{align}
\raisebox{-.5\height}{
 \begin{tikzpicture}
  \node (v0) at (0,0) {$\mathbf{0}_0^{2}$ };  
  \node (v1) at (6,0) {$\mathbf{1}_{0}^{2}$};  
  \node (vaux) at (3,0) {$\scriptscriptstyle{2}$};
  \draw  (v0) edge (vaux);
  \draw  (vaux) edge (v1);
%
  \node at (1.5,0.30) {$\substack{e_0 -x_1 \\e_0 - x_4}$};
  \node at (5,0.30) {$\substack{e_1- x_2 \\ e_1 -x_3   }$};
 \end{tikzpicture}
 }
\label{eq:SU2_on_2_final}
\end{align}
which is possible via further flop transitions.
\paragraph{Consistency of geometry.}
Before exploring the physics associated to the geometry \eqref{eq:SU2_on_2_final}, one verifies overall consistency.
\begin{compactitem}
 \item Firstly, one verifies the consistency condition \eqref{eq:consistency} on gluing curves of \eqref{eq:SU2_on_2_final}. With $J_\phi = \sum_i \phi_i S_i$, the truncated K\"ahler form, one obtains:
\begin{subequations}
\begin{align}
 \text{1st component:}\qquad S_0|_{S_1} \cdot J_\phi = S_1|_{S_0} \cdot J_\phi &= \phi_0 + \phi_1 \,, \\
 \text{2nd component:}\qquad S_0|_{S_1} \cdot J_\phi = S_1|_{S_0} \cdot J_\phi &= \phi_0 + \phi_1 \,.
\end{align}
\end{subequations}
\item Secondly, one verifies the Calabi-Yau condition \eqref{eq:CYcondition_gluing_curves} 
\begin{subequations}
\begin{align}
\text{1st component:}\qquad  (S_0|_{S_1})^2 + (S_1|_{S_0})^2  &=-2 \;,\\
 \text{2nd component:}\qquad  (S_0|_{S_1})^2 + (S_1|_{S_0})^2  &=-2 \;, 
\end{align}
\end{subequations}
which is consistent for genus $g=0$.
\item Lastly, the fibre intersections are computed to be
\begin{align}
-
\begin{pmatrix}
 f_0 \cdot K_{S_0}
 & f_0 \cdot S_1|_{S_0} \\
 f_1 \cdot S_0|_{S_1} &
 f_1 \cdot K_{S_1}
\end{pmatrix}
= 
\begin{pmatrix}
  2 & -2 \\ -2 & 2
\end{pmatrix}
 = C_{\widehat{A}^{(1)}_1}  \;,
\end{align}
which is consistent with 6d origin of an $\surmL(2)$ gauge algebra.
\end{compactitem}
\paragraph{Prepotential via intersection numbers.}
The triple intersection numbers for the compact surfaces in \eqref{eq:SU2_on_2_final} give rise to the following cubic part of the prepotential:
\begin{align}
 6\Fcal_{\trun} \equiv J_\phi^3   = 6 \phi_0^3-6 \phi_1 \phi_0^2-6 \phi_1^2 \phi_0+6 \phi_1^3 \,.
 \label{eq:Prepot_SU2_on_2}
\end{align}
Besides the compact surfaces, one may also add non-compact ones that introduce deformation parameters. For this, one may parametrise the K\"ahler form as follows
\begin{subequations}
\label{eq:Kahler_SU2_on_2_ansatz}
\begin{align}
    J|_{S_0} &= \phi_0 K_{S_0} +\phi_1 S_1|_{S_0}  
    + a_0 f_0 + b_0 e_0 + \sum_{i=0}^1 M_{i} x_i \,,\\
    J|_{S_1} &= \phi_0 S_0|_{S_1} + \phi_1 K_{S_1}   
    + a_1 f_1 + b_1 e_1 + \sum_{i=2}^3 M_{i} x_i \,.
\end{align}
\end{subequations}
recalling that $x_0\equiv x_4$ and $i=4\sim0$.
The introduced parameters can be determined via the following two requirements:
\begin{compactitem}
 \item The volume of the blowup curves match the physical mass terms given in \eqref{eq:prepot_affine_A}. Based on \eqref{eq:Kahler_SU2_on_2_ansatz}, one computes 
 \begin{subequations}
 \label{eq:hyper1_SU2_on_2}
 \begin{alignat}{2}
\vol(x_1) &= \phi_0 - \phi_1 + M_{1} 
& &\stackrel{!}{=} \phi_0 - \phi_1 + m_{0,1} \;,\\
\vol(e_0 - x_1) &= \phi_0 + \phi_1 - M_{1} - a_0
& \quad &\stackrel{!}{=} \phi_0 + \phi_1 - m_{0,1} \;,\\
\vol(x_2) &= - \phi_0 + \phi_1 +M_{2}
& &\stackrel{!}{=} -\phi_0 + \phi_1 + m_{0,1} \;,\\
\vol(e_1 - x_2) &= \phi_0 + \phi_1 - M_{2} - a_1 
& &\stackrel{!}{=} \phi_0 + \phi_1 - m_{0,1} \;,
 \end{alignat}
 \end{subequations}
 which is identified as the contributions of the first bifundamental with mass parameter $m_{0,1}$.
 The second bifundamental, with parameter $m_{1,0}$, contributes as 
 \begin{subequations}
 \label{eq:hyper2_SU2_on_2}
 \begin{alignat}{2}
\vol(x_4) &= \phi_0 - \phi_1 + M_{4} 
& &\stackrel{!}{=} \phi_0 - \phi_1 + m_{1,0} \;,\\
\vol(e_0 - x_4) &= \phi_0 + \phi_1 - M_{4} - a_0
& \quad &\stackrel{!}{=} \phi_0 + \phi_1 - m_{1,0} \;,\\
\vol(x_3) &= - \phi_0 + \phi_1 +M_{3}
& &\stackrel{!}{=} -\phi_0 + \phi_1 + m_{1,0} \;,\\
\vol(e_1 - x_3) &= \phi_0 + \phi_1 - M_{3} - a_1 
& &\stackrel{!}{=} \phi_0 + \phi_1 - m_{1,0} \;.
 \end{alignat}
\end{subequations} 
The set of linear equations \eqref{eq:hyper1_SU2_on_2}, \eqref{eq:hyper2_SU2_on_2} are straightforwardly solved by
\begin{align}
    M_{1}=M_{2} = m_{0,1}
    \;,\quad 
    M_{3}=M_{4} = m_{1,0}
    \;,\quad 
    a_0=a_1=0 \,.
\end{align}
 \item The effective gauge coupling \eqref{eq:effectice_coupling_geom} is geometrically given by
 \begin{align}
 \begin{aligned}
 (\tau_{\eff}) &=
     \begin{pmatrix}
      J|_{S_0} \cdot K_{S_0} &  J|_{S_0} \cdot S_1|_{S_0} \\
       J|_{S_1} \cdot S_0|_{S_1} &  J|_{S_1} \cdot K_{S_1}
     \end{pmatrix} \\
     &= 
     \begin{pmatrix}
      6 \phi_0 - 2 \phi_1 - (m_{0,1} +m_{1,0}) - 2 b_0 & -2(\phi_0+\phi_1) + m_{0,1}+m_{1,0} \\
       -2(\phi_0+\phi_1) + m_{0,1}+m_{1,0} & 6 \phi_1 - 2 \phi_0 - (m_{0,1} +m_{1,0}) - 2 b_1
     \end{pmatrix}\\
     &\stackrel{!}{=}\left( \frac{\partial^2 \Fcal_{5d}}{\partial \phi_i \partial \phi_j }\right) 
     \end{aligned}
 \end{align}
 such that comparing to the field theory expectation \eqref{eq:effective_coupling} imposes
 \begin{align}
     b_0 = - \frac{1}{g_0^2} \;,\quad 
      b_1 = - \frac{1}{g_1^2} \;.
 \end{align}
\end{compactitem}
Therefore, the K\"ahler form may be written as 
\begin{subequations}
\begin{align}
    J &= - \sum_{i=0}^1 \frac{1}{g_i^2} F_i  
    + \sum_{i=0}^1 \phi_i S_i
    + \sum_{i=0}^1 m_{i,i+1} N_{i,i+1} \;,
\end{align}
where the non-compact surfaces $F_i$, $N_{i,i+1}$ restrict to the compact surfaces as follows:
\begin{align}
    F_i|_{S_j} = \delta_{ij} e_i
    \;, \qquad 
    \begin{cases}
    N_{0,1}|_{S_0} = x_1 \\
    N_{0,1}|_{S_1} = x_2 \\
    \end{cases}
    \;, \qquad 
    \begin{cases}
    N_{1,0}|_{S_0} = x_4 \\
    N_{1,0}|_{S_1} = x_3 \\
    \end{cases}
    \,.
\end{align}
\label{eq:solution_SU2_on_2}
\end{subequations}
Based on \eqref{eq:solution_SU2_on_2} one computes the volume of the elliptic fibre
\begin{align}
    \vol(f_{\mathrm{ell}}) 
    = \vol(f_0)+\vol(f_1)
    = \frac{1}{g_0^2}+ \frac{1}{g_1^2} = \tau \; ,
\end{align}
which equals the modular parameter $\tau$.

Finally, the triple intersection numbers of the compact surfaces with the non-compact ones are
\begin{subequations}
\label{eq:intersections_new_SU2_on_2}
\begin{align}
    F_i \cdot S_j \cdot S_k &= 
    \begin{cases}
    F_i|_{S_j} \cdot K_{S_j} = -2 \delta_{i,j} \,, &\text{for } j=k  \,, \\
    F_i|_{S_j} \cdot S_k|_{S_j} = 0 \,,  &\text{for } j\neq k \,, 
\end{cases}
\\
    F_i \cdot F_j \cdot S_k &=0\,,  \qquad \forall i,j,k \,, \\
    N_{i,i+1} \cdot S_j \cdot S_k
    &=
    \begin{cases}
    N_{i,i+1}|_{S_j} \cdot K_{S_j} = 
    -1\cdot(\delta_{i,j} + \delta_{i+1,j}) \,,  &\text{for } j=k \,, \\
   N_{i,i+1}|_{S_j} \cdot S_{j+1}|_{S_j} = \delta_{i,j} + \delta_{i+1,j} \,,  &\text{for } k= j+1\sim j-1\,, 
    \end{cases}
    \\
    N_{i,i+1} \cdot N_{j,j+1} \cdot S_k &=
    N_{i,i+1}|_{S_k} \cdot N_{j,j+1}|_{S_k}
    = \delta_{i,k}\delta_{j,k} +\delta_{i+1,k}\delta_{j+1,k} 
    \;,\\
    N_{i,i+1} \cdot F_j \cdot S_k &=0 \,,  \qquad \forall i,j,k \;.
\end{align}
\end{subequations}
These intersection numbers give rise to the geometric prepotential $6 \Fcal_{\geom} = J^3$ which matches \eqref{eq:prepot_affine_A_phase}. Therefore, the geometry \eqref{eq:SU2_on_2_final} describes the $\widehat{A}_1$ KK-theory \eqref{eq:5d_affine_A-type_quiver} in the phase \eqref{eq:phase_choice_affine_A}.
%
%
\subsubsection{SU(3)}
\label{sec:SU3_on_2}
Consider a $-2$ curve with an $\surmL(3)$ gauge algebra and its untwisted circle compactification.
\paragraph{Geometry.}
Following \cite{Bhardwaj:2019fzv}, the relevant geometry is given by
\begin{align}
\raisebox{-.5\height}{
 \begin{tikzpicture}
  \node (v0) at (0,0) {$\mathbf{0}_0^{6}$ };  
  \node (v1) at (6,0) {$\mathbf{1}_{2}$};  
  \node (v2) at (3,-2) {$\mathbf{2}_{4}$};
  \draw  (v0) edge (v1);
  \draw  (v1) edge (v2);
  \draw  (v0) edge (v2);
%
  \node at (2,0.25) {$\scriptstyle{e_0}$};
  \node at (5,0.25) {$\scriptstyle{e_1} $};
  \node at (5.25,-0.85) {$\scriptstyle{h_1}$};
  \node at (4,-1.75) {$\scriptstyle{e_2}$};
  \node at (0.25,-0.85) {$\scriptstyle{e_0-\sum_{i=1}^6 x_i}$};
  \node at (2,-1.75) {$\scriptstyle{h_2}$};
 \end{tikzpicture}
 }
   \label{eq:SU3_on_2}
\end{align}
In $S_0=\FF_0^6$, any of the $e_0 - x_i$ curves has self-intersection $(e_0 - x_i)^2=-1$; hence, can be flopped. So, one may choose
\begin{align}
\raisebox{-.5\height}{
 \begin{tikzpicture}
  \node (v0) at (0,0) {$\mathbf{0}_0^{2}$ };  
  \node (v1) at (6,0) {$\mathbf{1}_{2}$};  
  \node (v2) at (3,-2) {$\mathbf{2}_{4}^4$};
  \draw  (v0) edge (v1);
  \draw  (v1) edge (v2);
  \draw  (v0) edge (v2);
%
  \node at (2,0.25) {$\scriptstyle{e_0}$};
  \node at (5,0.25) {$\scriptstyle{e_1} $};
  \node at (5.25,-0.85) {$\scriptstyle{h_1}$};
  \node at (4,-1.75) {$\scriptstyle{e_2}$};
  \node at (0.25,-0.85) {$\scriptstyle{e_0- \sum_{i=5}^6 x_i}$};
  \node at (1.5,-1.75) {$\scriptstyle{h_2-\sum_{i=1}^4 x_i}$};
 \end{tikzpicture}
 }
\end{align}
i.e. the blowups $x_{1,2,3,4}$ have been flopped.
Applying the isomorphism \eqref{eq:iso_Fn_Fn+1} separately for $x_{1,2,3,4}$ yields
\begin{align}
\raisebox{-.5\height}{
 \begin{tikzpicture}
  \node (v0) at (0,0) {$\mathbf{0}_0^{2}$ };  
  \node (v1) at (6,0) {$\mathbf{1}_{2}$};  
  \node (v2) at (3,-2) {$\mathbf{2}_{0}^4$};
  \draw  (v0) edge (v1);
  \draw  (v1) edge (v2);
  \draw  (v0) edge (v2);
%
  \node at (2,0.25) {$\scriptstyle{e_0}$};
  \node at (5,0.25) {$\scriptstyle{e_1 }$};
  \node at (5.25,-0.85) {$\scriptstyle{h_1}$};
  \node at (4.5,-1.75) {$\scriptstyle{e_2-\sum_{i=1}^4 x_i}$};
  \node at (0.25,-0.85) {$\scriptstyle{e_0- \sum_{i=5}^6 x_i}$};
  \node at (1.5,-1.75) {$\scriptstyle{h_2=e_2}$};
 \end{tikzpicture}
 }
\end{align}
In $\FF_0^4$, any of the $e_2 -x_i$ curves has self-intersection $(e_2 -x_i)^2=-1$; hence, can be flopped. One may choose
\begin{align}
\raisebox{-.5\height}{
 \begin{tikzpicture}
  \node (v0) at (0,0) {$\mathbf{0}_0^{2}$ };  
  \node (v1) at (6,0) {$\mathbf{1}_{2}^2$};  
  \node (v2) at (3,-2) {$\mathbf{2}_{0}^2$};
  \draw  (v0) edge (v1);
  \draw  (v1) edge (v2);
  \draw  (v0) edge (v2);
%
  \node at (2,0.25) {$\scriptstyle{e_0}$};
  \node at (5,0.25) {$\scriptstyle{e_1 }$};
  \node at (6,-0.85) {$\scriptstyle{h_1-\sum_{i=1}^2 x_i}$};
  \node at (4.5,-1.75) {$\scriptstyle{e_2-\sum_{i=3}^4 x_i}$};
  \node at (0.25,-0.85) {$\scriptstyle{e_0- \sum_{i=5}^6 x_i}$};
  \node at (2,-1.75) {$\scriptstyle{e_2}$};
 \end{tikzpicture}
 }
\end{align}
i.e.\ $x_{1,2}$ have been flopped.
Applying the isomorphism \eqref{eq:iso_Fn_Fn+1} separately for $x_1$ and $x_2$ yields
\begin{align}
\raisebox{-.5\height}{
 \begin{tikzpicture}
  \node (v0) at (0,0) {$\mathbf{0}_0^{2}$ };  
  \node (v1) at (6,0) {$\mathbf{1}_{0}^2$};  
  \node (v2) at (3,-2) {$\mathbf{2}_{0}^2$};
  \draw  (v0) edge (v1);
  \draw  (v1) edge (v2);
  \draw  (v0) edge (v2);
%
  \node at (2,0.25) {$\scriptstyle{e_0}$};
  \node at (4.5,0.25) {$\scriptstyle{e_1 - \sum_{i=1}^2 x_i}$};
  \node at (5.5,-0.85) {$\scriptstyle{h_1=e_1}$};
  \node at (4.5,-1.75) {$\scriptstyle{e_2-\sum_{i=3}^4 x_i}$};
  \node at (0.25,-0.85) {$\scriptstyle{e_0- \sum_{i=5}^6 x_i}$};
  \node at (2,-1.75) {$\scriptstyle{e_2}$};
 \end{tikzpicture}
 }
\end{align}
Lastly, the blowups can be distributed symmetrically to yield
\begin{align}
\raisebox{-.5\height}{
 \begin{tikzpicture}
  \node (v0) at (0,0) {$\mathbf{0}_0^{2}$ };  
  \node (v1) at (6,0) {$\mathbf{1}_{0}^2$};  
  \node (v2) at (3,-2) {$\mathbf{2}_{0}^2$};
  \draw  (v0) edge (v1);
  \draw  (v1) edge (v2);
  \draw  (v0) edge (v2);
%
  \node at (2,0.25) {$\scriptstyle{e_0-x_1}$};
  \node at (4.5,0.25) {$\scriptstyle{e_1 - x_2}$};
  \node at (5.5,-0.85) {$\scriptstyle{e_1-x_3}$};
  \node at (4.5,-1.75) {$\scriptstyle{e_2- x_4}$};
  \node at (0.25,-0.85) {$\scriptstyle{e_0- x_6}$};
  \node at (1.75,-1.75) {$\scriptstyle{e_2-x_5}$};
 \end{tikzpicture}
 }
 \label{eq:SU3_on_2_final}
\end{align}
which can be achieved via further flop transitions.
\paragraph{Consistency of geometry.}
Next, the consistency of the geometry \eqref{eq:SU3_on_2_final} is verified.
\begin{compactitem}
 \item Firstly, the consistency condition \eqref{eq:consistency} for the gluing curves in \eqref{eq:SU3_on_2_final} reads
\begin{subequations}
\begin{align}
 S_0|_{S_1} \cdot J_\phi = S_1|_{S_0} \cdot J_\phi &= \phi_0 + \phi_1 \;,\\
 S_0|_{S_2} \cdot J_\phi = S_2|_{S_0} \cdot J_\phi &=  \phi_0 + \phi_2 \;,\\
 S_1|_{S_2} \cdot J_\phi = S_2|_{S_1} \cdot J_\phi &= \phi_1 + \phi_2 \;,
\end{align}
\end{subequations}
where $J_\phi= \sum_i \phi_i S_i$ denotes the truncated K\"ahler form for \eqref{eq:SU3_on_2_final}.
\item Secondly, one verifies the Calabi-Yau condition \eqref{eq:CYcondition_gluing_curves} and finds
\begin{subequations}
\begin{align}
 (S_0|_{S_1})^2 + (S_1|_{S_0})^2  &=-2 \;,\\
 (S_0|_{S_2})^2 + (S_2|_{S_0})^2  &= -2\;, \\
 (S_1|_{S_2})^2 + (S_2|_{S_1})^2  &=-2 \;,
\end{align}
\end{subequations}
which is consistent for genus $g=0$.
\item Lastly, the fibre intersections read
\begin{align}
-
\begin{pmatrix}
 f_0 \cdot K_{S_0} & f_0 \cdot S_1|_{S_0}  & f_0 \cdot S_2|_{S_0} \\
 f_1 \cdot S_0|_{S_1} &  f_1 \cdot K_{S_1} & f_1 \cdot S_2|_{S_1} \\
 f_2 \cdot S_0|_{S_2} & f_2 \cdot S_1|_{S_2} &  f_2 \cdot K_{S_2} 
\end{pmatrix}
= 
\begin{pmatrix}
  2 & -1 & -1 \\ -1 & 2 & -1 \\ -1 & -1& 2
\end{pmatrix}
 = C_{\widehat{A}^{(1)}_2} \;,
\end{align}
which is consistent with 6d origin of an $\surmL(3)$ gauge algebra.
\end{compactitem}
\paragraph{Prepotential via intersection numbers.}
For the geometry \eqref{eq:SU3_on_2_final}, one finds the following cubic part of the prepotential from the intersection numbers of the compact surface:
\begin{align}
 6\Fcal_{\trun} \equiv J_\phi^3 = 6 \phi_0^3-3 \phi_1 \phi_0^2-3 \phi_2 \phi_0^2-3 \phi_1^2 \phi_0-3 \phi_2^2 \phi_0+6 \phi_1^3+6 \phi_2^3-3 \phi_1 \phi_2^2-3 \phi_1^2 \phi_2 \,.
 \label{eq:Prepot_SU3_on_2}
\end{align}
Besides the compact surfaces, one may also add non-compact ones that introduce deformation parameters. For this, one may parametrise the K\"ahler form as follows
\begin{subequations}
\label{eq:Kahler_SU3_on_2_ansatz}
\begin{align}
    J|_{S_0} &= \phi_0 K_{S_0} +\phi_1 S_1|_{S_0}  +\phi_2 S_2|_{S_0}
    + a_0 f_0 + b_0 e_0 + \sum_{i=0}^1 M_{i} x_i \,,\\
    J|_{S_1} &= \phi_0 S_0|_{S_1} + \phi_1 K_{S_1}  +  \phi_2 S_2|_{S_1} 
    + a_1 f_1 + b_1 e_1 + \sum_{i=2}^3 M_{i} x_i \,, \\
    J|_{S_2} &= \phi_0 S_0|_{S_2} + \phi_1 S_1|_{S_2} + \phi_2 K_{S_2}   
    + a_2 f_2 + b_2 e_2 + \sum_{i=4}^5 M_{i} x_i \,,
\end{align}
\end{subequations}
recalling that $x_0 \equiv x_6$ and $i=6\sim0$.
The additional parameters are determined by two requirements:
\begin{compactitem}
 \item The volume of the blowup curves match the physical mass terms given in \eqref{eq:prepot_affine_A}. Based on \eqref{eq:Kahler_SU3_on_2_ansatz}, one computes 
 \begin{subequations}
 \label{eq:hyper1_SU3_on_2}
 \begin{alignat}{2}
\vol(x_6) &= \phi_0 - \phi_2 + M_{6} 
& &\stackrel{!}{=} \phi_0 - \phi_2 + m_{2,0} \;,\\
\vol(e_0 - x_6) &= \phi_0 + \phi_2 - M_{6} - a_0
& \quad &\stackrel{!}{=} \phi_0 + \phi_2 - m_{2,0} \;,\\
\vol(x_5) &= - \phi_0 + \phi_2 +M_{5}
& &\stackrel{!}{=} -\phi_0 + \phi_2 + m_{2,0} \;,\\
\vol(e_2 - x_5) &= \phi_0 + \phi_2 - M_{5} - a_2 
& &\stackrel{!}{=} \phi_0 + \phi_2 - m_{2,0} \;,
 \end{alignat}
 \end{subequations}
 which is identified as the contributions of the $\surm(2)_2\times \surm(2)_0$ bifundamental with mass parameter $m_{2,0}$.
 The $\surm(2)_0\times\surm(2)_1$ bifundamental hypermultiplets, with mass parameters $m_{0,1}$, contributes as 
 \begin{subequations}
 \label{eq:hyper2_SU3_on_2}
 \begin{alignat}{2}
\vol(x_1) &= \phi_0 - \phi_1 + M_{1} 
& &\stackrel{!}{=} \phi_0 - \phi_1 + m_{0,1} \;,\\
\vol(e_0 - x_1) &= \phi_0 + \phi_1 - M_{1} - a_0
& \quad &\stackrel{!}{=} \phi_0 + \phi_1 - m_{0,1} \;,\\
\vol(x_2) &= - \phi_0 + \phi_1 +M_{2}
& &\stackrel{!}{=} -\phi_0 + \phi_1 + m_{0,1} \;,\\
\vol(e_1 - x_2) &= \phi_0 + \phi_1 - M_{2} - a_1 
& &\stackrel{!}{=} \phi_0 + \phi_1 - m_{0,1} \;.
 \end{alignat}
\end{subequations} 
The remaining blowups have volumes given by
\begin{subequations}
\label{eq:hyper3_SU3_on_2}
 \begin{alignat}{2}
\vol(x_3) &= \phi_1 - \phi_2 + M_{3} 
& &\stackrel{!}{=} \phi_1 - \phi_2 + m_{1,2} \;,\\
\vol(e_1 - x_3) &= \phi_1 + \phi_2 - M_{3} - a_1
& \quad &\stackrel{!}{=} \phi_1 + \phi_2 - m_{1,2} \;,\\
\vol(x_4) &= - \phi_1 + \phi_2 +M_{4}
& &\stackrel{!}{=} -\phi_1 + \phi_2 + m_{1,2} \;,\\
\vol(e_2 - x_4) &= \phi_1 + \phi_2 - M_{4} - a_2 
& &\stackrel{!}{=} \phi_1 + \phi_2 - m_{1,2} \;,
 \end{alignat}
\end{subequations} 
which identifies them as $\surm(2)_1\times\surm(2)_2$ bifundamental hypermultiplet with mass parameter $m_{1,2}$. The set of linear equations \eqref{eq:hyper1_SU3_on_2} -- \eqref{eq:hyper3_SU3_on_2} are straightforwardly solved by
\begin{align}
    M_{5}=M_{6} = m_{2,0}
    \;,\quad 
    M_{1}=M_{2} = m_{0,1}
    \;,\quad 
    M_{3}=M_{4} = m_{1,2}
    \;,\quad 
    a_0=a_1=a_2=0 \,.
\end{align}
 \item The effective gauge coupling \eqref{eq:effectice_coupling_geom} is geometrically given by
 \begin{align}
 \begin{aligned}
 (\tau_{\eff}) &=
     \begin{pmatrix}
     J|_{S_0} \cdot K_{S_0} &  J|_{S_0} \cdot S_1|_{S_0} &  J|_{S_0} \cdot S_2|_{S_0} \\
       J|_{S_1} \cdot S_0|_{S_1} &  J|_{S_1} \cdot K_{S_1} &  J|_{S_1} \cdot S_2|_{S_1} \\
       J|_{S_2} \cdot S_0|_{S_2}  &  J|_{S_2} \cdot S_1|_{S_2}&  J|_{S_2} \cdot K_{S_2}
     \end{pmatrix} 
     \stackrel{!}{=} \left(\frac{\partial^2 \Fcal_{5d}}{\partial\phi_i \partial\phi_j} \right)\\
       \text{with} \qquad
       J|_{S_0} \cdot K_{S_0} &= 6 \phi_0 -  \phi_1-  \phi_2 - (m_{0,1} +m_{2,0}) - 2 b_0\,,
      \\
       J|_{S_1} \cdot K_{S_1} &=6 \phi_1 -  \phi_0 -  \phi_2 - (m_{0,1} +m_{1,2}) - 2 b_1 \,,
      \\
      J|_{S_2} \cdot K_{S_2} &= 6 \phi_2 -  \phi_0 -  \phi_1 - (m_{1,2} +m_{2,0}) - 2 b_2 \,,
      \\
       J|_{S_0} \cdot S_1|_{S_0} &=  J|_{S_1} \cdot S_0|_{S_1} = -(\phi_0+\phi_1) + m_{0,1} \,,
      \\
       J|_{S_0} \cdot S_2|_{S_0} &= J|_{S_2} \cdot S_0|_{S_2} = -(\phi_0+\phi_2) + m_{2,0} \,,
      \\
       J|_{S_1} \cdot S_2|_{S_1} &= J|_{S_2} \cdot S_1|_{S_2} =  -(\phi_1+\phi_2) + m_{1,2} \,,
     \end{aligned}
 \end{align}
 such that comparing to the field theory expectation \eqref{eq:effective_coupling} imposes
 \begin{align}
     b_0 = - \frac{1}{g_0^2} \;,\quad 
      b_1 = - \frac{1}{g_1^2} \;,\quad 
      b_2 = - \frac{1}{g_2^2}\;.
 \end{align}
\end{compactitem}
Therefore, the K\"ahler form may be written as 
\begin{subequations}
\label{eq:solution_SU3_on_2}
\begin{align}
    J &= - \sum_{i=0}^2 \frac{1}{g_i^2} F_i  
    + \sum_{i=0}^2 \phi_i S_i
    + \sum_{i=0}^2 m_{i,i+1} N_{i,i+1} \;,
\end{align}
where the parameter $i+1=3\sim0$ is periodically identified.
The non-compact surfaces $F_i$, $N_{i,i+1}$ restrict to the compact surfaces as follows:
\begin{align}
    F_i|_{S_j} = \delta_{ij} e_i
    \;, \qquad 
    N_{i,i+1}|_{S_j}  = \delta_{i,j}\ x_{2i+1} + \delta_{i+1,j}\ x_{2i+2}
    \,.
\end{align}
\end{subequations}
Based on \eqref{eq:solution_SU3_on_2}, the volume of the elliptic fibre is given by
\begin{align}
    \vol(f_{\mathrm{ell}}) 
    = \sum_{i=0}^2 \vol(f_i)
    = \frac{1}{g_0^2}+ \frac{1}{g_1^2}+ \frac{1}{g_2^2} = \tau \; ,
\end{align}
which equals the modular parameter $\tau$.

Lastly, the triple intersection numbers of the compact surfaces with the non-compact ones are
\begin{subequations}
\label{eq:intersections_new_SU3_on_2}
\begin{align}
    F_i \cdot S_j \cdot S_k &= 
    \begin{cases}
    F_i|_{S_j} \cdot K_{S_j} = -2 \delta_{i,j}\,, &\text{for } j=k  \,,\\
    F_i|_{S_j} \cdot S_k|_{S_j} = 0 \,, &\text{for } j\neq k \,,
\end{cases}
 \\
    F_i \cdot F_j \cdot S_k &=0 \,, \qquad \forall i,j,k \,,\\
    N_{i,i+1} \cdot S_j \cdot S_k
    &=
    \begin{cases}
    N_{i,i+1}|_{S_j} \cdot K_{S_j} = 
    -1\cdot(\delta_{i,j} + \delta_{i+1,j})  \,, &\text{for } j=k  \,,\\
   N_{i,i+1}|_{S_j} \cdot S_{j+1}|_{S_j} = \delta_{i,j} \,, &\text{for } k= j+1 \,,\\
     N_{i,i+1}|_{S_j} \cdot S_{j-1}|_{S_j} =\delta_{i+1,j} \,, &\text{for } k= j-1 \,, \\
    N_{i,i+1}|_{S_j} \cdot S_{k}|_{S_j} = 0 \,, &\text{for } k\nin \{j,j\pm1 \} \,,
    \end{cases}
    \\
    N_{i,i+1} \cdot N_{j,j+1} \cdot S_k &=
    N_{i,i+1}|_{S_k} \cdot N_{j,j+1}|_{S_k}
    = \delta_{i,k}\delta_{j,k} +\delta_{i+1,k}\delta_{j+1,k}  
    \,,\\
    N_{i,i+1} \cdot F_j \cdot S_k \,,&=0 \qquad \forall i,j,k \,.
\end{align}
\end{subequations}
These intersection numbers give rise to the geometric prepotential $6 \Fcal_{\geom} =  J^3$ which matches \eqref{eq:prepot_affine_A_phase}. Therefore, the geometry \eqref{eq:SU3_on_2_final} describes the $\widehat{A}_2$ KK-theory \eqref{eq:5d_affine_A-type_quiver} in the phase \eqref{eq:phase_choice_affine_A}.
%
\subsubsection{\texorpdfstring{$\surm(\mathrm{even})$}{SU(even)}}
\label{sec:SUeven_on_2}
For the general case, one proceeds in an analogous manner to Sections \ref{sec:SU2_on_2}--\ref{sec:SU3_on_2}.
\paragraph{Geometry.}
The geometric description is based on \cite{Bhardwaj:2019fzv}  
\begin{align}
\raisebox{-.5\height}{
 \begin{tikzpicture}
  \node (v0) at (0,0) {$\mathbf{0}_0^{4n}$};  
  \node (v1) at (3,2){$\mathbf{1}_{2}$};  
  \node (v2) at (6,2) {$\mathbf{2}_4$};
  \node (v3) at (10,2) {$\mathbf{(n{-}1)}_{2n{-}2}$};
  \node (v4) at (13,0) {$\mathbf{n}_{2n}$};
  \node (v5) at (10,-2) {$\mathbf{(n{+}1)}_{2n{+}2}$};
  \node (v6) at (6,-2) {$\mathbf{(2n{-}2)}_{4n{-}4}$};
  \node (v7) at (3,-2) {$\mathbf{(2n{-}1)}_{4n{-}2}$};
%
  \node (g1) at (8,2) {$\cdots$};
  \node (g2) at (8,-2) {$\cdots$};
    \draw  (v0) edge (v1);
    \draw  (v1) edge (v2);
    \draw  (v3) edge (v4);
    \draw  (v4) edge (v5);
    \draw  (v6) edge (v7);
    \draw  (v7) edge (v0);
    \draw (v2) edge (g1);
    \draw (g1) edge (v3);
    \draw (v6) edge (g2);
    \draw (g2) edge (v5);
%
  \node at (0.5,0.75) {$\scriptstyle{e_0}$};
  \node at (2,1.75) {$\scriptstyle{e_1}$};
  \node at (3.5,2.25) {$\scriptstyle{ h_1}$};
  \node at (5.5,2.25) {$\scriptstyle{ e_2}$};
  \node at (6.5,2.25) {$\scriptstyle{ h_2}$};
  \node at (8.75,2.25) {$\scriptstyle{e_{n{-}1}}$};
  \node at (11.25,1.55) {$\scriptstyle{h_{n{-}1}}$};
  \node at (12.5,0.75) {$\scriptstyle{e_n}$};
  \node at (12.5,-0.85) {$\scriptstyle{h_n}$};  
  \node at (8.75,-2.25) {$\scriptstyle{h_{n{+}1}}$};
  \node at (11.25,-1.55) {$\scriptstyle{e_{n{+}1}}$};
  \node at (7.5,-2.25) {$\scriptstyle{ e_{2n{-}2}}$};
  \node at (5,-2.35) {$\scriptstyle{ h_{2n{-}2}}$};
  \node at (4,-2.35) {$\scriptstyle{ e_{2n{-}1}}$};
  \node at (0.25,-0.85) {$\scriptstyle{ e_0-\sum_{i=1}^{4n} x_i}$};
  \node at (1.75,-1.5) {$\scriptstyle{ h_{2n{-}1}}$};
 \end{tikzpicture}
 }
\end{align}
and all $4n$ blowups are inside the $0$-th compact surface $S_0=\FF_0^{4n}$.
Since all the blowups are at the gluing of $S_0$ with $S_{2n-1} = \FF_{4n-2}$, one can flop $4n-2$ out of them and obtains
\begin{align}
\raisebox{-.5\height}{
 \begin{tikzpicture}
  \node (v0) at (0,0) {$\mathbf{0}_0^{2}$};  
  \node (v1) at (3,2){$\mathbf{1}_{2}$};  
  \node (v2) at (6,2) {$\mathbf{2}_4$};
  \node (v3) at (10,2) {$\mathbf{(n{-}1)}_{2n{-}2}$};
  \node (v4) at (13,0) {$\mathbf{n}_{2n}$};
  \node (v5) at (10,-2) {$\mathbf{(n{+}1)}_{2n{+}2}$};
  \node (v6) at (6,-2) {$\mathbf{(2n{-}2)}_{4n{-}4}$};
  \node (v7) at (3,-2) {$\mathbf{(2n{-}1)}_{4n{-}2}^{4n{-2}}$};
%
  \node (g1) at (8,2) {$\cdots$};
  \node (g2) at (8,-2) {$\cdots$};
    \draw  (v0) edge (v1);
    \draw  (v1) edge (v2);
    \draw  (v3) edge (v4);
    \draw  (v4) edge (v5);
    \draw  (v6) edge (v7);
    \draw  (v7) edge (v0);
    \draw (v2) edge (g1);
    \draw (g1) edge (v3);
    \draw (v6) edge (g2);
    \draw (g2) edge (v5);
%
  \node at (0.5,0.75) {$\scriptstyle{e_0}$};
  \node at (2,1.75) {$\scriptstyle{e_1}$};
  \node at (3.5,2.25) {$\scriptstyle{ h_1}$};
  \node at (5.5,2.25) {$\scriptstyle{ e_2}$};
  \node at (6.5,2.25) {$\scriptstyle{ h_2}$};
  \node at (8.75,2.25) {$\scriptstyle{e_{n{-}1}}$};
  \node at (11.25,1.55) {$\scriptstyle{h_{n{-}1}}$};
  \node at (12.5,0.75) {$\scriptstyle{e_n}$};
  \node at (12.5,-0.85) {$\scriptstyle{h_n}$};  
  \node at (8.75,-2.25) {$\scriptstyle{h_{n{+}1}}$};
  \node at (11.25,-1.55) {$\scriptstyle{e_{n{+}1}}$};
  \node at (7.5,-2.25) {$\scriptstyle{ e_{2n{-}2}}$};
  \node at (5,-2.35) {$\scriptstyle{ h_{2n{-}2}}$};
  \node at (4,-2.35) {$\scriptstyle{ e_{2n{-}1}}$};
  \node at (0.25,-0.85) {$\scriptstyle{ e_0-\sum_{i=4n{-}1}^{4n} x_i}$};
  \node at (1.25,-1.5) {$\scriptstyle{ h_{2n{-}1}}-\sum_{i=1}^{4n{-}2} x_i$};
 \end{tikzpicture}
 }
\end{align}
Next, applying the isomorphism \eqref{eq:iso_Fn_Fn+1} $4n-2$ times reduces $\FF_{4n-2}^{4n-2} \to \FF_{0}^{4n-2}$. As a consequence, all these $4n-2$ blowups are in the gluing curve to $S_{2n-2} =\FF_{4n-4}$. In detail,
\begin{align}
\raisebox{-.5\height}{
 \begin{tikzpicture}
  \node (v0) at (0,0) {$\mathbf{0}_0^{2}$};  
  \node (v1) at (3,2){$\mathbf{1}_{2}$};  
  \node (v2) at (6,2) {$\mathbf{2}_4$};
  \node (v3) at (10,2) {$\mathbf{(n{-}1)}_{2n{-}2}$};
  \node (v4) at (13,0) {$\mathbf{n}_{2n}$};
  \node (v5) at (10,-2) {$\mathbf{(n{+}1)}_{2n{+}2}$};
  \node (v6) at (6,-2) {$\mathbf{(2n{-}2)}_{4n{-}4}$};
  \node (v7) at (3,-2) {$\mathbf{(2n{-}1)}_{0}^{4n{-2}}$};
%
  \node (g1) at (8,2) {$\cdots$};
  \node (g2) at (8,-2) {$\cdots$};
    \draw  (v0) edge (v1);
    \draw  (v1) edge (v2);
    \draw  (v3) edge (v4);
    \draw  (v4) edge (v5);
    \draw  (v6) edge (v7);
    \draw  (v7) edge (v0);
    \draw (v2) edge (g1);
    \draw (g1) edge (v3);
    \draw (v6) edge (g2);
    \draw (g2) edge (v5);
%
  \node at (0.5,0.75) {$\scriptstyle{e_0}$};
  \node at (2,1.75) {$\scriptstyle{e_1}$};
  \node at (3.5,2.25) {$\scriptstyle{ h_1}$};
  \node at (5.5,2.25) {$\scriptstyle{ e_2}$};
  \node at (6.5,2.25) {$\scriptstyle{ h_2}$};
  \node at (8.75,2.25) {$\scriptstyle{e_{n{-}1}}$};
  \node at (11.25,1.55) {$\scriptstyle{h_{n{-}1}}$};
  \node at (12.5,0.75) {$\scriptstyle{e_n}$};
  \node at (12.5,-0.85) {$\scriptstyle{h_n}$};  
  \node at (8.75,-2.25) {$\scriptstyle{h_{n{+}1}}$};
  \node at (11.25,-1.55) {$\scriptstyle{e_{n{+}1}}$};
  \node at (7.5,-2.25) {$\scriptstyle{ e_{2n{-}2}}$};
  \node at (5,-2.35) {$\scriptstyle{ h_{2n{-}2}}$};
  \node at (3.5,-2.45) {$\scriptstyle{ e_{2n{-}1}-\sum_{i=1}^{4n{-}3} x_i }$};
  \node at (0.25,-0.85) {$\scriptstyle{ e_0- \sum_{i=4n{-}1}^{4n}x_i}$};
  \node at (1.25,-1.5) {$\scriptstyle{ e_{2n{-}1}}$};
 \end{tikzpicture}
 }
\end{align}
and, again, one can flop $4n-4$ blowups into $S_{2n-2}=\FF_{4n-4}$. This leads to
\begin{align}
\raisebox{-.5\height}{
 \begin{tikzpicture}
  \node (v0) at (0,0) {$\mathbf{0}_0^{2}$};  
  \node (v1) at (3,2){$\mathbf{1}_{2}$};  
  \node (v2) at (6,2) {$\mathbf{2}_4$};
  \node (v3) at (10,2) {$\mathbf{(n{-}1)}_{2n{-}2}$};
  \node (v4) at (13,0) {$\mathbf{n}_{2n}$};
  \node (v5) at (10,-2) {$\mathbf{(n{+}1)}_{2n{+}2}$};
  \node (v6) at (6,-2) {$\mathbf{(2n{-}2)}_{4n{-}4}^{4n{-}4}$};
  \node (v7) at (3,-2) {$\mathbf{(2n{-}1)}_{0}^{2}$};
%
  \node (g1) at (8,2) {$\cdots$};
  \node (g2) at (8,-2) {$\cdots$};
    \draw  (v0) edge (v1);
    \draw  (v1) edge (v2);
    \draw  (v3) edge (v4);
    \draw  (v4) edge (v5);
    \draw  (v6) edge (v7);
    \draw  (v7) edge (v0);
    \draw (v2) edge (g1);
    \draw (g1) edge (v3);
    \draw (v6) edge (g2);
    \draw (g2) edge (v5);
%
  \node at (0.5,0.75) {$\scriptstyle{e_0}$};
  \node at (2,1.75) {$\scriptstyle{e_1}$};
  \node at (3.5,2.25) {$\scriptstyle{ h_1}$};
  \node at (5.5,2.25) {$\scriptstyle{ e_2}$};
  \node at (6.5,2.25) {$\scriptstyle{ h_2}$};
  \node at (8.75,2.25) {$\scriptstyle{e_{n{-}1}}$};
  \node at (11.25,1.55) {$\scriptstyle{h_{n{-}1}}$};
  \node at (12.5,0.75) {$\scriptstyle{e_n}$};
  \node at (12.5,-0.85) {$\scriptstyle{h_n}$};  
  \node at (8.75,-2.25) {$\scriptstyle{h_{n{+}1}}$};
  \node at (11.25,-1.55) {$\scriptstyle{e_{n{+}1}}$};
  \node at (7.5,-2.25) {$\scriptstyle{ e_{2n{-}2}}$};
  \node at (5.25,-1.55) {$\scriptstyle{ h_{2n{-}2}-\sum_{i=1}^{{4n-4}} x_i  }$};
  \node at (3.5,-2.45) {$\scriptstyle{ e_{2n{-}1}-\sum_{i={4n{-}3}}^{{4n{-}2}} x_i }$};
  \node at (0.25,-0.85) {$\scriptstyle{ e_0- \sum_{i={4n-1}}^{{4n}} x_i }$};
  \node at (1.25,-1.5) {$\scriptstyle{ e_{2n{-}1}}$};
 \end{tikzpicture}
 }
\end{align}
and applying the isomorphism \eqref{eq:iso_Fn_Fn+1} $4n-4$ times reduces $\FF_{4n-4}^{4n-4} \to \FF_{0}^{4n-4}$. Thus, the geometry becomes
\begin{align}
\raisebox{-.5\height}{
 \begin{tikzpicture}
  \node (v0) at (0,0) {$\mathbf{0}_0^{2}$};  
  \node (v1) at (3,2){$\mathbf{1}_{2}$};  
  \node (v2) at (6,2) {$\mathbf{2}_4$};
  \node (v3) at (10,2) {$\mathbf{(n{-}1)}_{2n{-}2}$};
  \node (v4) at (13,0) {$\mathbf{n}_{2n}$};
  \node (v5) at (10,-2) {$\mathbf{(n{+}1)}_{2n{+}2}$};
  \node (v6) at (6,-2) {$\mathbf{(2n{-}2)}_{0}^{4n{-}4}$};
  \node (v7) at (3,-2) {$\mathbf{(2n{-}1)}_{0}^{2}$};
%
  \node (g1) at (8,2) {$\cdots$};
  \node (g2) at (8,-2) {$\cdots$};
    \draw  (v0) edge (v1);
    \draw  (v1) edge (v2);
    \draw  (v3) edge (v4);
    \draw  (v4) edge (v5);
    \draw  (v6) edge (v7);
    \draw  (v7) edge (v0);
    \draw (v2) edge (g1);
    \draw (g1) edge (v3);
    \draw (v6) edge (g2);
    \draw (g2) edge (v5);
%
  \node at (0.5,0.75) {$\scriptstyle{e_0}$};
  \node at (2,1.75) {$\scriptstyle{e_1}$};
  \node at (3.5,2.25) {$\scriptstyle{ h_1}$};
  \node at (5.5,2.25) {$\scriptstyle{ e_2}$};
  \node at (6.5,2.25) {$\scriptstyle{ h_2}$};
  \node at (8.75,2.25) {$\scriptstyle{e_{n{-}1}}$};
  \node at (11.25,1.55) {$\scriptstyle{h_{n{-}1}}$};
  \node at (12.5,0.75) {$\scriptstyle{e_n}$};
  \node at (12.5,-0.85) {$\scriptstyle{h_n}$};  
  \node at (8.75,-2.25) {$\scriptstyle{h_{n{+}1}}$};
  \node at (11.25,-1.55) {$\scriptstyle{e_{n{+}1}}$};
  \node at (7.25,-2.35) {$\scriptstyle{ e_{2n{-}2} -\sum_{i=1}^{4n{-}4} x_i }$};
  \node at (5,-1.55) {$\scriptstyle{ e_{2n{-}2}  }$};
  \node at (4,-2.35) {$\scriptstyle{ e_{2n{-}1}-\sum_{i={4n-3}}^{{4n-2}} x_i }$};
  \node at (0.25,-0.85) {$\scriptstyle{ e_0-\sum_{i={4n-1}}^{4n} x_i}$};
  \node at (1.25,-1.5) {$\scriptstyle{ e_{2n{-}1}}$};
 \end{tikzpicture}
 }
\end{align}
and one realises that this processes of flop transition and isomorphism \eqref{eq:iso_Fn_Fn+1} can be repeated systematically for all compact surfaces $S_i$. 
Therefore, one can reduce all surfaces to $\FF_0^2$ and all blowups can be distributed symmetrically as follows:
\begin{align}
\raisebox{-.5\height}{
 \begin{tikzpicture}
  \node (v0) at (0,0) {$\mathbf{0}_0^{2}$};  
  \node (v1) at (3,2){$\mathbf{1}_{0}^2$};  
  \node (v2) at (6,2) {$\mathbf{2}_0^2$};
  \node (v3) at (10,2) {$\mathbf{(n{-}1)}_{0}^2$};
  \node (v4) at (13,0) {$\mathbf{n}_{0}^2$};
  \node (v5) at (10,-2) {$\mathbf{(n{+}1)}_{0}^2$};
  \node (v6) at (6,-2) {$\mathbf{(2n{-}2)}_{0}^{2}$};
  \node (v7) at (3,-2) {$\mathbf{(2n{-}1)}_{0}^{2}$};
%
  \node (g1) at (8,2) {$\cdots$};
  \node (g2) at (8,-2) {$\cdots$};
    \draw  (v0) edge (v1);
    \draw  (v1) edge (v2);
    \draw  (v3) edge (v4);
    \draw  (v4) edge (v5);
    \draw  (v6) edge (v7);
    \draw  (v7) edge (v0);
    \draw (v2) edge (g1);
    \draw (g1) edge (v3);
    \draw (v6) edge (g2);
    \draw (g2) edge (v5);
%
  \node at (0.5,0.75) {$\scriptstyle{e_0-x_1}$};
  \node at (2,1.75) {$\scriptstyle{e_1}-x_{2}$};
  \node at (3.5,2.35) {$\scriptstyle{ e_1-x_{3}}$};
  \node at (5.25,2.35) {$\scriptstyle{ e_2-x_{4}}$};
  \node at (6.75,2.35) {$\scriptstyle{ e_2-x_{5}}$};
  \node at (8.75,2.35) {$\scriptstyle{e_{n{-}1} -x_{2n{-}2}}$};
  \node at (11.5,1.55) {$\scriptstyle{e_{n{-}1} -x_{2n{-}1} }$};
  \node at (12.5,0.75) {$\scriptstyle{e_n -x_{2n} }$};
  \node at (12.5,-0.85) {$\scriptstyle{e_n -x_{2n{+}1} }  $};  
  \node at (8.75,-1.65) {$\scriptstyle{e_{n{+}1} -x_{2n{+}3} }$};
  \node at (11.5,-1.55) {$\scriptstyle{e_{n{+}1} -x_{2n{+}2} }$};
  \node at (7.25,-2.35) {$\scriptstyle{ e_{2n{-}2} - x_{4n{-}4} }$};
  \node at (5,-1.65) {$\scriptstyle{ e_{2n{-}2} -x_{4n{-}3} }$};
  \node at (4,-2.35) {$\scriptstyle{ e_{2n{-}1}- x_{4n{-}2} }$};
  \node at (0.25,-0.85) {$\scriptstyle{ e_0-x_{4n}}$};
  \node at (1.25,-1.5) {$\scriptstyle{ e_{2n{-}1} -x_{4n{-}1}}$};
 \end{tikzpicture}
 }
 \label{eq:geometry_SU_even_final}
\end{align}
which is a fully symmetric arrangement of $\FF_0^2$ surfaces glued along $e-x$ curves to the neighbouring compact surfaces.
\paragraph{Prepotential via intersection numbers.}
The claim is as follows:
\begin{compactitem}
\item The compact surfaces $S_i$ in \eqref{eq:geometry_SU_even_final} give rise to the truncated K\"ahler form
\begin{align}
 J_\phi = \sum_{i=0}^{2n-1} \phi_i S_i \;,
\end{align}
whose intersection numbers induce the (truncated) geometric prepotential
\begin{align}
 6\Fcal_{\trun} \equiv  J_\phi^3
 = \sum_{i=0}^{2n-1} 6 \phi_i^3
 - \sum_{i=0}^{2n-1} 3 \phi_{i} \phi_{i+1}^2 
 - \sum_{i=0}^{2n-1} 3 \phi_{i} \phi_{i-1}^2 
\end{align}
with the identification $\phi_{2n} \equiv \phi_0$.
\item One can add $2n$ non-compact surfaces $F_{i}$ for $i=0,1,\ldots, 2n-1$ which restrict to the compact surfaces in \eqref{eq:geometry_SU_even_final} as follows:
\begin{align}
 F_i|_{S_j} = \delta_{ij} e_i  \,.
\end{align}
and further $2n$ non-compact surfaces $N_{i,i+1}$ for $i=0,1,\ldots, 2n-1$ which are characterised by
\begin{align}
    N_{i,i+1} |_{S_j} = \delta_{i,j}\ x_{2i+1} + \delta_{i+1,j}\ x_{2i+2}  \,.
\end{align}
The K\"ahler form is supplemented via
\begin{align}
  J = - \sum_{i=0}^{2n-1}  \frac{1}{g_i^2}  F_i +  \sum_{i=0}^{2n-1} \phi_i S_i
  + \sum_{i=0}^{2n-1} m_{i,i+1} N_{i,i+1} \;,
\end{align}
and the associated prepotential becomes
\begin{align}
 6\Fcal_{\geom} \equiv  J^3
 =  6\Fcal_{\trun} 
 +
 \sum_{i=0}^{2n-1} 
 \left( 
 \frac{6}{g_i^2} \phi_i^2 
 -3 m_{i,i+1} (\phi_i^2 - 2\phi_i \phi_{i+1}+\phi_{i+1}^2 )
 - 3 m_{i,i+1}^2 (\phi_i +\phi_{i+1}) 
 \right) 
\end{align}
with the identification $\phi_{2n} \equiv \phi_0$. Hence, the geometric result agrees with prepotential \eqref{eq:prepot_affine_A_phase} of the affine quiver gauge theory \eqref{eq:5d_affine_A-type_quiver} in the phase \eqref{eq:phase_choice_affine_A}.
\end{compactitem}
\subsubsection{\texorpdfstring{$\surm(\mathrm{odd})$}{SU(odd)}}
The geometry is given by \cite{Bhardwaj:2019fzv}
\begin{align}
\raisebox{-.5\height}{
 \begin{tikzpicture}
  \node (v0) at (0,0) {$\mathbf{0}_0^{4n+2}$};  
  \node (v1) at (3,2){$\mathbf{1}_{2}$};  
  \node (v2) at (6,2) {$\mathbf{2}_4$};
  \node (v3) at (10,2) {$\mathbf{n}_{2n}$};
  \node (g1) at (8,2) {$\cdots$};
  \node (g2) at (8,-2) {$\cdots$};
  \node (v5) at (10,-2) {$\mathbf{(n{+}1)}_{2n+2}$};
  \node (v6) at (6,-2) {$\mathbf{(2n-1)}_{4n-2}$};
  \node (v7) at (3,-2) {$\mathbf{(2n)}_{4n}$};
  \draw  (v0) edge (v1);
  \draw  (v1) edge (v2);
  \draw  (v3) edge (v5);
  \draw  (v6) edge (v7);
  \draw  (v7) edge (v0);
    \draw (v2) edge (g1);
    \draw (g1) edge (v3);
    \draw (v6) edge (g2);
    \draw (g2) edge (v5);
%
  \node at (0.5,0.75) {$\scriptstyle{e_0}$};
  \node at (2,1.75) {$\scriptstyle{e_1}$};
  \node at (3.5,2.25) {$\scriptstyle{ h_1}$};
  \node at (5.5,2.25) {$\scriptstyle{ e_2}$};
  \node at (6.5,2.25) {$\scriptstyle{ h_2}$};
  \node at (8.75,2.25) {$\scriptstyle{e_{n}}$};
  \node at (10.5,1.55) {$\scriptstyle{h_{n}}$};
  \node at (8.75,-2.25) {$\scriptstyle{h_{n{+}1}}$};
  \node at (10.5,-1.55) {$\scriptstyle{e_{n{+}1}}$};
  \node at (7.5,-2.25) {$\scriptstyle{ e_{2n{-}1}}$};
  \node at (5,-2.35) {$\scriptstyle{ h_{2n{-}1}}$};
  \node at (4,-2.35) {$\scriptstyle{ e_{2n}}$};
  \node at (1.75,-1.5) {$\scriptstyle{ h_{2n}}$};
  \node at (0.25,-0.85) {$\scriptstyle{e_0-\sum_{i=1}^{4n{+}2} x_i}$};
 \end{tikzpicture}
 }
\end{align}
and one observes that the same reasoning as for $\surm(2n)$  applies. In brief, the $0$-th surface $S_0=\FF_0^{4n+2}$ carries all $4n+2$ blowups. Due to the degree $0$, all curves of the form $e_0-x_i$ can be flopped into the surface $S_{2n}=\FF_{4n}$. The isomorphism \eqref{eq:iso_Fn_Fn+1} suggests that one flops as many blowups into a given Hirzebruch surface as its degree. Thus, $4n$ blowups are flopped into $S_{2n}$ such that the repeated application of the isomorphism \eqref{eq:iso_Fn_Fn+1} yields $\FF_{4n}^{4n} \to \FF_{0}^{4n}$, but all the blowups are then in the gluing curves to $S_{2n-1}= \FF_{2n-2}$. Repeating this process for all other compact surfaces yields the claim.
%
%
\subsection{\texorpdfstring{$-3$ curve: $\surm(3)$ with $\mathbb{Z}_2$ twist}{-3 curve: SU(3) with Z2 twist}}
\label{sec:SU3_on_3_Z2_twist}
Consider a $-3$ curve with an $\surmL(3)$ gauge algebra such that the 6d theory has no hypermultiplets, also known as the 6d minimal $\surm(3)$ SCFT. The circle compactification with a $\Z_2$ outer automorphism twist on $\surmL(3)$
\begin{align}
 \raisebox{-.5\height}{
 \begin{tikzpicture}
\node (a1) at (0,0) {$\mathbf{3}$};
  \node at (0,0.35) {$\scriptstyle{ \surmL(3)^{(2)}}$};
   \end{tikzpicture}
 }
 \end{align}
is known to be dual to a 5d $\surm(3)$ gauge theory with Chern-Simons level $\kappa=9$ \cite{Hayashi:2018lyv,Jefferson:2018irk}.
\paragraph{5d description.}
The 5d gauge theory description is based on the prepotential \eqref{eq:F}, which can be evaluated using Appendix \ref{app:SU_roots_weights}.
Restricting to the Weyl chamber, i.e.\ $\langle \phi,\alpha_i \rangle \geq 0$ for $i=1,2$, and choosing $\kappa=9$, one finds: 
\begin{align}
 6\Fcal_{5d} 
 &=\frac{6}{g^2} \left( \phi_0^2 - \phi_0 \phi_1 + \phi_1^2  \right) 
 + 8 \phi_0^3+24 \phi_0^2 \phi_1 -30 \phi_0 \phi_1^2 +8 \phi_1^3 \,,
 \label{eq:prepot_field_theory_SU3_on_3_twist}
\end{align}
which has been computed before in \cite{Hayashi:2018lyv}. 
\paragraph{Geometry.}
According to \cite{Bhardwaj:2019fzv}, the geometry of the twisted circle compactification is described by
\begin{align}
 \raisebox{-.5\height}{
 \begin{tikzpicture}
  \node (v0) at (0,0) {$\mathbf{0}_{10}$ };  
  \node (v1) at (3,0) {$\mathbf{1}_{0}$};  
  \draw  (v0) edge (v1);
%
  \node at (0.5,0.25) {$\scriptstyle{e_0}$};
  \node at (2.25,0.25) {$\scriptstyle{4e_1+f_1}$};
 \end{tikzpicture}
 }
 \label{eq:A2on3curve}
\end{align}
and one computes the fibre intersections
\begin{align} 
\label{eq:-3Z26d}
   -\begin{pmatrix}
     f_0 \cdot K_{S_0} & f_0 \cdot S_1|_{S_0} \\
     f_1 \cdot S_0|_{S_1} & f_1 \cdot K_{S_1}
   \end{pmatrix} 
   =\begin{pmatrix}
     2 & -1 \\ -4 & 2
   \end{pmatrix}
   =C_{\widehat{A}_2^{(2)}}
\end{align}
which is consistent with the 6d origin of a $-3$ curve with an $\surmL(3)$ algebra and $\Z_2$ twist.
The associated truncated K\"ahler form $J_\phi = \sum_{i=0}^1 \phi_i S_i$  to \eqref{eq:A2on3curve} gives rise to the following cubic part of the prepotential
\begin{align}
\label{eq:prepot_geom_SU3_Z2_cubic}
 6 \Fcal_{\trun} \equiv J_\phi^3 
 &= 8  \phi_0^3 +  8 \phi_1^3 
 + 24  \phi_0^2 \phi_1
 - 30  \phi_0 \phi_1^2  \,.
\end{align}
In order to include the gauge coupling terms, it is instructive to note that 
\begin{align} \label{eq:-3Z25d}
   -\begin{pmatrix}
     f_0 \cdot K_{S_0} & f_0 \cdot S_1|_{S_0} \\
     e_1 \cdot S_0|_{S_1} & e_1 \cdot K_{S_1}
   \end{pmatrix} 
   =\begin{pmatrix}
     2 & -1 \\ -1 & 2
   \end{pmatrix}
   =C_{A_2}
   \quad \Rightarrow \quad 
   \begin{cases}
   -J_\phi\cdot f_0 &= \langle \alpha_1,\phi\rangle \,,\\
   -J_\phi\cdot e_1 &= \langle \alpha_2,\phi\rangle\,,
   \end{cases}
\end{align}
which identifies $f_0$ and $e_1$ with the simple roots $\alpha_i$ of the 5d gauge algebra $\surmL(3)$, see Appendix \ref{app:SU_roots_weights}.
One proceeds by adding a non-compact surface $F$ to the K\"ahler form
\begin{align}
 J =-\frac{1}{g^2} F + \phi_0 S_0+ \phi_1 S_1
 \label{eq:Ansatz_J_SU3_on_3_with_Z2}
\end{align}
which is characterised by the restrictions to the compact surfaces
\begin{align}
\begin{cases}
 F|_{S_0}= a_0 e_0  + b_0 f_0 \,,\\
 F|_{S_1}= a_1 e_1  + b_1 f_1\,.
\end{cases}
\end{align}
The parameters $a_i$, $b_i$ are determined from the effective coupling \eqref{eq:effective_coupling}.
Imposing that
\begin{align}
\label{eq:solve_SU3_on_3_with_Z2_initial}
    \begin{pmatrix}
       J|_{S_0}\cdot K_{S_0} & J|_{S_0} \cdot S_1|_{S_0} \\
      J|_{S_1} \cdot S_0|_{S_1} &  J|_{S_1}\cdot K_{S_1}
    \end{pmatrix}
    \stackrel{!}{=} \left( \frac{\partial^2 \Fcal_{5d}}{\partial \phi_i \partial \phi_i }\right)\equiv (\tau_{\eff}) 
\end{align}
is equivalent to imposing that the intersection numbers of $F$ with the $S_i$,
\begin{align}
\label{eq:solve_SU3_on_3_with_Z2}
\begin{pmatrix}
  F|_{S_0} \cdot  K_{S_0} & F|_{S_0} \cdot S_1|_{S_0} \\
  F|_{S_1} \cdot S_0|_{S_1} & F|_{S_1} \cdot  K_{S_1}~,
\end{pmatrix}
 &=
 \begin{pmatrix}
  8 a_0   -2 b_0  & - 10 a_0 +b_0 \\
  a_1   + 4b_1  & -2 a_1  - 2 b_1
 \end{pmatrix} 
  \stackrel{!}{=}
  \begin{pmatrix}  2 & -1 \\ -1 & 2 \end{pmatrix} \equiv - h_{ij}~,
\end{align}
equal the negative inverse metric tensor $h_{ij}$ of the 5d gauge group $\surm(3)$.
For simply laced Lie algebras, $h_{ij} $ equals the Cartan matrix. To see that \eqref{eq:solve_SU3_on_3_with_Z2_initial} is equivalent to \eqref{eq:solve_SU3_on_3_with_Z2}, one recalls that, firstly, the cubic parts of prepotential \eqref{eq:prepot_field_theory_SU3_on_3_twist} are reproduced by \eqref{eq:prepot_geom_SU3_Z2_cubic} and, secondly, there are no other deformation parameters or non-compact surfaces.
Solving the system of linear equations \eqref{eq:solve_SU3_on_3_with_Z2}, one finds  
\begin{align}
 \begin{cases}
 a_0=0\, , &  b_0=1 \,, \\
 a_1=1 \, , & b_1=0 \,,
 \end{cases}
 \qquad 
 \longrightarrow
 \qquad 
 \begin{cases}
F|_{S_0} = f_0 \,, \\
F|_{S_1} = e_1 \,.
 \end{cases}
 \label{eq:Expect_J_SU3_on_3_with_Z2}
\end{align}
One observes that $F|_{S_i}$ restricts to $f_0$ and $e_1$ respectively, because these act as simple roots \eqref{eq:-3Z25d}. Moreover, the prefactor of $1$ equals $\frac{2}{\langle\alpha_i,\alpha_i \rangle}$ for roots of $A_2$.

As a result, the K\"ahler form \eqref{eq:Ansatz_J_SU3_on_3_with_Z2} leads to the following geometric prepotential:
\begin{align}
 6 \Fcal_{\geom} \equiv J^3 
 =  \frac{6}{g^2} \left( \phi_0^2 - \phi_0 \phi_1 + \phi_1^2  \right)
 +8  \phi_0^3 +  8 \phi_1^3  + 24  \phi_0^2 \phi_1 - 30  \phi_0 \phi_1^2  \,.
 \label{eq:prepot_full_SU3_on_3_twist}
\end{align}
which matches the field theory result \eqref{eq:prepot_field_theory_SU3_on_3_twist}.
In addition, consider the elliptic fibre 
\begin{align}
 f_{\mathrm{ell}}= \sum_{i=0}^1 d_i f_i  =2 f_0 +f_1 \,,
\end{align}
with $d_i =$ Coxeter numbers of $A_2^{(2)}$.
Computing the volume with respect to \eqref{eq:Ansatz_J_SU3_on_3_with_Z2} yields
\begin{align}
\label{eq:elliptic_fibre_SU3_on3}
\begin{cases}
 \vol(f_0) &= 2 \phi_0 - \phi_1  \\
 \vol(f_1) &= \frac{1}{g^2} - 4\phi_0 +2 \phi_1
\end{cases}
\qquad \Rightarrow \qquad 
 \vol( f_{\mathrm{ell}}) =   \frac{1}{g^2} \,.
\end{align}
The geometry \eqref{eq:A2on3curve} displays the 6d properties manifestly, like the fibre intersection \eqref{eq:-3Z26d} and the elliptic fibre \eqref{eq:elliptic_fibre_SU3_on3}.
On the other hand, the 5d description is based on a simple exchange of fibre and base in the $\FF_0$ in \eqref{eq:A2on3curve}, which becomes apparent in the intersection properties \eqref{eq:-3Z25d}. 
Consequently, the 6d and 5d  geometric frame are related by a fibre-base duality.
%
%
\subsection{\texorpdfstring{$-4$ curve: $\sorm(8)$ with $\mathbb{Z}_3$ twist}{-4 curve: SO(8) with Z3 twist}}
\label{sec:SO8_on_4_Z3_twist}
The 6d minimal $\sorm(8)$ SCFT is realised by a $-4$ curve supporting a $\sormL(8)$ gauge algebra. 
The circle compactification twisted by a $\Z_3$ outer automorphism on the  $\sormL(8)$ algebra 
\begin{align}
 \raisebox{-.5\height}{
 \begin{tikzpicture}
\node (a1) at (0,0) {$\mathbf{4}$};
  \node at (0,0.35) {$\scriptstyle{ \sormL(8)^{(3)}}$};
   \end{tikzpicture}
 }
 \end{align}
is known to be dual to a 5d $\surm(4)$ gauge theory with Chern-Simons level  $\kappa=8$\cite{Razamat:2018gro}.
\paragraph{5d description.}
The 5d gauge theory description is based on the prepotential \eqref{eq:F}, which can be evaluated using Appendix \ref{app:SU_roots_weights}. 
Restricting to the Weyl-chamber of $A_3$, i.e.\ $\langle \phi,\alpha_i \rangle \geq 0$ for $i=1,2,3$, and specialising to $\kappa=8$ yields
\begin{align}
\begin{aligned}
  6\Fcal_{5d} &=
   \frac{6}{g^2} \left( \phi_0^2  - \phi_0 \phi_1 + \phi_1^2  - \phi_1 \phi_2 + \phi_2^2     \right) \\
 &\quad +8 \phi_0^3+24 \phi_1 \phi_0^2-30 \phi_1^2 \phi_0+8 \phi_1^3+8 \phi_2^3-24 \phi_1 \phi_2^2+18 \phi_1^2 \phi_2 \,.
\end{aligned}
\label{eq:prepot_field_theory_SO8_on_4_twist}
 \end{align}
\paragraph{Geometry.}
As argued in \cite{Bhardwaj:2019fzv}, the geometry is described by
\begin{align}
 \raisebox{-.5\height}{
 \begin{tikzpicture}
  \node (v0) at (0,0) {$\mathbf{0}_{10}$ };  
  \node (v1) at (3,0) {$\mathbf{1}_{8}$};  
  \node (v2) at (6,0) {$\mathbf{2}_{0}$};
  \draw  (v0) edge (v1);
  \draw  (v1) edge (v2);
%
  \node at (0.5,0.25) {$\scriptstyle{e_0}$};
  \node at (2.5,0.25) {$\scriptstyle{h_1}$};
  \node at (3.5,0.25) {$\scriptstyle{e_1}$};
  \node at (5,0.25) {$\scriptstyle{3e_2+f_2}$};
 \end{tikzpicture}
 }
 \label{eq:geom_SO8_on_4_with_Z3}
\end{align}
and one verifies that 
\begin{align} \label{eq:-4Z36d}
    -f_i \cdot S_j = \begin{pmatrix}
      2 & -1 & 0 \\ -1 & 2 & -1 \\ 0 & -3 & 2
    \end{pmatrix}
\end{align}
which equals the affine Cartan matrix of $\widehat{D}_4^{(3)}$ up to similarity transformation. Again, this is consistent with the 6d $\sormL(8)$ gauge algebra together with the $\Z_3$ twist.

Based on the truncated K\"ahler form $ J_\phi = \sum_{i=0}^2 \phi_i S_i$, 
the triple intersection numbers of the compact surfaces $S_i$ induce the following cubic part of the prepotential
\begin{align}
 6 \Fcal_{\trun} 
 &= 8 \phi_0^3   +  8 \phi_1^3   +  8 \phi_2^3 
 + 24 \phi_0^2 \phi_1
 -30 \phi_0 \phi_1^2 
 + 18 \phi_1^2 \phi_2
 -24 \phi_1 \phi_2^2 \,.
\end{align}
In order to include the only gauge theory parameter, the coupling $g$, it is useful to notice 
\begin{align} \label{eq:-4Z35d}
    \begin{pmatrix}
 f_0 \cdot  K_{S_0} & f_0 \cdot S_1|_{S_0} & f_0 \cdot S_2|_{S_0}\\
 f_1 \cdot S_0|_{S_1} & f_1 \cdot  K_{S_1} & f_1 \cdot S_2|_{S_1} \\
  e_2 \cdot S_0|_{S_2} & e_2 \cdot S_1|_{S_2} & e_2 \cdot  K_{S_2}
 \end{pmatrix} = \begin{pmatrix}
      2 & -1 & 0 \\ -1 & 2 & -1 \\ 0 & -1 & 2
    \end{pmatrix}
     = C_{A_3}
     \quad \Rightarrow \quad 
     \begin{cases}
     -J_\phi \cdot f_0 &= \langle \alpha_1 ,\phi \rangle \\
      -J_\phi \cdot f_1 &= \langle \alpha_2 ,\phi \rangle \\
       -J_\phi \cdot e_2 &= \langle \alpha_3 ,\phi \rangle \\
     \end{cases}
\end{align}
which identifies $f_0$, $f_1$, and $e_2$ as acting as simple roots $\alpha_i$ of the 5d gauge algebra $\surmL(4)$, see Appendix \ref{app:SU_roots_weights}.
Incorporating the gauge coupling contributions can be achieved by adding a non-compact surface $F$ to the K\"ahler form
\begin{align}
 J =-\frac{1}{g^2} F + \phi_0 S_0+ \phi_1 S_1+ \phi_2 S_2 \,,
 \label{eq:Ansatz_J_SO8_on_4_with_Z3}
\end{align}
which is characterised by its restrictions to the compact surfaces $S_i$, i.e.
\begin{align}
 F|_{S_i}= a_i e_i  + b_i f_i \qquad \text{for } i=0,1,2 \;. 
\end{align}
The parameters $a_i$, $b_i$ are determined by matching the effective gauge coupling as follows:
\begin{align}
\label{eq:solve_SO8_on_4_with_Z3_initial}
\left( J \cdot S_i \cdot S_j\right)
 \stackrel{!}{=} \left( \frac{\partial^2 \Fcal_{5d}}{ \partial \phi_i \partial \phi_j} \right) \equiv (\tau_{\eff}) \,,
\end{align}
i.e.\ combining \eqref{eq:effective_coupling} and \eqref{eq:effectice_coupling_geom}.
Since the triple intersection numbers of the compact surfaces reproduce the cubic part of the prepotential \eqref{eq:prepot_field_theory_SO8_on_4_twist}, and as there are no other deformation parameters or non-compact surfaces, the set of conditions \eqref{eq:solve_SO8_on_4_with_Z3_initial} is equivalent to
\begin{align}
\label{eq:solve_SO8_on_4_with_Z3}
\begin{aligned}
\left( F \cdot S_i \cdot S_j\right)
&=
\begin{pmatrix}
  8 a_0 -2 b_0 &  b_0-10 a_0 & 0\\
 8 b_1+b_1  & 6 a_1  -2 b_1 & b_1 -8 a_1 \\
  0 & a_2   +3 b_2 & -2 a_2 -2 b_2
\end{pmatrix} 
%
\stackrel{!}{=}
\begin{pmatrix} -2 & 1 & 0 \\ 1 & -2 & 1 \\ 0 & 1 & -2 \end{pmatrix} 
\equiv -h_{ij}
\,,
\end{aligned}
\end{align}
where $h_{ij}$ is the inverse metric tensor of the 5d $\surm(4)$ gauge group, see Appendix \ref{app:SU_roots_weights}. Again, for the simply laces $A_3$ algebra, $h_{ij}$ equals the Cartan matrix.
The system of linear equations \eqref{eq:solve_SO8_on_4_with_Z3} is solved by
\begin{align}
 \begin{cases}
 a_0=0\, , &  b_0=1 \,, \\
 a_1=0\, , &  b_1=1 \,,\\
 a_2=1\, , &  b_2=0 \,,
 \end{cases}
 \qquad 
 \longrightarrow
 \qquad 
 \begin{cases}
F|_{S_0} = f_0  \,,\\
F|_{S_1} = f_1 \,,\\
F|_{S_2} = e_2 \,.
 \end{cases} 
 \label{eq:Expect_J_SO8_on_4_with_Z3}
\end{align}
One observes that $F|_{S_i}$ restricts to $f_0$, $f_1$, and $e_2$ respectively, because these act as simple roots \eqref{eq:-4Z35d}. Also, note that the prefactor of $1$ equals $\frac{2}{\langle\alpha_i,\alpha_i \rangle}$ for roots of $A_3$.

The  K\"ahler form \eqref{eq:Ansatz_J_SO8_on_4_with_Z3} induces the following geometric prepotential:
\begin{align}
\begin{aligned}
 6\Fcal_{\geom} \equiv  J^3
 &=
 \frac{6}{g^2} \left( \phi_0^2 - \phi_0 \phi_1 + \phi_1^2 - \phi_1\phi_2 +\phi_2^2  \right)
 +8 \phi_0^3   +  8 \phi_1^3   +  8 \phi_2^3  \\
 &\qquad + 24 \phi_0^2 \phi_1
 -30 \phi_0 \phi_1^2 
 + 18 \phi_1^2 \phi_2
 -24 \phi_1 \phi_2^2 \,.
 \end{aligned}
 \label{eq:prepot_full_SO8_on_4_twist}
\end{align}
Consequently, the geometric result \eqref{eq:prepot_full_SO8_on_4_twist} matches the field theory expectation \eqref{eq:prepot_field_theory_SO8_on_4_twist}.
Moreover, consider the elliptic fibre
\begin{align}
 f_{\mathrm{ell}}= \sum_{i=0}^2 d_i f_i  =f_0 +2f_1+ f_2 \,,
\end{align}
with $d_i =$ Coxeter numbers of $D_4^{(3)}$.
The fibre volumes of \eqref{eq:geom_SO8_on_4_with_Z3} with respect to \eqref{eq:Ansatz_J_SO8_on_4_with_Z3} are
\begin{align}
\label{eq:elliptic_fibre_SO8_on_4}
\begin{cases}
 \vol(f_0) &= 2 \phi_0 - \phi_1  \,,\\
 \vol(f_1) &= -\phi_0 + 2\phi_1 - \phi_2 \,,\\
 \vol(f_2) &= \frac{1}{g_0^2} - 3\phi_1  + 2\phi_2 \,,
\end{cases}
\qquad \Rightarrow \qquad 
 \vol( f_{\mathrm{ell}}) =   \frac{1}{g^2} \,.
\end{align}
The geometry \eqref{eq:geom_SO8_on_4_with_Z3} had been derived to display the 6d properties manifestly, like the fibre intersection \eqref{eq:-4Z36d} and the elliptic fibre \eqref{eq:elliptic_fibre_SO8_on_4}.
On the other hand, the 5d gauge theory is more conveniently described by simply exchanging the fibre and base in the $\FF_0$ in \eqref{eq:geom_SO8_on_4_with_Z3}. As a result, the 5d gauge algebra becomes manifest in the intersection properties \eqref{eq:-4Z35d}. 
This is another example in which the 6d and 5d  geometric frame are related by a fibre-base duality.
%
\section{Fibre-base duality: 6d to 5d}
\label{sec:fibre-base_6d-5d}
The examples considered in Section \ref{sec:instructive} followed a transparent pattern: the compactified 6d theory is geometrically encoded in a collection of Hirzebruch surfaces such that their fibres encode the elliptic fibre from the F-theory construction. The geometry for the dual 5d theory, on the other hand, has been obtained via fibre-base duality from the 6d frame. In this section, fibre-base duality between the 6d frame and one particular 5d frame is investigate for rank 1 and rank 2 KK theories of Table \ref{tab:KK_theories}. As it turns out, not all examples admit a 6d-5d fibre-base duality.
\subsection{\texorpdfstring{6d $\Ncal=(2,0)$ $A_1$ -- 5d $\surm(2)_{\theta=0}$ + 1Adj}{6d A1 - 5d SU(2) + 1Adj} }
\label{sec:SU2_1Adj}
To begin with, consider a $-2$ curve with an $\surmL(1)$ algebra
\begin{align}
\label{eq:6d_M-string}
    \raisebox{-.5\height}{
 \begin{tikzpicture}
\node (a2) at (0,0) {$\mathbf{2}$};
  \node at (0,0.35) {$\scriptstyle{ \surmL(1)^{(1)}}$};
  \end{tikzpicture}
 }
\end{align}
which is known as $\Ncal=(2,0)$ $A_1$ theory or 6d rank-1 M-string theory. The 5d KK theory is known to be a $\surm(2)$ gauge theory with one adjoint hypermultiplet \cite{Douglas:2010iu,Lambert:2010iw}.
\paragraph{5d description.}
The prepotential \eqref{eq:F} is computed by restricting to the Weyl chamber $\phi\geq 0$ and choosing, for instance,  the phase
\begin{align}
    2\phi +m \geq 0 \geq -2\phi+m  \,,
    \label{eq:phase_choice_SU2_1Adj}
\end{align}
where $m$ is the mass parameter of the adjoint hypermultiplet. The prepotential in this phase reads
\begin{align}
    6 \Fcal_{5d} = 6 \frac{1}{g^2} \phi^2 - 6 m^2 \phi 
    \label{eq:PrePot_SU2_1Adj}
\end{align}
and $g$ denotes the gauge coupling.
\paragraph{Geometry.}
As detailed in \cite{Bhardwaj:2019fzv}, the geometric description for \eqref{eq:6d_M-string} is given by
\begin{align}
\raisebox{-.5\height}{
\begin{tikzpicture}
  \node (v0) at (0,0) {$\FF_{0}^{1+1}$};  
\draw (v0) to [out=140-100,in=220+100,looseness=6] (v0);
    \node at (1.5,0.7) {$\scriptstyle{e-x}$};
        \node at (1.5,-0.7) {$\scriptstyle{e-y}$};
 \end{tikzpicture}}
\end{align}
and one confirms that 
\begin{align}
    J_\phi = \phi S 
    \quad \Rightarrow  \quad 
    \Fcal_{\trun} = \frac{1}{3!} J_\phi^3 = 0 
    \qquad \text{and} \qquad 
    f\cdot S =0 \,.
    \label{eq:fibre_SU2+Adj}
\end{align}
The last equation is compatible with the trivial 6d gauge algebra. 
In view of a 5d gauge theory description, it is convenient to notice that 
\begin{align}
    -J_\phi \cdot e = 2 \phi  = \langle \alpha, \phi \rangle 
    \label{eq:root_via_volume_SU2+Adj}
\end{align}
such that the base $e$ is identified with the simple root $\alpha$ of $A_1$, see Appendix \ref{app:SU_roots_weights}.
To include the gauge theory parameter, one may parametrise the K\"ahler form as 
\begin{align}
    J|_{S} = \phi K_{S} + a e + b f   + c x + d y
    \label{eq:Kahler_form_SU2_1Adj_try}
\end{align}
and determine the coefficients as follows:
\begin{compactitem}
 \item The volume of curves with self-intersection $-1$ need to match the BPS masses. It is instructive to determine the truncated volume of the blowups 
 \begin{align}
     -J_\phi \cdot x = \langle0,\phi \rangle =0
 \end{align}
 such that $x$ is identified as the trivial weight in the adjoint representation of $\surm(2)$. Recalling \eqref{eq:root_via_volume_SU2+Adj}, one finds that the following $-1$ curves furnish the adjoint representation:
 \begin{align}
 \raisebox{-.5\height}{
\begin{tikzpicture}
\node[draw,circle,inner sep=0.8pt,fill,black]  (a1) at (0,0) {};
 \node[draw,circle,inner sep=0.8pt,fill,black]  (a2) at (0,-1) {};
 \node[draw,circle,inner sep=0.8pt,fill,black]  (a3) at (0,-2) {};
 \draw (a1)--(a2)--(a3);
 \draw[red,dashed] (-0.2,-1.5)--(4,-1.5);
 \node [right=1ex of a1] {$ \vol(e+x)  = \langle \alpha,\phi \rangle + m$};
 \node [right=1ex of a2] {$ \vol( x)  =  \langle 0,\phi \rangle +m$};
 \node [right=1ex of a3] {$\vol (e-x)  = -  (\langle -\alpha,\phi \rangle +m)$};
 \node at (-0.5,-0.5) {$\scriptstyle{\alpha \; \downarrow}$};
 \node at (-0.5,-1.5) {$\scriptstyle{\alpha\; \downarrow}$};
\end{tikzpicture}
}
     \qquad \text{and}\qquad 
     \vol(e-x) = \vol(e-y) 
  \end{align}
  and the volume conditions are solved by
  \begin{align}
      c=d=m \;, \quad  b=0 \,.
  \end{align}
   Note that these volumes motivate the choice of the phase \eqref{eq:phase_choice_SU2_1Adj}.
  \item The effective gauge coupling is geometrically given by
  \begin{align}
      J \cdot S \cdot S = -2 a \stackrel{!}{= }\frac{\partial^2 \Fcal}{\partial^2 \phi } = \frac{2}{g^2}
       \qquad \Rightarrow \qquad a = -\frac{1}{g^2}
  \end{align}
  and needs to match the field theory expression \eqref{eq:effective_coupling}.
\end{compactitem}
 The K\"ahler form can now be written as 
 \begin{align}
     J= -\frac{1}{g^2} F + \phi S + m N
 \end{align}
 where the non-compact surfaces $F$ and $N$ are glued to $S$ via
 \begin{align}
    F|_{S} = e  
    \qquad \text{and} \qquad 
    N|_{S} = x+y
    \,.
 \end{align} 
 It is reassuring to see that the non-compact divisor $N$ associated to the mass $m$ of the adjoint hypermultiplet has gluing curves $x$ and $y$ with the compact surface $S$, as expected in \cite{Bhardwaj:2019fzv}.
  In conclusion, the geometric prepotential $\Fcal_{\geom} = \frac{1}{3!} J^3$ agrees with the field theory expectation \eqref{eq:PrePot_SU2_1Adj}. Moreover, the basic curves have volumes given by
  \begin{align}
      \vol(e) =2 \phi 
      \;, \quad 
      \vol(f) =\frac{1}{g^2}
      \;, \quad
      \vol(x) = \vol(y) = m \,.
  \end{align}
 Here, fibre-base duality is clearly at play. The 6d elliptic fibre $f$ has volume given by the coupling, while the fibre for the 5d frame is $e$, because it acts as a root of $\surmL(2)$. 
%
\subsection{\texorpdfstring{6d $\surm(3)$ +6 F with $\Z_2$ twist -- 5d $\sprm(2)$ + 3 $\Lambda^2$}{6d SU(3) w/ Z2 -- 5d Sp(2) + 3Lambda2}}
\label{sec:SU3_on_2_Z2_twist}
The $\Z_2$-twisted circle compactification of a $-2$ curve with an $\surmL(3)$ gauge algebra 
\begin{align}
    \raisebox{-.5\height}{
 \begin{tikzpicture}
\node (a1) at (0,0) {$\mathbf{2}$};
  \node at (0,0.35) {$\scriptstyle{ \surmL(3)^{(2)}}$};
   \end{tikzpicture}
 }
 \label{eq:6d_SU3_on_2_twist_Z2}
\end{align}
is known to be dual to a 5d $\sprm(2)$ gauge theory with three hypermultiplets transforming in the rank-2 anti-symmetric representation $\Lambda^2\equiv [0,1]_C$ of $\sprm(2)$ \cite{Hayashi:2015vhy}.
\paragraph{5d description.}
The 5d gauge theory description is based on the prepotential \eqref{eq:F} and can be evaluated using Appendix \ref{app:Sp_roots_weights}.
Since the 5d theory has non-trivial matter content, restricting to the Weyl-chamber of $C_2$, i.e.\ $\langle \phi,\alpha_i \rangle \geq 0$ for $i=1,2$, is insufficient for reaching a single, well-defined 5d phase. 
A suitable choice of phase is given by
\begin{align}
    \raisebox{-.5\height}{
\begin{tikzpicture}
\node[draw,circle,inner sep=0.8pt,fill,black]  (v1) at (0,0) {};
 \node[draw,circle,inner sep=0.8pt,fill,black]  (v2) at (0,-1) {};
 \node[draw,circle,inner sep=0.8pt,fill,black]  (v3) at (0,-2) {};
 \node[draw,circle,inner sep=0.8pt,fill,black]  (v4) at (0,-3) {};
 \node[draw,circle,inner sep=0.8pt,fill,black]  (v5) at (0,-4) {};
 \draw (v1)--(v2)--(v3)--(v4)--(v5);
 \draw[red,dashed] (-0.2,-3.5)--(3,-3.5);
 \node[red] at (3.5,-3.5) {$\substack{\text{phase}\\ \text{choice}}$};
 \node [right=1ex of v1] {$ \langle \phi,v_1 \rangle +m_f \geq 0$};
 \node [right=1ex of v2] {$\langle \phi,v_2 \rangle +m_f \geq 0$};
 \node [right=1ex of v3] {$ \langle \phi,v_3 \rangle +m_f \geq 0$};
 \node [right=1ex of v4] {$\langle \phi,v_4 \rangle +m_f \geq 0$};
 \node [right=1ex of v5] {$ \langle \phi,v_5 \rangle +m_f\leq 0$};
 \node at (-0.5,-0.5) {$\scriptstyle{\alpha_2 \; \downarrow}$};
 \node at (-0.5,-1.5) {$\scriptstyle{\alpha_1\; \downarrow}$};
 \node at (-0.5,-2.5) {$\scriptstyle{\alpha_1 \; \downarrow}$};
 \node at (-0.5,-3.5) {$\scriptstyle{\alpha_2 \; \downarrow}$};
\end{tikzpicture}
}
\label{eq:phase_choice_Sp2_Lambda2}
\end{align}
where $v_i \in [0,1]_C$. In the phase \eqref{eq:phase_choice_Sp2_Lambda2}, the prepotential becomes
\begin{align}
\begin{aligned}
  6\Fcal_{5d} &=
  \frac{6 }{g^2 } \left(   2\phi_0^2 -2 \phi_0 \phi_1 + \phi_1^2 \right)
  + 8 \phi_0^3
  +12 \phi_1 \phi_0^2
  -18 \phi_1^2 \phi_0+5 \phi_1^3   
  \\
  &\quad   -3\sum_{f=1}^3 m_f \left( 4 \phi_0^2 -4 \phi_0 \phi_1+ \phi_1^2\right)
   -3 \sum_{f=1}^3 m_f^2 \phi_1 \,,
\end{aligned}
\label{eq:prepot_field_theory_SU3_on_2_twist}
 \end{align}
 where $g$ denotes the gauge coupling and $m_f$ are the mass parameter of the three hypermultiplets. 
\paragraph{Geometry.}
The $\Z_2$ twisted circle compactification of $\surm(3)$ on a $-2$ curve \eqref{eq:6d_SU3_on_2_twist_Z2} is geometrically described by \cite{Jefferson:2018irk}
\begin{align}
\label{eq:geom_SU3_on_2_Z2_6d}
\raisebox{-.5\height}{
\begin{tikzpicture}
  \node (v0) at (0,0) {$\mathbf{0}_{6}$};  
  \node (v1) at (4,0){$\mathbf{1}_{0}^3$};  
  \draw  (v0) edge (v1);
  \node at (0.5,0.2) {$\scriptstyle{e_0}$};
  \node at (2.5,0.2) {$\scriptstyle{4e_1+2f_1-2\sum_i x_i}$};
 \end{tikzpicture}}
\end{align}
for which one computes that the fibre intersections
\begin{align}
\label{eq:6d_fibre_SU3_on_2_Z2_twist}
    -\begin{pmatrix}
      f_0 \cdot K_{S_0} & f_0 \cdot S_1|_{S_0} \\
       f_1 \cdot S_0|_{S_1} &  f_1 \cdot K_{S_1} \\
    \end{pmatrix}
    = \begin{pmatrix}
      2 & -1 \\ -4 & 2
    \end{pmatrix}
    \equiv  C_{\widehat{A}_2^{(2)}}
\end{align}
yield the affine Cartan matrix of the twisted algebra $\widehat{A}_2^{(2)}$. Hence, \eqref{eq:geom_SU3_on_2_Z2_6d} describes the 6d duality frame of an $A_2$ algebra with a $\Z_2$ twist. 
The truncated K\"ahler form $J_\phi = \sum_i \phi_i S_i$ induces the following truncated prepotential
\begin{align}
6\Fcal_\trun=
8\phi_0^3+12\phi_0^2\phi_1-18\phi_0\phi_1^2+5\phi_1^3\,,
\label{SU3on-2withZ2}
\end{align}
which agrees with \eqref{eq:prepot_field_theory_SU3_on_2_twist}.

The parameters of the 5d $\sprm(2)$ gauge theory, one coupling $g$ and three mass parameters $m_f$, have to be incorporated into the K\"ahler form for a complete description. 
Before proceeding, it is useful to recall that the dual 5d frame can be geometrically realised by an $\FF_0$ isomorphism $e \leftrightarrow f$, see for instance \cite{Bhardwaj:2020gyu,Kim:2020hhh}. The immediate consequence is that 
\begin{align}
    - \begin{pmatrix}
      f_0 \cdot K_{S_0} & f_0 \cdot S_1|_{S_0} \\
       e_1 \cdot S_0|_{S_1} &  e_1 \cdot K_{S_1} \\
    \end{pmatrix}
    = \begin{pmatrix}
      2 & -1 \\ -2 & 2
    \end{pmatrix}
    \equiv  C_{C_2}
    \quad 
    \Rightarrow
    \quad
    \begin{cases}
    -J_\phi \cdot f_0 = \langle\alpha_1,\phi \rangle \\
    -J_\phi \cdot e_1 = \langle\alpha_2,\phi  \rangle 
    \end{cases}
    \label{eq:Sp2_3Lambda_roots}
\end{align}
i.e.\ $f_0$ and $e_1$ correspond to the simple roots $\alpha_{1,2}$ of $C_2$, see Appendix \ref{app:Sp_roots_weights}.
To begin with, the K\"ahler form can be extend to include all gauge theory parameters  
\begin{align}
\label{eq:J_Sp2_3Lambda}
     J = - \frac{1}{g^2} F +\sum_{i=0}^1 \phi_i S_i  + \sum_{f=1}^3 m_f N_f \,.
 \end{align}
The gluing curves of the non-compact surfaces $F$, $N_f$ with the compact $S_i$ are determined by parametrising \eqref{eq:J_Sp2_3Lambda} restricted to the $S_i$ as follows:
\begin{align}
\label{eq:J_Sp2_3Lambda_try}
\begin{aligned}
    J|_{S_0} &= \phi_0 K_{S_0} + \phi_1 S_1|_{S_0} + a_0 f_0 + b_0 e_0 \,, \\
    J|_{S_1} &=  \phi_0 S_0|_{S_1}+ \phi_1 K_{S_1} + a_1 f_1 + b_1 e_1 + \sum_{f=1}^3 c_f x_f \,,
    \end{aligned}
 \end{align}
and the appearing parameters are fully fixed  by physical requirements.
\begin{compactitem}
 \item The fundamental BPS particles need to be identified. It is helpful to evaluate the (truncated) volume of the blowups $x_i$
 \begin{align}
     -J_\phi \cdot x_f = -2\phi_0 +\phi_1 = \langle v_4 , \phi\rangle \,,
 \end{align}
where $v_4=(-2,1) \in [0,1]_C$, see Appendix \ref{app:Sp_roots_weights}. Recalling \eqref{eq:Sp2_3Lambda_roots}, the full set of $[0,1]_C$ weights is recovered via acting with the $C_2$ roots, i.e.\ adding $f_0$ or $e_1$ to the blowups $x_f$. 
 The resulting $-1$ curves have to have volumes given by the BPS masses as follows:
 \begin{align}
    \raisebox{-.5\height}{
\begin{tikzpicture}
\node[draw,circle,inner sep=0.8pt,fill,black]  (v1) at (0,0) {};
 \node[draw,circle,inner sep=0.8pt,fill,black]  (v2) at (0,-1) {};
 \node[draw,circle,inner sep=0.8pt,fill,black]  (v3) at (0,-2) {};
 \node[draw,circle,inner sep=0.8pt,fill,black]  (v4) at (0,-3) {};
 \node[draw,circle,inner sep=0.8pt,fill,black]  (v5) at (0,-4) {};
 \draw (v1)--(v2)--(v3)--(v4)--(v5);
 \draw[red,dashed] (-0.2,-3.5)--(5,-3.5);
 \node [right=1ex of v1] {$  \vol (2f_0 +e_1+ x_f)  \stackrel{!}{=} \langle v_1, \phi\rangle +m_f$};
 \node [right=1ex of v2] {$\vol (2f_0+ x_f) \stackrel{!}{=} \langle v_2 , \phi\rangle +m_f $};
 \node [right=1ex of v3] {$  \vol( f_0+ x_f ) \stackrel{!}{=} \langle v_3 , \phi\rangle+m_f $};
 \node [right=1ex of v4] {$ \vol (x_f)  \stackrel{!}{=} \langle v_4 , \phi\rangle +m_f $};
 \node [right=1ex of v5] {$   \vol (e_1-x_f)   \stackrel{!}{=} -\left( \langle v_5 , \phi\rangle+m_f  \right)$};
 \node at (-0.5,-0.5) {$\scriptstyle{\alpha_2 \; \downarrow}$};
 \node at (-0.5,-1.5) {$\scriptstyle{\alpha_1\; \downarrow}$};
 \node at (-0.5,-2.5) {$\scriptstyle{\alpha_1 \; \downarrow}$};
 \node at (-0.5,-3.5) {$\scriptstyle{\alpha_2 \; \downarrow}$};
\end{tikzpicture}
}
\quad \Rightarrow \quad 
    \begin{cases}
     c_f=m_f  \;, &f=1,2,3 \,,\\
     a_1=b_0=0 \;,
    \end{cases}
     \label{eq:blowup_x_Sp2+Lambda2}
\end{align}
 which in retrospect motivates the phase \eqref{eq:phase_choice_Sp2_Lambda2}.
 \item The effective coupling from geometry \eqref{eq:effectice_coupling_geom} has to match the field theory expectation \eqref{eq:effective_coupling}.
 The resulting linear equations are solved by
 \begin{align}
     a_0 = 2\sum_{f=1}^3 m_f - 2\frac{1}{g^2}
     \;, \quad 
     b_1 = -\frac{1}{g^2}
      \,.
 \end{align}
\end{compactitem}
 As a result, the non-compact surfaces $F$, $N_f$ in \eqref{eq:J_Sp2_3Lambda} restrict to the compact surfaces as follows
 \begin{align}
     \begin{cases}
     F|_{S_0} &= 2 f_0 \\
     F|_{S_1} &=  e_1
     \end{cases}
     \quad \text{and} \qquad 
     \begin{cases}
     N_f|_{S_0} &= 2 f_0 \\
     N_f|_{S_1} &=  x_f
     \end{cases}
     \; \text{for } f=1,2,3\,.
 \end{align}
One observes that $F|_{S_i}$ restricts to $2f_0$ and $e_1$ respectively, because $f_0$, $e_1$ act as simple roots \eqref{eq:Sp2_3Lambda_roots}. Also, the prefactors are equal to $\frac{2}{\langle\alpha_i,\alpha_i \rangle}=2,1$ for the roots of $C_2$.

 Computing the volumes of the basic curves yields
 \begin{alignat}{2}
 \begin{aligned}
     \vol(f_0) &= 2\phi_0 -\phi_1& \qquad 
     \vol(f_1) &= -2\phi_0 +2\phi_1 \\
     \vol(e_1) &= -4\phi_0 +2\phi_1 +\frac{1}{g^2} & \qquad 
     \vol(x_j) &= -2\phi_0 +\phi_1 +m_j \,.
     \end{aligned}
 \end{alignat}
 Moreover, one can keep track of the volume of the elliptic fibre, which in the 6d frame \eqref{eq:geom_SU3_on_2_Z2_6d} is given by
 \begin{align}
     f_{\mathrm{ell}} = d_0 f _0 + d_1 f_1 = 2f_0 + f_1
     \qquad 
    \vol(f_{\mathrm{ell}}) = \frac{1}{g^2}
 \end{align}
 with $d_i =$ Coxeter numbers of $A_2^{(2)}$.
%
\subsection{\texorpdfstring{6d $\surm(3)$ +12 F with $\Z_2$ twist -- 5d $G_2$ + 6 F}{6d SU(3) + 12 F w Z2 -- 5d G2 +6F}}
\label{sec:SU3_on_1_Z2_twist}
The 6d theory given by a $-1$ curve supporting an $\surmL(3)$ gauge algebra has 12 fundamental hypermultiplets. Consider the $\Z_2$ twisted circle compactification
\begin{align}
     \raisebox{-.5\height}{
 \begin{tikzpicture}
\node (a1) at (0,0) {$\mathbf{1}$};
  \node at (0,0.35) {$\scriptstyle{ \surmL(3)^{(2)}}$};
   \end{tikzpicture}
 }
 \label{eq:6d_SU3_on_1_Z2_twist}
\end{align}
where $\Z_2$ acts as outer automorphism on $\surmL(3)$. The resulting KK theory is known to have three different 5d gauge theory descriptions; among them is a $G_2$ theory \cite{Jefferson:2018irk}.
\paragraph{5d description.}
The 5d $G_2$ gauge theory with six fundamental hypermultiplets is based on the prepotential \eqref{eq:F}, which can be evaluated using Appendix \ref{app:G2_roots_weights}.
Since the 5d theory has non-trivial matter content, restricting to the Weyl-chamber of $G_2$, i.e.\ $\langle \phi,\alpha_i \rangle \geq 0$ for $i=1,2$, is insufficient for reaching a single, well-defined 5d phase. 
A suitable choice of phase is given by
\begin{align}
    \raisebox{-.5\height}{
\begin{tikzpicture}
\node[draw,circle,inner sep=0.8pt,fill,black]  (v1) at (0,-2) {};
 \node  (aux) at (0,-3) {$\vdots$};
 \node[draw,circle,inner sep=0.8pt,fill,black]  (v5) at (0,-4) {};
 \node[draw,circle,inner sep=0.8pt,fill,black]  (v6) at (0,-5) {};
 \node[draw,circle,inner sep=0.8pt,fill,black]  (v7) at (0,-6) {};
 \draw (v1)--(aux)--(v5)--(v6)--(v7);
 \draw[red,dashed] (-0.2,-4.5)--(3,-4.5);
 \node[red] at (3.5,-4.5) {$\substack{\text{phase}\\ \text{choice}}$};
 \node [right=1ex of v1] {$\langle \phi,w_1 \rangle +m_f  \geq 0$};
 \node [right=1ex of v5] {$\langle \phi,w_5 \rangle +m_f \geq 0$};
 \node [right=1ex of v6] {$\langle \phi,w_6 \rangle +m_f  \leq 0$};
 \node [right=1ex of v7] {$\langle \phi,w_7 \rangle +m_f  \leq 0$};
 \node at (-0.5,-2.5) {$\scriptstyle{\alpha_1 \; \downarrow}$};
 \node at (-0.5,-4.5) {$\scriptstyle{\alpha_2\; \downarrow}$};
 \node at (-0.5,-5.5) {$\scriptstyle{\alpha_1 \; \downarrow}$};
\end{tikzpicture}
}
\label{eq:phase_choice_G2_6F}
\end{align}
where $w_i \in [1,0]_{G_2}$, see Appendix \ref{app:G2_roots_weights}. In the phase \eqref{eq:phase_choice_G2_6F}, the prepotential becomes
\begin{align}
\begin{aligned}
  6\Fcal_{5d} &=
  \frac{6 }{g^2 } \left(   3\phi_1^2 -3 \phi_1 \phi_2 + \phi_2^2 \right)
  +8 \phi_1^3-6 \phi_2^2 \phi_1+2 \phi_2^3   \\
  &\quad -3 \sum_{f=1}^6 m_f  ( 4\phi_1^2- 4\phi_1 \phi_2 + \phi_2^2 )
  -3 \sum_{f=1}^6 m_f^2 \phi_2
    \,,
\end{aligned}
\label{eq:prepot_field_theory_G2_6F}
 \end{align}
 where $g$ is the gauge coupling and $m_f$ are the mass parameter.
 \paragraph{Geometry.}
The geometric description for \eqref{eq:6d_SU3_on_1_Z2_twist} is given by \cite{Bhardwaj:2019jtr}
\begin{align}
 \raisebox{-.5\height}{
 \begin{tikzpicture}
  \node (v0) at (0,0) {$\mathbf{2}_{0}^6$ };  
  \node (v1) at (5,0) {$\mathbf{1}_{2}$};  
  \draw  (v0) edge (v1);
%
  \node at (1.25,0.25) {$\scriptstyle{3e+4f-2\sum_i x_i}$};
  \node at (4.5,0.25) {$\scriptstyle{e}$};
 \end{tikzpicture}
 }
 \label{eq:geom_SU3_on_1_Z2_6d_frame}
\end{align}
and one verifies that
\begin{align}
\label{eq:6d_fibre_SU3_on_1_Z2_twist}
    -\begin{pmatrix}
      f_1 \cdot K_{S_1} & f_1 \cdot S_2|_{S_1} \\
      f_2 \cdot S_1|_{S_2}  &f_2 \cdot K_{S_2} 
    \end{pmatrix}
    = \begin{pmatrix}
      2 & -1 \\ -4 & 2
    \end{pmatrix}
    \equiv  C_{\widehat{A}_2^{(2)}}
\end{align}
yields the affine Cartan matrix of $A_2^{(2)}$, which is consistent with the 6d gauge algebra $\surmL(3)$ accompanied by a $\Z_2$ twist.

The parameters of the 5d $G_2$ gauge theory, one coupling $g$ and six mass parameters $m_f$, have to be incorporated into the K\"ahler form 
\begin{align}
    J= - \frac{1}{g^2}F+ \sum_{i=1}^2 \phi_i S_i +\sum_{f=1}^6 m_f  N_f 
    \label{eq:Kahler_form_G2_6F}
\end{align}
for a complete description. 
Before proceeding, it is useful to recall that the dual 5d frame can be geometrically realised by an $\FF_0$ isomorphism $e \leftrightarrow f$ on $S_2$, see for instance \cite{Bhardwaj:2019fzv, Kim:2020hhh}. The immediate consequence is that 
\begin{align}
    - \begin{pmatrix}
      f_1 \cdot K_{S_1} & f_1 \cdot S_2|_{S_1} \\
       e_2 \cdot S_1|_{S_2} &  e_2 \cdot K_{S_2} \\
    \end{pmatrix}
    = \begin{pmatrix}
      2 & -1 \\ -3 & 2
    \end{pmatrix}
    \equiv  C_{G_2}
    \quad 
    \Leftrightarrow
    \quad
    \begin{cases}
    -J_\phi \cdot f_1 = \langle\alpha_1,\phi \rangle \\
    -J_\phi \cdot e_2 = \langle\alpha_2,\phi  \rangle 
    \end{cases}
    \label{eq:G2_6F_roots}
\end{align}
i.e.\ $f_1$ and $e_2$ correspond to the simple roots $\alpha_{1,2}$ of $G_2$, see Appendix \ref{app:G2_roots_weights}.

The full K\"ahler form \eqref{eq:Kahler_form_G2_6F} restricted to the compact surfaces $S_i$ can be parametrised as 
\begin{align}
\label{eq:Kahler_form_G2_6F_try}
\begin{aligned}
    J|_{S_1} &= \phi_1 K_{S_1} + \phi_2 S_2|_{S_1} +a_1 e_1 + b_1 f_1 \,,\\
    J|_{S_1} &=  \phi_1 S_1|_{S_2} + \phi_2 K_{S_2} +a_2 e_2 + b_2 f_2  
    + \sum_{f=1}^6 c_f x_f \,.
    \end{aligned}
\end{align}
The parameters are determined as follows:
\begin{compactitem}
 \item The volume of $-1$ curves need to reproduce the masses of fundamental BPS particles. In detail, the truncated volume of the blowups are give by
 \begin{align}
     -J_\phi \cdot x_i = -2\phi_1+ \phi_2  = \langle w_5 ,\phi \rangle
 \end{align}
 where $w_5 \in [1,0]_{G_2}$, see Appendix \ref{app:G2_roots_weights}. Recalling \eqref{eq:G2_6F_roots}, the remaining weights can be constructed by acting with the simple roots of $G_2$, i.e.\ adding $f_1$ or $e_2$ to the blowups $x_f$.  The resulting $-1$ curves need to have volumes that match the BPS masses as follows:
\begin{align}
     \raisebox{-.5\height}{
\begin{tikzpicture}
\node[draw,circle,inner sep=0.8pt,fill,black]  (v1) at (0,0) {};
 \node[draw,circle,inner sep=0.8pt,fill,black]  (v2) at (0,-1) {};
 \node[draw,circle,inner sep=0.8pt,fill,black]  (v3) at (0,-2) {};
 \node[draw,circle,inner sep=0.8pt,fill,black]  (v4) at (0,-3) {};
 \node[draw,circle,inner sep=0.8pt,fill,black]  (v5) at (0,-4) {};
 \node[draw,circle,inner sep=0.8pt,fill,black]  (v6) at (0,-5) {};
 \node[draw,circle,inner sep=0.8pt,fill,black]  (v7) at (0,-6) {};
 \draw (v1)--(v2)--(v3)--(v4)--(v5)--(v6)--(v7);
 \draw[red,dashed] (-0.2,-4.5)--(5,-4.5);
 \node [right=1ex of v1] {$  \vol (3f_1 +e_2+ x_f)  = \langle w_1, \phi\rangle +m_f$};
 \node [right=1ex of v2] {$ \vol (2f_1 +e_2+ x_f)  = \langle w_2, \phi\rangle +m_f$};
 \node [right=1ex of v3] {$  \vol (2f_1+ x_f)  = \langle w_3 , \phi\rangle+m_f $};
 \node [right=1ex of v4] {$\vol( f_1+ x_f ) = \langle w_4 , \phi\rangle+m_f $};
  \node [right=1ex of v5] {$  \vol( x_f)  = \langle w_5 , \phi\rangle+m_f$};
  \node [right=1ex of v6] {$\vol (e_2-x_f)   = - \left( \langle w_6 , \phi\rangle +m_f \right)$};
  \node [right=1ex of v7] {$\vol (f_1+e_2-x_f)   = - \left( \langle w_7 , \phi\rangle  +m_f \right) $};
 \node at (-0.5,-0.5) {$\scriptstyle{\alpha_1 \; \downarrow}$};
 \node at (-0.5,-1.5) {$\scriptstyle{\alpha_2\; \downarrow}$};
 \node at (-0.5,-2.5) {$\scriptstyle{\alpha_1 \; \downarrow}$};
 \node at (-0.5,-3.5) {$\scriptstyle{\alpha_1 \; \downarrow}$};
 \node at (-0.5,-4.5) {$\scriptstyle{\alpha_2\; \downarrow}$};
 \node at (-0.5,-5.5) {$\scriptstyle{\alpha_1 \; \downarrow}$};
\end{tikzpicture}
}
\quad \Rightarrow \quad 
     \begin{cases}
     c_f=M_f \;, & f=1,\ldots,6  \,,\\
     a_1 = b_2 =0 \;.
     \end{cases}
     \label{eq:weights_via_volumes_G2+F}
 \end{align}
 Note that this identification motivates the phase \eqref{eq:phase_choice_G2_6F}.
\item The geometric effective coupling \eqref{eq:effectice_coupling_geom}
needs to match the field theory expectation \eqref{eq:effective_coupling}. The arising linear equations are solved by
\begin{align}
    b_1 =-\frac{3}{g^2} +2\sum_{f=1}^6 m_f
    \;, \quad
    a_2 =-\frac{1}{g^2} \,.
\end{align}
\end{compactitem}
As a result, the non-compact surfaces $F$ and $N_f$ in the K\"ahler form \eqref{eq:Kahler_form_G2_6F}  are glued to the compact surfaces as follows:
\begin{align}
F|_{S_i} =\begin{cases} 
3 f_1 \;, & i=1 \,,\\
 e_2\;,  & i=2 \,,\\
\end{cases} 
\qquad 
N_f |_{S_i} =
\begin{cases}
2f_1 \;, & i=1 \,, \\
x_f\;,  & i=2\,.
\end{cases}
\end{align}
An immediate observation is that $F|_{S_i}$ restricts to $3f_1$ and $e_2$ respectively, because $f_1$, $e_2$ act as simple roots \eqref{eq:G2_6F_roots}. Also, the prefactors are equal to $\frac{2}{\langle\alpha_i,\alpha_i \rangle}=3,1$ for the roots of $G_2$.

Moreover, one can keep track of the volume of the elliptic fibre, which in the 6d frame \eqref{eq:geom_SU3_on_1_Z2_6d_frame} is given by
 \begin{align}
     f_{\mathrm{ell}} = d_0 f_1 + d_1 f_2 = 2f_1 + f_2
     \qquad 
    \vol(f_{\mathrm{ell}}) = \frac{1}{g^2}
 \end{align}
 with $d_i =$ Coxeter numbers of $A_2^{(2)}$.
%
\subsection{\texorpdfstring{6d $\sprm(1)$ + 10 F -- 5d $\sprm(2)$ +10F}{6d SU(3) +10F - Sp(2) +10F} }
\label{sec:Sp1_on_1_no_twist}
The 6d theory corresponding to a $-1$ curves with an $\sprmL(1)$ gauge algebra has $10$ fundamental hypermultiplets. The untwisted circle compactification of
\begin{align}
     \raisebox{-.5\height}{
 \begin{tikzpicture}
  \node (a1) at (0,0) {$\mathbf{1}$};
  \node at (0,0.35) {$\scriptstyle{ \sprmL(1)^{(1)}}$};
   \end{tikzpicture}
 }
 \label{eq:6d_sp(1)_on_1_no_twist}
\end{align}
is known to have two 5d gauge theory descriptions \cite{Hayashi:2015zka}: $\surm(3)_{0}$ with 10 fundamentals as well as $\sprm(2)$ with $10$ fundamental hypermultiplets.
In this section, the 5d $\sprm(2)$ theory is considered, because its geometric description is related to the 6d theory via fibre-base duality. 
\paragraph{5d description.}
The 5d gauge theory description is based on the prepotential \eqref{eq:F}, which can be evaluated via Appendix \ref{app:Sp_roots_weights}.
Besides restricting to the Weyl-chamber of $C_2$, one needs to choose a suitable phase, such as
\begin{align}
    \raisebox{-.5\height}{
\begin{tikzpicture}
\node[draw,circle,inner sep=0.8pt,fill,black]  (w1) at (0,0) {};
 \node[draw,circle,inner sep=0.8pt,fill,black]  (w2) at (0,-1) {};
 \node[draw,circle,inner sep=0.8pt,fill,black]  (w3) at (0,-2) {};
 \node[draw,circle,inner sep=0.8pt,fill,black]  (w4) at (0,-3) {};
 \draw (w1)--(w2)--(w3)--(w4);
 \draw[red,dashed] (-0.2,-1.5)--(3,-1.5);
 \node[red] at (3.5,-1.5) {$\substack{\text{phase}\\ \text{choice}}$};
 \node [right=1ex of w1] {$\langle \phi,w_1 \rangle +m_f \geq 0$};
 \node [right=1ex of w2] {$ \langle \phi,w_2 \rangle +m_f \geq 0$};
 \node [right=1ex of w3] {$\langle \phi,w_3 \rangle +m_f \leq 0$};
 \node [right=1ex of w4] {$ \langle \phi,w_4 \rangle +m_f  \leq 0$};
 \node at (-0.5,-0.5) {$\scriptstyle{\alpha_1 \; \downarrow}$};
 \node at (-0.5,-1.5) {$\scriptstyle{\alpha_2\; \downarrow}$};
 \node at (-0.5,-2.5) {$\scriptstyle{\alpha_1 \; \downarrow}$};
\end{tikzpicture}
}
\label{eq:phase_choice_Sp2_10F}
\end{align}
where $w_i \in [1,0]_C$, see Appendix \ref{app:Sp_roots_weights}.
The prepotential becomes
\begin{align}
  6\Fcal_{5d} &=
   \frac{6}{g^2} \left( 2\phi_1^2  - 2\phi_1 \phi_2 + \phi_2^2 \right)
 +8 \phi_1^3 
 -18 \phi_1^2 \phi_2
 +12 \phi_1 \phi_2^2
 -2 \phi_2^3 
 -3 \sum_{f=1}^{10} m_f \phi_1^2
 \,,
\label{eq:prepot_field_theory_Sp2_10F}
 \end{align}
with $g$ the gauge coupling, and $m_f$ the mass parameters of the fundamental hypermultiplets.
\paragraph{Geometry.}
Geometrically, the theory \eqref{eq:6d_sp(1)_on_1_no_twist} is described by \cite{Bhardwaj:2019fzv}
\begin{align}
 \raisebox{-.5\height}{
 \begin{tikzpicture}
  \node (v0) at (0,0) {$\mathbf{1}_{0}$ };  
  \node (v1) at (4,0) {$\mathbf{2}_{1}^{10}$};  
  \draw  (v0) edge (v1);
%
  \node at (1,0.25) {$\scriptstyle{2e_1+f_1}$};
  \node at (2.75,0.25) {$\scriptstyle{2h_2-\sum_{i=1}^{10} x_i }$};
 \end{tikzpicture}
 }
 \label{eq:geom_Sp1_on_1_curve_no_twist_6d}
\end{align}
and one verifies straightforwardly that 
\begin{align}
\label{eq:6d_fibre_Sp1_on_1_no_twist}
    -\begin{pmatrix}
      f_1 \cdot K_{S_1} & f_1 \cdot S_2|_{S_1} \\
      f_2 \cdot S_1|_{S_2}  &f_2 \cdot K_{S_2} 
    \end{pmatrix}
    = \begin{pmatrix}
      2 & -2 \\ -2 & 2
    \end{pmatrix} 
    \equiv  C_{\widetilde{A}_1^{(1)}}
\end{align}
which is consistent with the 6d gauge algebra $A_1 \cong C_1$ without any twist.

In view of the 5d description, a useful observation on the geometry \eqref{eq:geom_Sp1_on_1_curve_no_twist_6d} is 
\begin{align}
\label{eq:roots_via_volume_Sp2_10F}
    -\begin{pmatrix}
      e_1 \cdot K_{S_1} & e_1 \cdot S_2|_{S_1} \\
      f_2 \cdot S_1|_{S_2}  &f_2 \cdot K_{S_2} 
    \end{pmatrix}
    = \begin{pmatrix}
      2 & -1 \\ -2 & 2
    \end{pmatrix} 
    \equiv  C_{C_2}
    \quad \Rightarrow \quad 
    \begin{cases}
    -J_\phi \cdot e_1 &= \langle \alpha_1 ,\phi \rangle \\
    -J_\phi \cdot f_2 &= \langle \alpha_2 ,\phi \rangle 
    \end{cases}
\end{align}
which identifies $e_1$, $f_2$ as the simple roots $\alpha_i$ of the 5d $\sprm(2)$ gauge group, see Appendix \ref{app:Sp_roots_weights}. 

In order the include all gauge theory parameters, one gauge coupling $g$ and ten mass parameters $m_i$, one may parametrise the K\"ahler form as
\begin{align}
\label{eq:Kahler_form_Sp2_10F}
    J= - \frac{1}{g^2}F+ \sum_{i=1}^2 \phi_i S_i + \sum_{f=1}^{10} \frac{m_f}{2}  N_f \,.
\end{align}
The gluing curves of the non-compact $F$, $N_f$ with the compact $S_i$ are determined by parametrising the restrictions of \eqref{eq:Kahler_form_Sp2_10F} via
\begin{align}
\label{eq:Kahler_form_Sp2_10F_try}
\begin{aligned}
    J|_{S_1} &= \phi_1 K_{S_1} + \phi_2 S_2|_{S_1} +a_1 e_1 + b_1 f_1 \,,\\
    J|_{S_2} &=  \phi_1 S_1|_{S_2} + \phi_2 K_{S_2} +a_2 e_2 + b_2 f_2  
    + \sum_{f=1}^{10} c_f x_f\,.
    \end{aligned}
\end{align}
The parameters are determined as follows:
\begin{compactitem}
 \item The volume of $-1$ curves should match the BPS masses. It is instructive to consider the blowups $x_i$
 \begin{align}
      - J_\phi \cdot x_f = -\phi_1 +\phi_2 = \langle w_2 ,\phi \rangle  \,,
 \end{align}
 where $w_2 \in [1,0]_C$, see Appendix \ref{app:Sp_roots_weights}. Recalling \eqref{eq:roots_via_volume_Sp2_10F}, the remaining weights are realised by adding $e_1$ or $f_2$ to the blowups. The volumes of the resulting $-1$ curves are required to match the BPS masses
\begin{align}
    \raisebox{-.5\height}{
\begin{tikzpicture}
\node[draw,circle,inner sep=0.8pt,fill,black]  (w1) at (0,0) {};
 \node[draw,circle,inner sep=0.8pt,fill,black]  (w2) at (0,-1) {};
 \node[draw,circle,inner sep=0.8pt,fill,black]  (w3) at (0,-2) {};
 \node[draw,circle,inner sep=0.8pt,fill,black]  (w4) at (0,-3) {};
 \draw (w1)--(w2)--(w3)--(w4);
 \draw[red,dashed] (-0.2,-1.5)--(5,-1.5);
 \node [right=1ex of w1] {$\vol(e_1 +x_f) \stackrel{!}{=} \langle w_1 ,\phi \rangle +m_f$};
 \node [right=1ex of w2] {$\vol(x_f) \stackrel{!}{=} \langle w_2 ,\phi \rangle +m_f$};
 \node [right=1ex of w3] {$\vol(f_2 -x_f) \stackrel{!}{=} -
 \left(\langle w_3 ,\phi \rangle +m_f\right)$};
 \node [right=1ex of w4] {$ \vol(e_1 +f_2-x_f) \stackrel{!}{=} -\left( \langle w_4 ,\phi \rangle +m_f \right)$};
 \node at (-0.5,-0.5) {$\scriptstyle{\alpha_1 \; \downarrow}$};
 \node at (-0.5,-1.5) {$\scriptstyle{\alpha_2\; \downarrow}$};
 \node at (-0.5,-2.5) {$\scriptstyle{\alpha_1 \; \downarrow}$};
\end{tikzpicture}
}
\quad \Rightarrow \quad 
   \begin{cases}
  c_f=m_f\,,  & f=1,\ldots,10 \,, \\
  b_1 = a_2 =0 \,.
   \end{cases}
\end{align}
Again, the volume the $-1$ define the phase of the theory, which motivates \eqref{eq:phase_choice_Sp2_10F}.
\item The geometric effective coupling \eqref{eq:effectice_coupling_geom} needs to match the field theory expectation \eqref{eq:effective_coupling}. The arising linear equations are solved by
\begin{align}
    a_1=  -\frac{2}{g^2}
    \;, \quad
    b_2 = -\frac{1}{g^2}- \frac{1}{2} \sum_{f=1}^{10} m_f
     \,.
\end{align}
\end{compactitem}
In summary, the non-compact surfaces $F$ and $N_f$ in \eqref{eq:Kahler_form_Sp2_10F} are glued to the $S_i$ as follows:
\begin{align}
F|_{S_i} = \begin{cases}
2e_1\,, & i=1 \,,\\
f_2 \,, & i=2\,,
\end{cases}  
\qquad 
N_f |_{S_i} =
\begin{cases}
0 \,,& i=1 \,,\\
2x_f -f_2 \,, & i=2 \,.
\end{cases}
\end{align}
Again, the $F|_{S_i}$ restrict to $2e_1$, and $f_2$ respectively, because the $e_1$, $f_2$ act as simple roots \eqref{eq:roots_via_volume_Sp2_10F}. The prefactors equal $\frac{2}{\langle\alpha_i,\alpha_i \rangle}=2,1$ for the roots of $C_2$.
In addition, one may keep track of the elliptic fibre associated to \eqref{eq:geom_Sp1_on_1_curve_no_twist_6d} and compute its volume to be
\begin{align}
    \vol(f_{\mathrm{ell}}) = \vol(f_1)+\vol(f_2) = \frac{2}{g^2} \;.
\end{align}
\subsection{\texorpdfstring{6d Rank-2 E-string -- 5d $\sprm(2)$ + 1$
\Lambda^2$ + 8F}{Rank-2 E-string -- 5d Sp(2) + 1AS + 8F} }
The rank-2 E-string is given by a $-1$ curve with an $\sprmL(0)$ algebra intersecting a $-2$ curve with a trivial $\surmL(1)$ algebra
\begin{align}
     \raisebox{-.5\height}{
 \begin{tikzpicture}
  \node (a1) at (0,0) {$\mathbf{1}$};
  \node at (0,0.35) {$\scriptstyle{ \sprmL(0)^{(1)}}$};
  \node  (a2) at (1.5,0) {$ \mathbf{2}$};
  \node at (1.5,0.35) {$\scriptstyle{ \surmL(1)^{(1)}}$};
  \draw  (a1) edge (a2);
   \end{tikzpicture}
 }
 \label{eq:geom_6d_rank2_E-string}
\end{align}
The 5d reduction has two gauge theory descriptions \cite{Jefferson:2017ahm}: $\surm(3)_{\frac{3}{2}}$ with nine fundamentals and $\sprm(2)$ with one rank-2 anti-symmetric and eight fundamental hypermultiplets. In this section, the $\sprm(2)$ theory is consider because of the natural fibre-base duality to the 6d frame.

\paragraph{5d description.}
The 5d gauge theory description is based on the prepotential \eqref{eq:F}, which is evaluated via the details provided in Appendix \ref{app:Sp_roots_weights}.
Since the 5d theory has non-trivial matter content, restricting to the Weyl-chamber of $C_2$, i.e.\ $\langle \phi,\alpha_i \rangle \geq 0$ for $i=1,2$, is insufficient for reaching a single, well-defined 5d phase. 
A suitable choice of phase is given by
\begin{align}
\raisebox{-.5\height}{
\begin{tikzpicture}
\node[draw,circle,inner sep=0.8pt,fill,black]  (w1) at (0,0) {};
 \node[draw,circle,inner sep=0.8pt,fill,black]  (w2) at (0,-1) {};
 \node[draw,circle,inner sep=0.8pt,fill,black]  (w3) at (0,-2) {};
 \node[draw,circle,inner sep=0.8pt,fill,black]  (w4) at (0,-3) {};
 \draw (w1)--(w2)--(w3)--(w4);
 \draw[red,dashed] (-0.2,-1.5)--(3,-1.5);
 \node[red] at (3.5,-1.5) {$\substack{\text{phase}\\ \text{choice}}$};
 \node [right=1ex of w1] {$\langle \phi,w_1 \rangle +m_f \geq 0$};
 \node [right=1ex of w2] {$ \langle \phi,w_2 \rangle +m_f \geq 0$};
 \node [right=1ex of w3] {$\langle \phi,w_3 \rangle +m_f \leq 0$};
 \node [right=1ex of w4] {$ \langle \phi,w_4 \rangle +m_f  \leq 0$};
 \node at (-0.5,-0.5) {$\scriptstyle{\alpha_1 \; \downarrow}$};
 \node at (-0.5,-1.5) {$\scriptstyle{\alpha_2\; \downarrow}$};
 \node at (-0.5,-2.5) {$\scriptstyle{\alpha_1 \; \downarrow}$};
\end{tikzpicture}
}
\qquad \text{and} \qquad 
   \raisebox{-.5\height}{
\begin{tikzpicture}
\node[draw,circle,inner sep=0.8pt,fill,black]  (v1) at (0,0) {};
 \node[draw,circle,inner sep=0.8pt,fill,black]  (v2) at (0,-1) {};
 \node[draw,circle,inner sep=0.8pt,fill,black]  (v3) at (0,-2) {};
 \node[draw,circle,inner sep=0.8pt,fill,black]  (v4) at (0,-3) {};
 \node[draw,circle,inner sep=0.8pt,fill,black]  (v5) at (0,-4) {};
 \draw (v1)--(v2)--(v3)--(v4)--(v5);
 \draw[red,dashed] (-0.2,-2.5)--(3,-2.5);
 \node[red] at (3.5,-2.5) {$\substack{\text{phase}\\ \text{choice}}$};
 \node [right=1ex of v1] {$ \langle \phi,v_1 \rangle +m_f \geq 0$};
 \node [right=1ex of v2] {$\langle \phi,v_2 \rangle +m_f \geq 0$};
 \node [right=1ex of v3] {$ \langle \phi,v_3 \rangle +m_f \geq 0$};
 \node [right=1ex of v4] {$\langle \phi,v_4 \rangle +m_f \leq 0$};
 \node [right=1ex of v5] {$ \langle \phi,v_5 \rangle +m_f\leq 0$};
 \node at (-0.5,-0.5) {$\scriptstyle{\alpha_2 \; \downarrow}$};
 \node at (-0.5,-1.5) {$\scriptstyle{\alpha_1\; \downarrow}$};
 \node at (-0.5,-2.5) {$\scriptstyle{\alpha_1 \; \downarrow}$};
 \node at (-0.5,-3.5) {$\scriptstyle{\alpha_2 \; \downarrow}$};
\end{tikzpicture}
}
\label{eq:phase_choice_Sp2_1Lambda2_8F}
\end{align}
with $w_i \in [1,0]_C$ and $v_i \in [0,1]_C$, see Appendix \ref{app:Sp_roots_weights}. In the phase \eqref{eq:phase_choice_Sp2_1Lambda2_8F}, the prepotential becomes
\begin{align}
\begin{aligned}
  6\Fcal_{5d} &=
  \frac{6 }{g^2 } \left(   2\phi_0^2 -2 \phi_0 \phi_1 + \phi_1^2 \right)
   -3 \sum_{f=1}^8 m_f^2 \phi_1
  -6 M^2 \phi_0
   \,,
\end{aligned}
\label{eq:prepot_field_theory_Sp2_1Lambda2_8F}
 \end{align}
where $g$ denotes the gauge coupling, $m_f$ are the mass parameters of the fundamental hypermultiplets, and $M$ is the mass parameter of the $\Lambda^2$ hypermultiplet.
\paragraph{Geometry.}
Utilising \cite{Bhardwaj:2019fzv}, the geometry from 6d setup \eqref{eq:geom_6d_rank2_E-string} is given by
\begin{align}
\raisebox{-.5\height}{
 \begin{tikzpicture}
  \node (v0) at (0,0) {$\mathbf{0}_{0}^{1+1}$};  
  \node (v1) at (4,0){$\mathbf{1}_{1}^{8}$}; 
  \draw  (v0) edge (v1);
    \draw (v0) to [out=140,in=220,looseness=6] (v0);
      \node at (-1.5,0.7) {$\scriptstyle{e_0-w}$};
        \node at (-1.5,-0.7) {$\scriptstyle{e_0-z}$};
  \node at (0.75,0.17) {$\scriptstyle{f_0}$};
  \node at (2.5,0.17) {$\scriptstyle{2e_1+3f_1-\sum_{i=1}^8 x_i}$};
 \end{tikzpicture}
 }
 \label{eq:geom_Sp0_on_1_SU1_on_2_6d}
\end{align}
and one computes immediately 
\begin{align}
\label{eq:roots_via_volumes_Sp2_8F_1Lambda}
    -\begin{pmatrix}
      e_0 \cdot K_{S_1} & e_0 \cdot S_2|_{S_1} \\
      f_1 \cdot S_1|_{S_2}  &f_1 \cdot K_{S_2} 
    \end{pmatrix}
    = \begin{pmatrix}
      2 & -1 \\ -2 & 2
    \end{pmatrix}
    \equiv  C_{C_2}
    \quad 
    \Rightarrow \quad 
    \begin{cases}
    -J_\phi \cdot e_0 &= \langle \alpha_1 ,\phi \rangle \\ 
        -J_\phi \cdot f_1 &= \langle \alpha_2 ,\phi \rangle 
    \end{cases}
\end{align}
which identifies the Cartan matrix of the 5d $\sprm(2)$ gauge group. Moreover, $\{e_0,\, f_1\}$ act as simple roots $\alpha_i$ of $C_2$, see Appendix \ref{app:Sp_roots_weights}. On the other hand, The $6d$ gauge algebras are trivial on both nodes of  \eqref{eq:geom_6d_rank2_E-string}. This piece of information is encoded in the elliptic fibre $f_{\rm ell}$ of the Calabi-Yau. One can readily confirm it by noticing that
\begin{align}
    -f_0\cdot S_i=K_{S_1}\cdot S_i=0\,,
\end{align}
where $K_{S_1}=-2e_1-3f_1-\sum_{i=1}^8 x_i$ is the canonical K\"ahler form of $\FF_1^8$, as well as the gluing curve identified with $f_0$ in the geometry \eqref{eq:geom_Sp0_on_1_SU1_on_2_6d}. Therefore the elliptic fibre is given by
\begin{align}
    f_{\rm ell}=f_0=-K_{S_1}\,.
\end{align}
The full K\"ahler form 
\begin{align}
    J= - \frac{1}{g^2}F+ \sum_{i=0}^1 \phi_i S_i 
    + \sum_{f=1}^8 \frac{m_f}{2}  N^F_f 
    +M  N^{\Lambda} 
    \label{eq:Kahler_form_Sp2_1Lambda_8F}
\end{align}
includes all gauge theory parameter $g$, $m_f$, and $M$. To derive the gluing curves of the non-compact surfaces with the compact surfaces, one parametrises \eqref{eq:Kahler_form_Sp2_1Lambda_8F} as follows:
\begin{align}
\label{eq:Kahler_form_Sp2_1Lambda_try}
\begin{aligned}
    J|_{S_0} &= \phi_0 K_{S_0} + \phi_1 S_1|_{S_0} +a_0 e_0 + b_0 f_0 + d_w w+ d_z z \,,\\
    J|_{S_1} &=  \phi_0 S_0|_{S_1} + \phi_1 K_{S_1} +a_1 e_1 + b_1 f_1  
    + \sum_{f=1}^8 c_f x_f \,.
    \end{aligned}
\end{align}
The parameters are determined as follows:
\begin{compactitem}
 \item The volumes of $-1$ curves should reproduce the masses of fundamental BPS particles. To get some insight, one computes the truncated volume of the blowups
 \begin{subequations}
 \begin{align}
     -J_\phi \cdot x_f &= -\phi_0+ \phi_1  = \langle w_2, \phi \rangle  \,,\\
     -J_\phi \cdot w &= -J_\phi \cdot z = 0  = \langle v_3, \phi \rangle \,,
 \end{align}
 \end{subequations}
 where $w_2 \in [1,0]_C$ and $v_3 \in [0,1]_C$, see Appendix \ref{app:Sp_roots_weights}. Recalling \eqref{eq:roots_via_volumes_Sp2_8F_1Lambda}, one constructs the remaining $-1$ curves that furnish the representations $[1,0]_C$ and $[0,1]_C$ by adding the base $e_0$ and fibre $f_1$ to the blowups. The volumes of the resulting curves need to satisfy
\begin{align}
\raisebox{-.5\height}{
\begin{tikzpicture}
\node[draw,circle,inner sep=0.8pt,fill,black]  (w1) at (0,0) {};
 \node[draw,circle,inner sep=0.8pt,fill,black]  (w2) at (0,-1) {};
 \node[draw,circle,inner sep=0.8pt,fill,black]  (w3) at (0,-2) {};
 \node[draw,circle,inner sep=0.8pt,fill,black]  (w4) at (0,-3) {};
 \draw (w1)--(w2)--(w3)--(w4);
 \draw[red,dashed] (-0.2,-1.5)--(5,-1.5);
 \node [right=1ex of w1] {$\vol (e_0 + x_f) \stackrel{!}{=} \langle w_1 ,\phi \rangle +m_f$};
 \node [right=1ex of w2] {$ \vol (x_f) \stackrel{!}{=} \langle w_2 ,\phi \rangle +m_f$};
 \node [right=1ex of w3] {$\vol (f_1-x_f) \stackrel{!}{=} - \left( \langle w_3 ,\phi \rangle +m_f\right)$};
 \node [right=1ex of w4] {$\vol (e_0 + f_1 -x_f)  \stackrel{!}{=} - \left(\langle w_4 ,\phi \rangle  +m_f \right)$};
 \node at (-0.5,-0.5) {$\scriptstyle{\alpha_1 \; \downarrow}$};
 \node at (-0.5,-1.5) {$\scriptstyle{\alpha_2\; \downarrow}$};
 \node at (-0.5,-2.5) {$\scriptstyle{\alpha_1 \; \downarrow}$};
\end{tikzpicture}
}
   \raisebox{-.5\height}{
\begin{tikzpicture}
\node[draw,circle,inner sep=0.8pt,fill,black]  (v1) at (0,0) {};
 \node[draw,circle,inner sep=0.8pt,fill,black]  (v2) at (0,-1) {};
 \node[draw,circle,inner sep=0.8pt,fill,black]  (v3) at (0,-2) {};
 \node[draw,circle,inner sep=0.8pt,fill,black]  (v4) at (0,-3) {};
 \node[draw,circle,inner sep=0.8pt,fill,black]  (v5) at (0,-4) {};
 \draw (v1)--(v2)--(v3)--(v4)--(v5);
 \draw[red,dashed] (-0.2,-2.5)--(5,-2.5);
 \node [right=1ex of v1] {$\vol (e_0 +f_1 + w) \stackrel{!}{=} \langle v_1 ,\phi \rangle +M$};
 \node [right=1ex of v2] {$\vol  (e_0 + w) \stackrel{!}{=} \langle v_2 ,\phi \rangle +M$};
 \node [right=1ex of v3] {$\vol  (w) \stackrel{!}{=} \langle v_3 ,\phi \rangle +M$};
 \node [right=1ex of v4] {$\vol  (e_0-w) \stackrel{!}{=} -\langle v_4 ,\phi \rangle +M$};
 \node [right=1ex of v5] {$\vol  (e_0+f_1-w) \stackrel{!}{=} -\left(\langle v_5 ,\phi \rangle +M\right)$};
 \node at (-0.5,-0.5) {$\scriptstyle{\alpha_2 \; \downarrow}$};
 \node at (-0.5,-1.5) {$\scriptstyle{\alpha_1\; \downarrow}$};
 \node at (-0.5,-2.5) {$\scriptstyle{\alpha_1 \; \downarrow}$};
 \node at (-0.5,-3.5) {$\scriptstyle{\alpha_2 \; \downarrow}$};
\end{tikzpicture}
}
 \label{eq:weights_via_volume_Sp2_1Lambda_8F}
\end{align}
such that one finds
\begin{align}
    b_0=a_1=0  
    \;, \qquad 
     c_f=m_f \;,  \quad f=1,\ldots 8 
     \;, \qquad
     d_w=d_z=M \;.
\end{align}
The volumes of these $-1$ curves define the phase of the theory, which motivates \eqref{eq:phase_choice_Sp2_1Lambda2_8F}
\item The geometric effective coupling \eqref{eq:effectice_coupling_geom} needs to match the field theory expectation \eqref{eq:effective_coupling}. The arising linear equations are solved by
\begin{align}
    a_0 =-\frac{2}{g^2}
    \;, \quad
    b_1 =-\frac{1}{g^2}-\frac{1}{2} \sum_{f=1}^8 m_f \,.
\end{align}
\end{compactitem}
As a result, the gluing curves of the non-compact surfaces $F$, $N^{F}_f$, and $N^{\Lambda}$,  in \eqref{eq:Kahler_form_Sp2_1Lambda_8F} with the compact surfaces $S_i$ are as follows:
\begin{align}
F|_{S_i} =\begin{cases} 
2 e_0\;, & i=0\,, \\
 f_1\;, & i=1 \,,\\
\end{cases}
\quad 
N^F_f |_{S_i} =
\begin{cases}
 0\;,  & i=0 \,,\\
 2x_{f}-f_1\;, & i=1\,,
\end{cases}
\quad 
N^{\Lambda} |_{S_i} =
\begin{cases}
w+z\;, & i=0 \,,\\
0\;, & i=1\,.
\end{cases}
\end{align}
An immediate observation is that $F|_{S_i}$ restrict to $2e_0$, $f_1$ respectively, because $\{e_0,\, f_1\}$ act as simple roots \eqref{eq:roots_via_volumes_Sp2_8F_1Lambda}. The prefactors equal $\frac{2}{\langle\alpha_i,\alpha_i \rangle}=2,1$ for the roots of $C_2$. Moreover, for the $N_f^F$ divisors restricted to $S_1$, the volumes of curves $2x_f-f_1$ have positive volumes ${\rm Vol}(2x_f-f_1)=m_f$, implying the curves are effective. In addition, one may keep track of the elliptic fibre associated to \eqref{eq:geom_Sp0_on_1_SU1_on_2_6d} and compute its volume to be
\begin{align}
    \vol(f_{\mathrm{ell}}) = \vol(f_0)=\vol(-K_{S_1}) = \frac{2}{g^2} \;.
\end{align}
The factor of $2$ would be due to the node of $-1$ curve in \eqref{eq:geom_6d_rank2_E-string}, which has been also observed in rank-1 E-string, see \eqref{eq:volume_of_ellipitic_fibre_of_Estring}.
%
\subsection{\texorpdfstring{6d $\Ncal=(2,0)$ $A_2$ -- 5d $\surm(3)_{0}$ +1Adj, }{6d A2 - 5d SU(3) +1Adj} }
The 6d $\Ncal=(2,0)$ $A_2$ theory is given by
\begin{align}
 \raisebox{-.5\height}{
 \begin{tikzpicture}
  \node (a1) at (0,0) {$\mathbf{2}$};
  \node at (0,0.35) {$\scriptstyle{ \surmL(1)^{(1)}}$};
  \node  (a2) at (1.5,0) {$ \mathbf{2}$};
  \node at (1.5,0.35) {$\scriptstyle{ \surmL(1)^{(1)}}$};
  \draw  (a1) edge (a2);
   \end{tikzpicture}
 }
 \label{eq:6d_su1_on_2_edge_su1_on_2}
\end{align}
and the untwisted circle reduction is known to be $\surm(3)_0$ with one hypermultiplet in the adjoint representation \cite{Douglas:2010iu,Lambert:2010iw}.
\paragraph{5d description.}
Since the 5d KK descriptions is known, the prepotential \eqref{eq:F} is derived using Appendix \ref{app:SU_roots_weights}.
In addition to restricting to the Weyl-chamber of $A_2$, i.e.\ $\langle \phi,\alpha_i \rangle \geq 0$ for $i=1,2$, the following phase is chosen 
\begin{align}
    \raisebox{-.5\height}{
\begin{tikzpicture}
 \node[draw,circle,inner sep=0.8pt,fill,black]  (a2) at (0,-1) {};
 \node[draw,circle,inner sep=0.8pt,fill,black]  (a3L) at (-1,-2) {};
 \node[draw,circle,inner sep=0.8pt,fill,black]  (a4L) at (-1,-3) {};
 \node[draw,circle,inner sep=0.8pt,fill,black]  (a5L) at (-1,-4) {};
 \node[draw,circle,inner sep=0.8pt,fill,black]  (a3R) at (1,-2) {};
 \node[draw,circle,inner sep=0.8pt,fill,black]  (a4R) at (1,-3) {};
 \node[draw,circle,inner sep=0.8pt,fill,black]  (a5R) at (1,-4) {};
 \node[draw,circle,inner sep=0.8pt,fill,black]  (a6) at (0,-5) {};
 \draw (a2)--(a3L)--(a4L)--(a5L)--(a6) (a2)--(a3R)--(a4R)--(a5R)--(a6);
 \draw[red,dashed] (-3.5,-3.5)--(-0.85,-3.5) (-0.3,-3.5)--(0.3,-3.5) (0.85,-3.5)--(4.25,-3.5);
 \node[red] at (4.75,-3.5) {$\substack{\text{phase}\\ \text{choice}}$};
 \node [right=1ex of a2] {$  \langle \phi,\alpha_1 +\alpha_2 \rangle +m_f  \geq 0$};
 \node [left=1ex of a3L] {$\langle \phi,\alpha_2 \rangle +m_f  \geq 0$};
 \node [right=1ex of a3R] {$\langle \phi,\alpha_1 \rangle +m_f  \geq 0$};
 \node [left=1ex of a4L] {$ \langle \phi,0 \rangle +m_f \geq 0 $};
 \node [right=1ex of a4R] {$ \langle \phi,0 \rangle +m_f \geq 0$};
 \node [left=1ex of a5L] {$ \langle \phi, -\alpha_2 \rangle +m_f  \leq 0$};
 \node [right=1ex of a5R] {$ \langle \phi, -\alpha_1 \rangle +m_f  \leq 0$};
 \node [right=1ex of a6] {$ \langle \phi,-(\alpha_1 +\alpha_2) \rangle +m_f  \leq 0 $};
 \node at (-0.75,-1.25) {$\scriptstyle{\alpha_1\; \swarrow}$};
 \node at (0.25,-1.75) {$\scriptstyle{\alpha_2\; \searrow}$};
 \node at (-0.6,-2.5) {$\scriptstyle{\downarrow \; \alpha_2  }$};
 \node at (-0.6,-3.5) {$\scriptstyle{\downarrow \;\alpha_2}$};
 \node at (0.6,-2.5) {$\scriptstyle{ \alpha_1 \; \downarrow }$};
 \node at (0.6,-3.5) {$\scriptstyle{\alpha_1 \; \downarrow}$};
 \node at (-0.75,-4.75) {$\scriptstyle{\alpha_1\; \searrow}$};
 \node at (0.25,-4.25) {$\scriptstyle{\alpha_2\; \swarrow}$};
\end{tikzpicture}
}
\label{eq:phase_choice_SU3_CS=0_1Adj}
\end{align}
with $\alpha_{1,2}$ the simple roots of $A_2$, see Appendix \ref{app:SU_roots_weights}. The prepotential becomes
\begin{align}
  6\Fcal_{5d} &=
   \frac{6}{g^2} \left( \phi_0^2  - \phi_0 \phi_1 + \phi_1^2 \right)
 -6  m^2 (\phi_0+\phi_1)
 \,,
\label{eq:prepot_field_theory_SU3_CS=0_1Adj}
 \end{align}
 with $g$ the gauge coupling and $m$ the mass parameter of the adjoint hypermultiplet.
\paragraph{Geometry.}
Utilising \cite{Bhardwaj:2019fzv}, the geometry from 6d setup \eqref{eq:6d_su1_on_2_edge_su1_on_2} is given by
\begin{align}
\raisebox{-.5\height}{
 \begin{tikzpicture}
  \node (v0) at (0,0) {$\mathbf{0}_{0}^{1+1}$};  
  \node (v1) at (3,0){$\mathbf{1}_{0}^{1+1}$};  
  \node (gh) at (1.5,0){$\scriptscriptstyle{2}$}; 
  \draw  (v0) edge (gh);
    \draw  (gh) edge (v1);
    \draw (v0) to [out=140,in=220,looseness=6] (v0);
      \node at (-1.5,0.7) {$\scriptstyle{e_0-x}$};
        \node at (-1.5,-0.7) {$\scriptstyle{e_0-y}$};
    \draw (v1) to [out=140-100,in=220+100,looseness=6] (v1);
    \node at (3+1.5,0.7) {$\scriptstyle{e_1-z}$};
        \node at (3+1.5,-0.7) {$\scriptstyle{e_1-w}$};
  \node at (0.75,0.3) {$\scriptstyle{\substack{f_0-x\\ x}}$};
  \node at (2.25,0.3) {$\scriptstyle{\substack{f_1-z\\ z}}$};
 \end{tikzpicture}
 }
 \label{eq:geom_SU1_on_2_SU1_on_2_6d}
\end{align}
such that one readily confirms that 
\begin{align}
 -f_i \cdot S_j = 0    \,,
\end{align}
which is reflecting the fact that the 6d gauge algebras are trivial.
In order to describe 5d $\surm(3)$ gauge theory via \eqref{eq:geom_SU1_on_2_SU1_on_2_6d}, it is useful to observe that 
\begin{align}
\label{eq:roots_via_volumes_SU3_CS=0_1Adj}
    -\begin{pmatrix}
      e_0 \cdot K_{S_0} & e_0 \cdot S_1|_{S_0} \\
      e_1 \cdot S_0|_{S_1}  &e_1 \cdot K_{S_1} 
    \end{pmatrix}
    = \begin{pmatrix}
      2 & -1 \\ -1 & 2
    \end{pmatrix} 
    \equiv  C_{A_2}
    \quad \Rightarrow \quad 
    \begin{cases}
    -J_\phi \cdot e_0 &= \langle \alpha_1 ,\phi \rangle \\
    -J_\phi \cdot e_1 &= \langle \alpha_2 ,\phi \rangle 
    \end{cases}
\end{align}
which identifies the $e_{0,1}$ as the simple roots $\alpha_i$ of the 5d $\surmL(3)$ gauge algebra, see Appendix \ref{app:SU_roots_weights}. As above, the two gauge theory parameter, one gauge coupling $g$ and one mass parameter, are introduced by suitably parametrising the K\"ahler form:
\begin{align}
\label{eq:Kahler_form_SU3_CS=0_1Adj}
    J= - \frac{1}{g^2}F+ \sum_{i=0}^1 \phi_i S_i +  m  N \,.
\end{align}
The intersections of the non-compact $F$, $N$ with the compact $S_i$ are determined by restricting \eqref{eq:Kahler_form_SU3_CS=0_1Adj} as 
\begin{align}
\label{eq:Kahler_form_SU3_CS=0_1Adj_try}
\begin{aligned}
    J|_{S_0} &= \phi_0 K_{S_0} + \phi_1 S_1|_{S_0} +a_0 e_0 + b_0 f_0 + c_x x + c_y y\,,\\
    J|_{S_1} &=  \phi_0 S_0|_{S_1} + \phi_1 K_{S_1} +a_1 e_1 + b_1 f_1  
    +  c_z z + c_w w \,.
    \end{aligned}
\end{align}
and determining the parameter as follows:
\begin{compactitem}
 \item Firstly, one identifies the $-1$ curves that give rise to the fundamental BPS particles. It is instructive  to evaluate the volumes of the blowups; in detail,
 \begin{align}
     -J_\phi \cdot x =  -J_\phi \cdot y = -J_\phi \cdot z  =-J_\phi \cdot w = 0 = \langle 0 ,\phi \rangle
 \end{align}
 where $0$ is a trivial weight in $[1,1]_A$, see Appendix \ref{app:SU_roots_weights}. Recalling \eqref{eq:roots_via_volumes_SU3_CS=0_1Adj}, the remaining weights are realised by the following $-1$ curves:
 \begin{align}
    \raisebox{-.5\height}{
\begin{tikzpicture}
 \node[draw,circle,inner sep=0.8pt,fill,black]  (a2) at (0,-1) {};
 \node[draw,circle,inner sep=0.8pt,fill,black]  (a3L) at (-1,-2) {};
 \node[draw,circle,inner sep=0.8pt,fill,black]  (a4L) at (-1,-3) {};
 \node[draw,circle,inner sep=0.8pt,fill,black]  (a5L) at (-1,-4) {};
 \node[draw,circle,inner sep=0.8pt,fill,black]  (a3R) at (1,-2) {};
 \node[draw,circle,inner sep=0.8pt,fill,black]  (a4R) at (1,-3) {};
 \node[draw,circle,inner sep=0.8pt,fill,black]  (a5R) at (1,-4) {};
 \node[draw,circle,inner sep=0.8pt,fill,black]  (a6) at (0,-5) {};
 \draw (a2)--(a3L)--(a4L)--(a5L)--(a6) (a2)--(a3R)--(a4R)--(a5R)--(a6);
 \draw[red,dashed] (-4.25,-3.5)--(-0.85,-3.5) (-0.3,-3.5)--(0.3,-3.5) (0.85,-3.5)--(4.25,-3.5);
 \node [right=1ex of a2] {$ \substack{ \vol (e_0 +e_1+ x)\\ \vol (e_0 +e_1+ z)} = \langle \alpha_1 +\alpha_2 ,\phi \rangle  +m$};
 \node [left=1ex of a3L] {$\vol (e_1+z)  = \langle \alpha_2,\phi \rangle +m$};
 \node [right=1ex of a3R] {$\vol (e_0+x)  = \langle \alpha_1,\phi \rangle +m$};
 \node [left=1ex of a4L] {$ \vol (z) = -\langle 0 ,\phi \rangle +m $};
 \node [right=1ex of a4R] {$\vol (x) = -\langle 0 ,\phi \rangle +m$};
 \node [left=1ex of a5L] {$ \vol (e_1-z) =- \left(\langle- \alpha_2,\phi \rangle +m\right) $};
 \node [right=1ex of a5R] {$ \vol (e_0-x) =- \left(\langle- \alpha_1,\phi \rangle +m\right)$};
 \node [right=1ex of a6] {$\substack{ \vol (e_0 +e_1- x) \\ \vol (e_0 +e_1- z)  }=-\left( \langle -(\alpha_1 +\alpha_2) ,\phi \rangle+m\right) $};
 \node at (-0.75,-1.25) {$\scriptstyle{\alpha_1\; \swarrow}$};
 \node at (0.25,-1.75) {$\scriptstyle{\alpha_2\; \searrow}$};
 \node at (-0.6,-2.5) {$\scriptstyle{\downarrow \; \alpha_2  }$};
 \node at (-0.6,-3.5) {$\scriptstyle{\downarrow \;\alpha_2}$};
 \node at (0.6,-2.5) {$\scriptstyle{ \alpha_1 \; \downarrow }$};
 \node at (0.6,-3.5) {$\scriptstyle{\alpha_1 \; \downarrow}$};
 \node at (-0.75,-4.75) {$\scriptstyle{\alpha_1\; \searrow}$};
 \node at (0.25,-4.25) {$\scriptstyle{\alpha_2\; \swarrow}$};
\end{tikzpicture}
}
\end{align}
and volumes have to match the BPS masses.
Additionally, the self-gluings in \eqref{eq:geom_SU1_on_2_SU1_on_2_6d} impose
\begin{align}
    \vol (e_0-x) = \vol (e_0-y) 
    \quad \text{and} \quad
 \vol (e_1-z)= \vol (e_1-w)
\end{align}
such that one finds
\begin{align}
     c_I = m  \;, \quad I=x,y,z,w 
     \;, \qquad 
   b_i=0 \;, \quad  i=0,1  \;.
\end{align}
One notes that the choices of volumes motivate the choice of phase \eqref{eq:phase_choice_SU3_CS=0_1Adj}.
\item Secondly, the geometric effective gauge coupling \eqref{eq:effectice_coupling_geom} needs to match the field theory expectation \eqref{eq:effective_coupling}. The arising linear equations are solved by
\begin{align}
    a_0=  a_1 = -\frac{1}{g^2}  \,.
\end{align}
\end{compactitem}
Consequently, the non-compact surfaces $F$, $N$ in \eqref{eq:Kahler_form_SU3_CS=0_1Adj} restrict to the compact $S_i$ as follows:
\begin{align}
F|_{S_i} =e_i  
\,,\quad 
N |_{S_i} =
\begin{cases}
x+y \;, & i=0 \;,\\
z+w \;, & i=1\,.
\end{cases}
\end{align}
One may notice, as above, that the $F|_{S_i}$ restrict to the $e_i$ respectively, because the $e_i$ act as simple roots \eqref{eq:roots_via_volumes_SU3_CS=0_1Adj}. The prefactor of $1$ equals $\frac{2}{\langle\alpha_i,\alpha_i \rangle}$ for the roots of $A_2$.
Moreover, the volumes of the fibres are given by
\begin{align}
    \vol(f_i) = \vol(f_{\mathrm{ell}})= \frac{1}{g^2} \;, \quad  i=0,1 \,,
\end{align}
which identifies the volume of the elliptic fibre.
%
\subsection{\texorpdfstring{6d $\Ncal=(2,0)$ $A_3$ with twist -- 5d $\sprm(2)_{0}$ + 1 Adj}{6d A3 with twist - Sp(2) +1Adj} }
Starting from the 6d $\Ncal=(2,0)$ $A_3$ theory, the twist by the permutation symmetry 
\begin{align}
\raisebox{-.5\height}{
 \begin{tikzpicture}
  \node (b1) at (0,0) {$\mathbf{2}$};
  \node at (0,0.35) {$\scriptstyle{ \surmL(1)^{(1)}}$};
  \node  (b2) at (1.5,0) {$ \mathbf{2}$};
  \node at (1.5,0.35) {$\scriptstyle{ \surmL(1)^{(1)}}$};
  \node (b3) at (3,0) {$\mathbf{2}$};
  \node at (3,0.35) {$\scriptstyle{ \surmL(1)^{(1)}}$};
  \draw  (b1) edge (b2);
  \draw  (b2) edge (b3);
\draw[->] (4.5,0)--(6.5,0);
  \node at (5.5,0.25) {\footnotesize{permutation}};
  \node at (5.5,-0.25) {\footnotesize{twist}};
  \node (a1) at (0+8,0) {$\mathbf{2}$};
  \node at (0+8,0.35) {$\scriptstyle{ \surmL(1)^{(1)}}$};
  \node  (a2) at (1.5+8,0) {$ \mathbf{2}$};
  \node at (1.5+8,0.35) {$\scriptstyle{ \surmL(1)^{(1)}}$};
  \node (gh) at (0.75+8,0) {$\scriptscriptstyle{2} $};
  \draw  (a1) edge (gh);
  \draw[->] (gh) edge (a2);
   \end{tikzpicture}
 }
 \label{eq:6d_su1_on_2_double_edge_su1_on_2}
\end{align}
leads to the 5d KK theory described by $\sprm(2)_{0}$  with one adjoint hypermultiplet \cite{Tachikawa:2011ch}.
Although the theory is known to be non-geometric, it is nevertheless interesting to explore the possibility of fibre-base duality. 
\paragraph{5d description.}
Since the 5d KK descriptions is known, the prepotential \eqref{eq:F} is derived using Appendix \ref{app:Sp_roots_weights}.
In addition to restricting to the Weyl-chamber of $C_2$, i.e.\ $\langle \phi,\alpha_i \rangle \geq 0$ for $i=1,2$, the following phase is chosen 
\begin{align}
    \raisebox{-.5\height}{
\begin{tikzpicture}
\node[draw,circle,inner sep=0.8pt,fill,black]  (a1) at (0,0) {};
 \node[draw,circle,inner sep=0.8pt,fill,black]  (a2) at (0,-1) {};
 \node[draw,circle,inner sep=0.8pt,fill,black]  (a3L) at (-1,-2) {};
 \node[draw,circle,inner sep=0.8pt,fill,black]  (a4L) at (-1,-3) {};
 \node[draw,circle,inner sep=0.8pt,fill,black]  (a5L) at (-1,-4) {};
 \node[draw,circle,inner sep=0.8pt,fill,black]  (a3R) at (1,-2) {};
 \node[draw,circle,inner sep=0.8pt,fill,black]  (a4R) at (1,-3) {};
 \node[draw,circle,inner sep=0.8pt,fill,black]  (a5R) at (1,-4) {};
 \node[draw,circle,inner sep=0.8pt,fill,black]  (a6) at (0,-5) {};
 \node[draw,circle,inner sep=0.8pt,fill,black]  (a7) at (0,-6) {};
 \draw (a1)--(a2)--(a3L)--(a4L)--(a5L)--(a6) (a2)--(a3R)--(a4R)--(a5R)--(a6)--(a7);
 \draw[red,dashed] (-3.5,-3.5)--(-0.85,-3.5) (-0.3,-3.5)--(0.3,-3.5) (0.85,-3.5)--(4.25,-3.5);
 \node[red] at (4.75,-3.5) {$\substack{\text{phase}\\ \text{choice}}$};
 \node [right=1ex of a1] {$\langle \phi,2\alpha_1 +\alpha_2 \rangle +m_f  \geq 0$};
 \node [right=1ex of a2] {$\langle \phi,\alpha_1 +\alpha_2 \rangle +m_f  \geq 0$};
 \node [left=1ex of a3L] {$\langle \phi,\alpha_2 \rangle +m_f  \geq 0$};
 \node [right=1ex of a3R] {$\langle \phi,\alpha_1 \rangle +m_f  \geq 0$};
 \node [left=1ex of a4L] {$\langle \phi,0 \rangle +m_f  \geq 0$};
 \node [right=1ex of a4R] {$\langle \phi,0 \rangle +m_f  \geq 0$};
 \node [left=1ex of a5L] {$\langle \phi, -\alpha_2 \rangle +m_f  \leq 0$};
 \node [right=1ex of a5R] {$\langle \phi, -\alpha_1 \rangle +m_f  \leq 0$};
 \node [right=1ex of a6] {$\langle \phi,-(\alpha_1 +\alpha_2) \rangle +m_f  \leq 0$};
  \node [right=1ex of a7] {$\langle \phi,-(2\alpha_1 +\alpha_2) \rangle +m_f  \leq 0$};
 \node at (-0.5,-0.5) {$\scriptstyle{\alpha_1 \; \downarrow}$};
 \node at (-0.75,-1.25) {$\scriptstyle{\alpha_1\; \swarrow}$};
 \node at (0.25,-1.75) {$\scriptstyle{\alpha_2\; \searrow}$};
 \node at (-0.6,-2.5) {$\scriptstyle{\downarrow \; \alpha_2  }$};
 \node at (-0.6,-3.5) {$\scriptstyle{\downarrow \;\alpha_2}$};
 \node at (0.6,-2.5) {$\scriptstyle{ \alpha_1 \; \downarrow }$};
 \node at (0.6,-3.5) {$\scriptstyle{\alpha_1 \; \downarrow}$};
 \node at (-0.75,-4.75) {$\scriptstyle{\alpha_1\; \searrow}$};
 \node at (0.25,-4.25) {$\scriptstyle{\alpha_2\; \swarrow}$};
 \node at (-0.5,-5.5) {$\scriptstyle{\alpha_1 \; \downarrow}$};
\end{tikzpicture}
}
\label{eq:phase_choice_Sp2_1Adj}
\end{align}
with $\alpha_{1,2}$ the simple roots of $C_2$, see Appendix \ref{app:Sp_roots_weights}. The prepotential becomes
\begin{align}
  6\Fcal_{5d} &=
   \frac{6}{g^2} \left(2 \phi_0^2  - 2\phi_0 \phi_1 + \phi_1^2 \right)
 -6  m^2 (\phi_0+\phi_1)
 \,,
\label{eq:prepot_field_theory_Sp2_1Adj}
 \end{align}
 with $g$ the gauge coupling and $m$ the mass parameter of the adjoint hypermultiplet.
\paragraph{Geometry.}
Following \cite{Bhardwaj:2019fzv}, the geometry derived from the 6d setup \eqref{eq:6d_su1_on_2_double_edge_su1_on_2} is given by
\begin{align}
\label{eq:geom_SU1_on_2_SU1_on_2_double_edge_6d}
\raisebox{-.5\height}{
 \begin{tikzpicture}
  \node (v0) at (0,0) {$\mathbf{0}_{0}^{1+1}$};  
  \node (v1) at (3,0){$\mathbf{1}_{0}^{1+1}$};  
  \node (gh) at (1.5,0){$\scriptscriptstyle{2}$}; 
  \draw  (v0) edge (gh);
    \draw  (gh) edge (v1);
    \draw (v0) to [out=140,in=220,looseness=6] (v0);
      \node at (-1.5,0.7) {$\scriptstyle{e_0-x}$};
        \node at (-1.5,-0.7) {$\scriptstyle{e_0-y}$};
    \draw (v1) to [out=140-100,in=220+100,looseness=6] (v1);
    \node at (3+1.5,0.7) {$\scriptstyle{e_1-z}$};
        \node at (3+1.5,-0.7) {$\scriptstyle{e_1-w}$};
  \node at (0.75,0.3) {$\scriptstyle{\substack{f_0-x\\ x}}$};
  \node at (2.25,0.3) {$\scriptstyle{\substack{2f_1-z\\ z}}$};
 \end{tikzpicture}
 }
\end{align}
such that one readily confirms that 
\begin{align}
 -f_i \cdot S_j = 0    \,,
\end{align}
which is reflecting the fact that the 6d gauge algebras are trivial.

For the 5d $\sprm(2)$ gauge theory, it is helpful to observe that 
\begin{align}
\label{eq:roots_via_volumes_Sp2_1Adj}
    -\begin{pmatrix}
      e_0 \cdot K_{S_0} & e_0 \cdot S_1|_{S_0} \\
      e_1 \cdot S_0|_{S_1}  &e_1 \cdot K_{S_1} 
    \end{pmatrix}
    = \begin{pmatrix}
      2 & -1 \\ -2 & 2
    \end{pmatrix} 
    \equiv  C_{C_2}
    \quad \Rightarrow \quad 
    \begin{cases}
    -J_\phi \cdot e_0 &= \langle \alpha_1 ,\phi \rangle \\
    -J_\phi \cdot e_1 &= \langle \alpha_2 ,\phi \rangle 
    \end{cases}
\end{align}
which identifies the $e_{0,1}$ as the simple roots $\alpha_{1,2}$ of the 5d $\sprmL(2)$ gauge algebra, see Appendix \ref{app:Sp_roots_weights}. As above, the two gauge theory parameter, one gauge coupling $g$ and one mass parameter, can be included in the K\"ahler form:
\begin{align}
\label{eq:Kahler_form_Sp2_1Adj}
    J= - \frac{1}{g^2}F+ \sum_{i=0}^1 \phi_i S_i +  m  N \,.
\end{align}
The gluing curves of the $F$, $N$ with the compact $S_i$ can be determined by parametrising \eqref{eq:Kahler_form_Sp2_1Adj} restricted to the $S_i$ as 
\begin{align}
\label{eq:Kahler_form_Sp2_1Adj_try}
\begin{aligned}
    J|_{S_0} &= \phi_0 K_{S_0} + \phi_1 S_1|_{S_0} +a_0 e_0 + b_0 f_0 + c_x x + c_y y\,,\\
    J|_{S_1} &=  \phi_0 S_0|_{S_1} + \phi_1 K_{S_1} +a_1 e_1 + b_1 f_1  
    +  c_z z + c_w w \,.
    \end{aligned}
\end{align}
The parameters are determined as follows:
\begin{compactitem}
 \item Firstly, one identifies the $-1$ curves that give rise to the fundamental BPS particles. It is instructive  to evaluate the volumes of the blowups; in detail,
 \begin{align}
     -J_\phi \cdot x =  -J_\phi \cdot y = -J_\phi \cdot z  =-J_\phi \cdot w = 0 = \langle 0 ,\phi \rangle
 \end{align}
 where $0$ is a trivial weight in $[2,0]_C$, see Appendix \ref{app:Sp_roots_weights}. Recalling \eqref{eq:roots_via_volumes_Sp2_1Adj}, the remaining weights are realised by the following $-1$ curves:
 \begin{align}
    \raisebox{-.5\height}{
\begin{tikzpicture}
\node[draw,circle,inner sep=0.8pt,fill,black]  (a1) at (0,0) {};
 \node[draw,circle,inner sep=0.8pt,fill,black]  (a2) at (0,-1) {};
 \node[draw,circle,inner sep=0.8pt,fill,black]  (a3L) at (-1,-2) {};
 \node[draw,circle,inner sep=0.8pt,fill,black]  (a4L) at (-1,-3) {};
 \node[draw,circle,inner sep=0.8pt,fill,black]  (a5L) at (-1,-4) {};
 \node[draw,circle,inner sep=0.8pt,fill,black]  (a3R) at (1,-2) {};
 \node[draw,circle,inner sep=0.8pt,fill,black]  (a4R) at (1,-3) {};
 \node[draw,circle,inner sep=0.8pt,fill,black]  (a5R) at (1,-4) {};
 \node[draw,circle,inner sep=0.8pt,fill,black]  (a6) at (0,-5) {};
 \node[draw,circle,inner sep=0.8pt,fill,black]  (a7) at (0,-6) {};
 \draw (a1)--(a2)--(a3L)--(a4L)--(a5L)--(a6) (a2)--(a3R)--(a4R)--(a5R)--(a6)--(a7);
 \draw[red,dashed] (-3.5,-3.5)--(-0.85,-3.5) (-0.3,-3.5)--(0.3,-3.5) (0.85,-3.5)--(4.25,-3.5);
 \node [right=1ex of a1] {$\substack{\vol (2e_0 +e_1+ x)\\  \vol (2e_0 +e_1+ z) }= \langle 2\alpha_1 +\alpha_2 ,\phi \rangle  +m$};
 \node [right=1ex of a2] {$\substack{ \vol (e_0 +e_1+ x) \\  \vol (e_0 +e_1+ z) }= \langle \alpha_1 +\alpha_2 ,\phi \rangle  +m$};
 \node [left=1ex of a3L] {$\vol (e_1+z) = \langle \alpha_2,\phi \rangle +m$};
 \node [right=1ex of a3R] {$\vol (e_0+x) = \langle \alpha_1,\phi \rangle +m$};
 \node [left=1ex of a4L] {$\vol (z) = \langle 0 ,\phi \rangle +m$};
 \node [right=1ex of a4R] {$\vol (x) = \langle 0 ,\phi \rangle +m$};
 \node [left=1ex of a5L] {$\vol (e_1-z) =- \left(\langle- \alpha_2,\phi \rangle +m\right)$};
 \node [right=1ex of a5R] {$\vol (e_0-x) =- \left(\langle- \alpha_1,\phi \rangle +m\right)$};
 \node [right=1ex of a6] {$\substack{ \vol (e_0 +e_1- x) \\ \vol (e_0 +e_1- z) } =-\left( \langle -(\alpha_1 +\alpha_2) ,\phi \rangle+m\right) $};
  \node [right=1ex of a7] {$\substack{ \vol (2e_0 +e_1- x) \\ \vol (2e_0 +e_1- z) } =-\left( \langle -(2\alpha_1 +\alpha_2) ,\phi \rangle+m\right)$};
 \node at (-0.5,-0.5) {$\scriptstyle{\alpha_1 \; \downarrow}$};
 \node at (-0.75,-1.25) {$\scriptstyle{\alpha_1\; \swarrow}$};
 \node at (0.25,-1.75) {$\scriptstyle{\alpha_2\; \searrow}$};
 \node at (-0.6,-2.5) {$\scriptstyle{\downarrow \; \alpha_2  }$};
 \node at (-0.6,-3.5) {$\scriptstyle{\downarrow \;\alpha_2}$};
 \node at (0.6,-2.5) {$\scriptstyle{ \alpha_1 \; \downarrow }$};
 \node at (0.6,-3.5) {$\scriptstyle{\alpha_1 \; \downarrow}$};
 \node at (-0.75,-4.75) {$\scriptstyle{\alpha_1\; \searrow}$};
 \node at (0.25,-4.25) {$\scriptstyle{\alpha_2\; \swarrow}$};
 \node at (-0.5,-5.5) {$\scriptstyle{\alpha_1 \; \downarrow}$};
\end{tikzpicture}
}
\end{align}
and the volumes are required to match the BPS masses. The self-gluings in \eqref{eq:geom_SU1_on_2_SU1_on_2_double_edge_6d} impose
\begin{align}
    \vol (e_0-x) = \vol (e_0-y) 
  \quad \text{and} \quad
 \vol (e_1-z)= \vol (e_1-w)
\end{align}
such that one finds
\begin{align}
      c_I = m \;, \quad I=x,y,z,w  
      \;, \qquad 
   b_i=0 \;, \quad  i=0,1  \;.
\end{align}
This identification of volumes motivates the choice of phase \eqref{eq:phase_choice_Sp2_1Adj}.
\item Secondly, the geometric effective gauge coupling \eqref{eq:effectice_coupling_geom} needs to match the field theory expectation \eqref{eq:effective_coupling}. The arising linear equations are solved by
\begin{align}
    a_0=   -\frac{2}{g^2}
    \;, \qquad 
    a_1 = -\frac{1}{g^2}
     \,.
\end{align}
\end{compactitem}
As a result, the non-compact surfaces $F$, $N$ in \eqref{eq:Kahler_form_Sp2_1Adj} restrict to the compact $S_i$ ones as follows:
\begin{align}
F|_{S_i} =
\begin{cases}
2 e_0  \;, & i=0 \,,\\
 e_1  \;, & i=1 \,,\\
\end{cases}
\,,\quad 
N |_{S_i} =
\begin{cases}
x+y \;, & i=0 \;,\\
z+w \;, & i=1\,.
\end{cases}
\end{align}
One realises that $F|_{S_i}$ restrict to $2e_0$ and $e_1$ respectively, because the $e_i$ act as simple roots \eqref{eq:roots_via_volumes_Sp2_1Adj}. The prefactors equal $\frac{2}{\langle\alpha_i,\alpha_i \rangle}=2,1$ for the roots of $C_2$.
Moreover, the volumes of the fibres are given by
\begin{align}
\begin{cases}
    \vol(f_0) &=  \frac{2}{g^2}   \,, \\
    \vol(f_1) &= \frac{1}{g^2} \,,
\end{cases}
\qquad \text{such that} \qquad  
\vol(f_{\mathrm{ell} } ) = \frac{1}{g^2}
\end{align}
which yields the volume of the elliptic fibre. The factor $2$ is due to the permutation twist in \eqref{eq:6d_su1_on_2_double_edge_su1_on_2}.
\subsection{\texorpdfstring{6d $\Ncal=(2,0)$ $D_4$ with twist -- 5d $G_2$ +1Adj}{6d D4 with twist -- 5d G2 +1Adj}}
\label{sec:6d_D4_with_twist}
The 6d  $\Ncal=(2,0)$ $D_4$ theory is geometrically realised by four $-2$ curves, supporting trivial $\surmL(1)$ algebras, which intersect in the pattern of the $D_4$ Dynkin diagram. The circle reduction twisted by permutations on the tensor multiplets
\begin{align}
\raisebox{-.2\height}{
 \begin{tikzpicture}
 \node (b1) at (0,0) {$\mathbf{2}$};
  \node at (0,0.35) {$\scriptstyle{ \surmL(1)^{(1)}}$};
  \node  (b2) at (1.5,0) {$ \mathbf{2}$};
  \node at (1.5,0.35) {$\scriptstyle{ \surmL(1)^{(1)}}$};
  \node (b3) at (3,0.75) {$\mathbf{2}$};
  \node at (3,0.35+0.75) {$\scriptstyle{ \surmL(1)^{(1)}}$};
  \node (b4) at (3,-0.75) {$\mathbf{2}$};
  \node at (3,0.35-0.75) {$\scriptstyle{ \surmL(1)^{(1)}}$};
  \draw  (b1) edge (b2);
  \draw  (b2) edge (b3);
  \draw  (b2) edge (b4);
  \draw[->] (4.5,0)--(6.5,0);
  \node at (5.5,0.25) {\footnotesize{permutation}};
  \node at (5.5,-0.25) {\footnotesize{twist}};
  \node (a1) at (0+8,0) {$\mathbf{2}$};
  \node at (0+8,0.35) {$\scriptstyle{ \surmL(1)^{(1)}}$};
  \node  (a2) at (1.5+8,0) {$ \mathbf{2}$};
  \node at (1.5+8,0.35) {$\scriptstyle{ \surmL(1)^{(1)}}$};
  \node (gh) at (0.75+8,0) {$\scriptscriptstyle{3} $};
  \draw  (a1) edge (gh);
  \draw[->] (gh) edge (a2);
   \end{tikzpicture}
 }
 \label{eq:geom_D4_twisted}
\end{align}
is known to admit a 5d descriptions as 
$G_2$ with one adjoint hypermultiplet \cite{Tachikawa:2011ch}.
\paragraph{5d description.}
Since the 5d KK descriptions is known, the prepotential \eqref{eq:F} is derived using Appendix \ref{app:G2_roots_weights}.
In addition to restricting to the Weyl-chamber of $G_2$, i.e.\ $\langle \phi,\alpha_i \rangle \geq 0$ for $i=1,2$, the following phase is chosen 
\begin{align}
    \raisebox{-.5\height}{
\begin{tikzpicture}
\node[draw,circle,inner sep=0.8pt,fill,black]  (a00) at (0,1) {};
\node (aux1) at (0,0) {$\vdots$};
 \node[draw,circle,inner sep=0.8pt,fill,black]  (a2) at (0,-1) {};
 \node[draw,circle,inner sep=0.8pt,fill,black]  (a3L) at (-1,-2) {};
 \node[draw,circle,inner sep=0.8pt,fill,black]  (a4L) at (-1,-3) {};
 \node[draw,circle,inner sep=0.8pt,fill,black]  (a5L) at (-1,-4) {};
 \node[draw,circle,inner sep=0.8pt,fill,black]  (a3R) at (1,-2) {};
 \node[draw,circle,inner sep=0.8pt,fill,black]  (a4R) at (1,-3) {};
 \node[draw,circle,inner sep=0.8pt,fill,black]  (a5R) at (1,-4) {};
 \node[draw,circle,inner sep=0.8pt,fill,black]  (a6) at (0,-5) {};
\node (aux2) at (0,-6) {$\vdots$};
 \node[draw,circle,inner sep=0.8pt,fill,black]  (a9) at (0,-7) {};
 \draw (a00)--(aux1)--(a2)--(a3L)--(a4L)--(a5L)--(a6) (a2)--(a3R)--(a4R)--(a5R)--(a6)--(aux2)--(a9);
 \draw[red,dashed] (-3.5,-3.5)--(-0.85,-3.5) (-0.3,-3.5)--(0.3,-3.5) (0.85,-3.5)--(4.25,-3.5);
 \node[red] at (4.75,-3.5) {$\substack{\text{phase}\\ \text{choice}}$};
 \node [right=1ex of a00] {$\langle \phi,3\alpha_1 +2\alpha_2 \rangle +m_f  \geq 0$};
 \node [right=1ex of a2] {$\langle \phi,\alpha_1 +\alpha_2 \rangle +m_f  \geq 0$};
 \node [left=1ex of a3L] {$\langle \phi,\alpha_2 \rangle +m_f  \geq 0$};
 \node [right=1ex of a3R] {$\langle \phi,\alpha_1 \rangle +m_f  \geq 0$};
 \node [left=1ex of a4L] {$\langle \phi,0 \rangle +m_f  \geq 0$};
 \node [right=1ex of a4R] {$\langle \phi,0 \rangle +m_f  \geq 0$};
 \node [left=1ex of a5L] {$\langle \phi, -\alpha_2 \rangle +m_f  \leq 0$};
 \node [right=1ex of a5R] {$\langle \phi, -\alpha_1 \rangle +m_f  \leq 0$};
 \node [right=1ex of a6] {$\langle \phi,-(\alpha_1 +\alpha_2) \rangle +m_f  \leq 0$};
  \node [right=1ex of a9] {$\langle \phi,-(3\alpha_1 +2\alpha_2) \rangle +m_f  \leq 0 $};
 \node at (-0.5,0.5) {$\scriptstyle{\alpha_2 \; \downarrow}$};
 \node at (-0.75,-1.25) {$\scriptstyle{\alpha_1\; \swarrow}$};
 \node at (0.25,-1.75) {$\scriptstyle{\alpha_2\; \searrow}$};
 \node at (-0.6,-2.5) {$\scriptstyle{\downarrow \; \alpha_2  }$};
 \node at (-0.6,-3.5) {$\scriptstyle{\downarrow \;\alpha_2}$};
 \node at (0.6,-2.5) {$\scriptstyle{ \alpha_1 \; \downarrow }$};
 \node at (0.6,-3.5) {$\scriptstyle{\alpha_1 \; \downarrow}$};
 \node at (-0.75,-4.75) {$\scriptstyle{\alpha_1\; \searrow}$};
 \node at (0.25,-4.25) {$\scriptstyle{\alpha_2\; \swarrow}$};
 \node at (-0.5,-6.5) {$\scriptstyle{\alpha_2 \; \downarrow}$};
\end{tikzpicture}
}
\label{eq:phase_choice_G2_1Adj}
\end{align}
with $\alpha_{1,2}$ the simple roots of $G_2$, see Appendix \ref{app:G2_roots_weights}. The prepotential becomes
\begin{align}
  6\Fcal_{5d} &=
   \frac{6}{g^2} \left( 3\phi_0^2  - 3 \phi_0 \phi_1 + \phi_1^2 \right)
 -6  m^2 (\phi_0+\phi_1)
 \,,
\label{eq:prepot_field_theory_G2_1Adj}
 \end{align}
 with $g$ the gauge coupling and $m$ the mass parameter of the adjoint hypermultiplet.
\paragraph{Geometry.}
The theory \eqref{eq:geom_D4_twisted} is geometrically realised by \cite{Bhardwaj:2019fzv}
\begin{align}
\raisebox{-.5\height}{
 \begin{tikzpicture}
  \node (v0) at (0,0) {$\mathbf{0}_{0}^{1+1}$};  
  \node (v1) at (3,0){$\mathbf{1}_{0}^{1+1}$};
  \node (gh) at (1.5,0){$\scriptscriptstyle{2}$};  
  \draw  (v0) edge (gh);
  \draw  (gh) edge (v1);
  \node at (0.65,0.2) {$\scriptstyle{f_0-x}$};
  \node at (0.65,-0.2) {$\scriptstyle{x}$};
  \node at (2.25,0.2) {$\scriptstyle{3f_1-x}$};
  \node at (2.25,-0.2) {$\scriptstyle{x}$};
  \draw (v0) to [out=135,in=225,looseness=3] (v0);
  \draw (v1) to [out=45,in=-45,looseness=3] (v1);
  \node at (-1.1,0.3) {$\scriptstyle{e_0-x}$};
  \node at (-1.1,-0.3) {$\scriptstyle{e_0-y}$};
  \node at (4.1,0.3) {$\scriptstyle{e_1-z}$};
  \node at (4.1,-0.3) {$\scriptstyle{e_1-w}$};
\end{tikzpicture}
 }
 \label{eq:geom_SU1_on_2_SU1_on_2_triple_edge_6d}
\end{align}
and due to the self-gluing curves, the truncated prepotential is trivially zero. Moreover, one verifies 
\begin{align}
 -f_i \cdot S_j = 0  \,,
\end{align}
which is reflecting the fact that the 6d gauge algebras are trivial.

For the 5d $G_2$ gauge theory, it is instructive to note that 
\begin{align}
\label{eq:roots_via_volumes_G2_1Adj}
    -\begin{pmatrix}
      e_0 \cdot K_{S_0} & e_0 \cdot S_1|_{S_0} \\
      e_1 \cdot S_0|_{S_1}  &e_1 \cdot K_{S_1} 
    \end{pmatrix}
    = \begin{pmatrix}
      2 & -1 \\ -3 & 2
    \end{pmatrix} 
    \equiv  C_{G_2}
    \quad \Rightarrow \quad 
    \begin{cases}
    -J_\phi \cdot e_0 &= \langle \alpha_1 ,\phi \rangle \\
    -J_\phi \cdot e_1 &= \langle \alpha_2 ,\phi \rangle 
    \end{cases}
\end{align}
which identifies the $e_{0,1}$ as the simple roots $\alpha_{1,2}$ of the 5d $G_2$ gauge algebra, see Appendix \ref{app:G2_roots_weights}. As above, the two gauge theory parameter, one gauge coupling $g$ and one mass parameter $m$, are introduced by suitably parametrising the K\"ahler form:
\begin{align}
\label{eq:Kahler_form_G2_1Adj}
    J= - \frac{1}{g^2}F+ \sum_{i=0}^1 \phi_i S_i +  m  N \,.
\end{align}
In order to determine the gluing curves of $F$, $N$ with the compact $S_i$, one restricts the K\"ahler form \eqref{eq:Kahler_form_G2_1Adj} to the $S_i$ and expands in a suitable base as follows:
\begin{align}
\label{eq:Kahler_form_G2_1Adj_try}
\begin{aligned}
    J|_{S_0} &= \phi_0 K_{S_0} + \phi_1 S_1|_{S_0} +a_0 e_0 + b_0 f_0 + c_x x + c_y y\,,\\
    J|_{S_1} &=  \phi_0 S_0|_{S_1} + \phi_1 K_{S_1} +a_1 e_1 + b_1 f_1  
    +  c_z z + c_w w \,.
    \end{aligned}
\end{align}
The parameters are determined as follows:
\begin{compactitem}
 \item Firstly, one identifies the $-1$ curves that give rise to the fundamental BPS particles. It is instructive  to evaluate the volumes of the blowups; in detail,
 \begin{align}
     -J_\phi \cdot x =  -J_\phi \cdot y = -J_\phi \cdot z  =-J_\phi \cdot w = 0 = \langle 0 ,\phi \rangle
 \end{align}
 where $0$ is a trivial weight in $[0,1]_{G_2}$, see Appendix \ref{app:G2_roots_weights}. Recalling \eqref{eq:roots_via_volumes_G2_1Adj}, the remaining weights are realised by the following $-1$ curves:
 \begin{align}
     \raisebox{-.5\height}{
\begin{tikzpicture}
\node[draw,circle,inner sep=0.8pt,fill,black]  (a00) at (0,2) {};
 \node[draw,circle,inner sep=0.8pt,fill,black]  (a0) at (0,1) {};
\node[draw,circle,inner sep=0.8pt,fill,black]  (a1) at (0,0) {};
 \node[draw,circle,inner sep=0.8pt,fill,black]  (a2) at (0,-1) {};
 \node[draw,circle,inner sep=0.8pt,fill,black]  (a3L) at (-1,-2) {};
 \node[draw,circle,inner sep=0.8pt,fill,black]  (a4L) at (-1,-3) {};
 \node[draw,circle,inner sep=0.8pt,fill,black]  (a5L) at (-1,-4) {};
 \node[draw,circle,inner sep=0.8pt,fill,black]  (a3R) at (1,-2) {};
 \node[draw,circle,inner sep=0.8pt,fill,black]  (a4R) at (1,-3) {};
 \node[draw,circle,inner sep=0.8pt,fill,black]  (a5R) at (1,-4) {};
 \node[draw,circle,inner sep=0.8pt,fill,black]  (a6) at (0,-5) {};
 \node[draw,circle,inner sep=0.8pt,fill,black]  (a7) at (0,-6) {};
 \node[draw,circle,inner sep=0.8pt,fill,black]  (a8) at (0,-7) {};
 \node[draw,circle,inner sep=0.8pt,fill,black]  (a9) at (0,-8) {};
 \draw (a00)--(a0)--(a1)--(a2)--(a3L)--(a4L)--(a5L)--(a6) (a2)--(a3R)--(a4R)--(a5R)--(a6)--(a7)--(a8)--(a9);
 \draw[red,dashed] (-3.5,-3.5)--(-0.85,-3.5) (-0.3,-3.5)--(0.3,-3.5) (0.85,-3.5)--(4.25,-3.5);
 \node [right=1ex of a00] {$\substack{\vol (3e_0 +2e_1+ x) \\ \vol (3e_0 +2e_1+ z)} = \langle 3\alpha_1 +2\alpha_2 ,\phi \rangle  +m$};
 \node [right=1ex of a0] {$\substack{\vol (3e_0 +e_1+ x)\\  \vol (3e_0 +e_1+ z) } = \langle 3\alpha_1 +\alpha_2 ,\phi \rangle  +m$};
 \node [right=1ex of a1] {$\substack{ \vol (2e_0 +e_1+ x)\\  \vol (2e_0 +e_1+ z) } = \langle 2\alpha_1 +\alpha_2 ,\phi \rangle  +m$};
 \node [right=1ex of a2] {$ \substack{ \vol (e_0 +e_1+ x)\\  \vol (e_0 +e_1+ z) } = \langle \alpha_1 +\alpha_2 ,\phi \rangle  +m$};
 \node [left=1ex of a3L] {$\vol (e_1+z)  = \langle \alpha_2,\phi \rangle +m$};
 \node [right=1ex of a3R] {$\vol (e_0+x)  = \langle \alpha_1,\phi \rangle +m$};
 \node [left=1ex of a4L] {$\vol (z) = \langle 0 ,\phi \rangle +m$};
 \node [right=1ex of a4R] {$\vol (x) = \langle 0 ,\phi \rangle +m$};
 \node [left=1ex of a5L] {$\vol (e_1-z)  =- \left(\langle- \alpha_2,\phi \rangle +m\right)$};
 \node [right=1ex of a5R] {$\vol (e_0-x)  =- \left(\langle- \alpha_1,\phi \rangle +m\right)$};
 \node [right=1ex of a6] {$\substack{ \vol (e_0 +e_1- x) \\ \vol (e_0 +e_1- z) } =-\left( \langle -(\alpha_1 +\alpha_2) ,\phi \rangle+m\right)$};
 \node [right=1ex of a7] {$\substack{\vol (2e_0 +e_1- x) \\ \vol (2e_0 +e_1- z) } =-\left( \langle -(2\alpha_1 +\alpha_2) ,\phi \rangle+m\right)$};
 \node [right=1ex of a8] {$\substack{\vol (3e_0 +e_1- x) \\ \vol (3e_0 +e_1- z)  }=-\left( \langle -(3\alpha_1 +\alpha_2) ,\phi \rangle+m\right)$};
 \node [right=1ex of a9] {$\substack{ \vol (3e_0 +2e_1- x) \\ \vol (3e_0 +2e_1- z) }  =-\left( \langle -(3\alpha_1 +2\alpha_2) ,\phi \rangle+m\right)$};
 \node at (-0.5,1.5) {$\scriptstyle{\alpha_2 \; \downarrow}$};
 \node at (-0.5,0.5) {$\scriptstyle{\alpha_1 \; \downarrow}$};
 \node at (-0.5,-0.5) {$\scriptstyle{\alpha_1 \; \downarrow}$};
 \node at (-0.75,-1.25) {$\scriptstyle{\alpha_1\; \swarrow}$};
 \node at (0.25,-1.75) {$\scriptstyle{\alpha_2\; \searrow}$};
 \node at (-0.6,-2.5) {$\scriptstyle{\downarrow \; \alpha_2  }$};
 \node at (-0.6,-3.5) {$\scriptstyle{\downarrow \;\alpha_2}$};
 \node at (0.6,-2.5) {$\scriptstyle{ \alpha_1 \; \downarrow }$};
 \node at (0.6,-3.5) {$\scriptstyle{\alpha_1 \; \downarrow}$};
 \node at (-0.75,-4.75) {$\scriptstyle{\alpha_1\; \searrow}$};
 \node at (0.25,-4.25) {$\scriptstyle{\alpha_2\; \swarrow}$};
 \node at (-0.5,-5.5) {$\scriptstyle{\alpha_1 \; \downarrow}$};
 \node at (-0.5,-6.5) {$\scriptstyle{\alpha_1 \; \downarrow}$};
 \node at (-0.5,-7.5) {$\scriptstyle{\alpha_2 \; \downarrow}$};
\end{tikzpicture}
}
 \end{align}
and the self-gluing curves in \eqref{eq:geom_SU1_on_2_SU1_on_2_triple_edge_6d} impose
\begin{align}
    \vol (e_0-x) = \vol (e_0-y) 
  \quad \text{and} \quad
 \vol (e_1-z)= \vol (e_1-w) \,.
\end{align}
These conditions are solved by
\begin{align}
    c_I = m \,, \quad I\in\{x,,y,z,w\} \,, \qquad 
    b_i=0 \,, \quad  i=0,1 \,,
\end{align}
 One notes that these volumes motivate the choice of phase \eqref{eq:phase_choice_G2_1Adj}.
\item Secondly, the geometric effective gauge coupling \eqref{eq:effectice_coupling_geom}
needs to match the field theory expectation \eqref{eq:effective_coupling}. The arising linear equations are solved by
\begin{align}
    a_0=   -\frac{3}{g^2}
    \;, \qquad 
    a_1 = -\frac{1}{g^2}
     \,.
\end{align}
\end{compactitem}
Consequently, the non-compact surfaces $F$ and $N$, introduced in \eqref{eq:Kahler_form_G2_1Adj}, are glued to the compact $S_i$ as follows:
\begin{align}
F|_{S_i} =
\begin{cases}
3 e_0  \;, & i=0\\
 e_1  \;, & i=1\\
\end{cases}
\,,\quad 
N |_{S_i} =
\begin{cases}
x+y \;, & i=0 \;,\\
z+w \;, & i=1\,.
\end{cases}
\end{align}
An immediate observation is that the $F|_{S_i}$ restrict to $3e_0$ and $e_1$ respectively, because the $e_i$ act as simple roots \eqref{eq:roots_via_volumes_G2_1Adj}. The prefactors equal $\frac{2}{\langle\alpha_i,\alpha_i \rangle}=3,1$ for the roots of $G_2$.
Moreover, the volumes of the fibres are given by
\begin{align}
\begin{cases}
    \vol(f_0) &=  \frac{3}{g^2}   \,, \\
    \vol(f_1) &= \frac{1}{g^2} \,,
\end{cases}
\qquad \text{such that} \qquad  
\vol(f_{\mathrm{ell} } ) = \frac{1}{g^2}
\end{align}
which yields the volume of the elliptic fibre. The factor $3$ is due to the permutation twist in \eqref{eq:geom_D4_twisted}.
\section{Fibre-base like duality: 5d to 5d}
\label{sec:fibre-base_5d-5d}
Inspecting Table \ref{tab:KK_theories}, it is clear that some twisted compactifications of 6d SCFTs may lead to several 5d descriptions. In Section \ref{sec:fibre-base_6d-5d}, only the 5d theories that are related by fibre-base duality to the 6d setup have been considered. Nevertheless, most of the 5d theories admit an honest geometric description. As detailed, for instance in \cite{Bhardwaj:2019jtr,Bhardwaj:2020gyu}, these geometric frames are in many cases related by some operations on the Hirzebruch surfaces and their gluing curves. 
In this section, two examples are considered for which the relation between two 5d frames is simply realised in terms of the $\FF_0$ isomorphism $e \leftrightarrow f$.
%
%
\subsection{\texorpdfstring{$\surm(3)_{\frac{3}{2}}$ +1Sym -- $\sprm(2)_{\pi}$ +1Adj}{SU(3), CS=3/2 +1Adj, Sp(2),theta=pi +1Adj} }
Starting from the 6d $\Ncal=(2,0)$ $A_4$ theory, the twist by the $\Z_2$ permutation symmetry 
\begin{align}
 \raisebox{-.5\height}{
 \begin{tikzpicture}
  \node (b1) at (0,0) {$\mathbf{2}$};
  \node at (0,0.35) {$\scriptstyle{ \surmL(1)^{(1)}}$};
  \node  (b2) at (1.5,0) {$ \mathbf{2}$};
  \node at (1.5,0.35) {$\scriptstyle{ \surmL(1)^{(1)}}$};
  \node (b3) at (3,0) {$\mathbf{2}$};
  \node at (3,0.35) {$\scriptstyle{ \surmL(1)^{(1)}}$};
  \node  (b4) at (4.5,0) {$ \mathbf{2}$};
  \node at (4.5,0.35) {$\scriptstyle{ \surmL(1)^{(1)}}$};
  \draw  (b1) edge (b2);
  \draw  (b2) edge (b3);
  \draw  (b3) edge (b4);
\draw[->] (6,0)--(8,0);
  \node at (7,0.25) {\footnotesize{permutation}};
  \node at (7,-0.25) {\footnotesize{twist}};
  \node (a1) at (0+10,0) {$\mathbf{2}$};
  \node at (0+10,0.35) {$\scriptstyle{ \surmL(1)^{(1)}}$};
  \node  (a2) at (1.5+10,0) {$ \mathbf{2}$};
  \node at (1.5+10,0.35) {$\scriptstyle{ \surmL(1)^{(1)}}$};
  \draw  (a1) edge (a2);
  \draw (a1) to [out=200,in=340,looseness=6] (a1);
   \end{tikzpicture}
 }
 \label{eq:geom_6d_A3_folded}
\end{align}
leads to a 5d KK theory with two known gauge theory descriptions:
$\surm(3)_{\frac{3}{2}}$ with one rank-2 symmetric hypermultiplet \cite{Jefferson:2017ahm} and $\sprm(2)_{\pi}$ with one adjoint hypermultiplet \cite{Tachikawa:2011ch}.

In this section, it is demonstrated that a suitable 5d frame for $\sprm(2)$ is related via a $\FF_0$ isomorphism $e\leftrightarrow f$ to a 5d frame for the $\surm(3)$ description. In other words, this mimics a fibre-base like duality between the two dual 5d theories.
\paragraph{5d $\sprm(2)$ description.}
Since the 5d KK descriptions is known, the prepotential \eqref{eq:F} is derived using Appendix \ref{app:Sp_roots_weights}.
In addition to restricting to the Weyl-chamber of $C_2$, i.e.\ $\langle \phi,\alpha_i \rangle \geq 0$ for $i=1,2$, the following phase is chosen 
\begin{align}
\raisebox{-.5\height}{
\begin{tikzpicture}
\node[draw,circle,inner sep=0.8pt,fill,black]  (a1) at (0,0) {};
 \node[draw,circle,inner sep=0.8pt,fill,black]  (a2) at (0,-1) {};
 \node[draw,circle,inner sep=0.8pt,fill,black]  (a3L) at (-1,-2) {};
 \node[draw,circle,inner sep=0.8pt,fill,black]  (a4L) at (-1,-3) {};
 \node[draw,circle,inner sep=0.8pt,fill,black]  (a5L) at (-1,-4) {};
 \node[draw,circle,inner sep=0.8pt,fill,black]  (a3R) at (1,-2) {};
 \node[draw,circle,inner sep=0.8pt,fill,black]  (a4R) at (1,-3) {};
 \node[draw,circle,inner sep=0.8pt,fill,black]  (a5R) at (1,-4) {};
 \node[draw,circle,inner sep=0.8pt,fill,black]  (a6) at (0,-5) {};
 \node[draw,circle,inner sep=0.8pt,fill,black]  (a7) at (0,-6) {};
 \draw (a1)--(a2)--(a3L)--(a4L)--(a5L)--(a6) (a2)--(a3R)--(a4R)--(a5R)--(a6)--(a7);
 \draw[red,dashed] (-0.2,-5.5)--(4.5,-5.5);
 \node[red] at (5,-5.5) {$\substack{\text{phase}\\ \text{choice}}$};
 \node [right=1ex of a1] {$\langle \phi,2\alpha_1 +\alpha_2 \rangle +m_f  \geq 0$};
 \node [right=1ex of a2] {$\langle \phi,\alpha_1 +\alpha_2 \rangle +m_f  \geq 0$};
 \node [left=1ex of a3L] {$\langle \phi,\alpha_2 \rangle +m_f  \geq 0$};
 \node [right=1ex of a3R] {$\langle \phi,\alpha_1 \rangle +m_f  \geq 0$};
 \node [left=1ex of a4L] {$\langle \phi,0 \rangle +m_f  \geq 0$};
 \node [right=1ex of a4R] {$\langle \phi,0 \rangle +m_f  \geq 0$};
 \node [left=1ex of a5L] {$\langle \phi, -\alpha_2 \rangle +m_f  \geq 0$};
 \node [right=1ex of a5R] {$\langle \phi, -\alpha_1 \rangle +m_f  \geq 0$};
 \node [right=1ex of a6] {$\langle \phi,-(\alpha_1 +\alpha_2) \rangle +m_f  \geq 0$};
  \node [right=1ex of a7] {$\langle \phi,-(2\alpha_1 +\alpha_2) \rangle +m_f  \leq 0$};
 \node at (-0.5,-0.5) {$\scriptstyle{\alpha_1 \; \downarrow}$};
 \node at (-0.75,-1.25) {$\scriptstyle{\alpha_1\; \swarrow}$};
 \node at (0.25,-1.75) {$\scriptstyle{\alpha_2\; \searrow}$};
 \node at (-0.6,-2.5) {$\scriptstyle{\downarrow \; \alpha_2  }$};
 \node at (-0.6,-3.5) {$\scriptstyle{\downarrow \;\alpha_2}$};
 \node at (0.6,-2.5) {$\scriptstyle{ \alpha_1 \; \downarrow }$};
 \node at (0.6,-3.5) {$\scriptstyle{\alpha_1 \; \downarrow}$};
 \node at (-0.75,-4.75) {$\scriptstyle{\alpha_1\; \searrow}$};
 \node at (0.25,-4.25) {$\scriptstyle{\alpha_2\; \swarrow}$};
 \node at (-0.5,-5.5) {$\scriptstyle{\alpha_1 \; \downarrow}$};
\end{tikzpicture}
}
\label{eq:phase_choice_Sp2_Pi_1Adj}
\end{align}
with $\alpha_{1,2}$ the simple roots of $C_2$, see Appendix \ref{app:Sp_roots_weights}. The prepotential becomes
\begin{align}
\begin{aligned}
  6\Fcal_{5d} &=
  12 \phi_1^2 \phi_2 -18 \phi_1 \phi_2^2 + 8 \phi_2^3
   +\frac{6}{g^2} \left(2 \phi_1^2  - 2\phi_1 \phi_2 + \phi_2^2 \right) \\
 &\qquad -6  m^2 \phi_1 
 -24 m \phi_1^2 
 +36 m \phi_1 \phi_2
 -18 m \phi_2^2
 \,,
 \end{aligned}
\label{eq:prepot_field_theory_Sp2_Pi_1Adj}
 \end{align}
 with $g$ the gauge coupling and $m$ the mass parameter of the adjoint hypermultiplet.
\paragraph{Geometry.}
The geometric proposal for \eqref{eq:geom_6d_A3_folded} in \cite{Bhardwaj:2019fzv} does not have a manifest fibre-base duality between the 6d and 5d frame. 
A convenient 5d geometry is given by \cite{Bhardwaj:2020gyu}
\begin{align}
\raisebox{-.5\height}{
 \begin{tikzpicture}
  \node (v0) at (0,0) {$\mathbf{1}_{6}^{1+1}$};  
  \node (v1) at (3,0){$\mathbf{2}_{0}$};  
  \draw  (v0) edge (v1);
    \draw (v0) to [out=140,in=220,looseness=6] (v0);
      \node at (-1.5,0.7) {$\scriptstyle{x}$};
        \node at (-1.5,-0.7) {$\scriptstyle{y}$};
%
  \node at (0.75,0.25) {$\scriptstyle{e_1}$};
  \node at (2.25,0.25) {$\scriptstyle{2e_2+f_2}$};
 \end{tikzpicture}
 }
 \label{eq:geom_su1_on_2_su1_on_2_loop_5d}
\end{align}
and one notices that 
\begin{align}
\label{eq:roots_via_volumes_Sp2_Pi_1Adj}
    -\begin{pmatrix}
      f_1 \cdot K_{S_1} & f_1 \cdot S_2|_{S_1} \\
      f_2 \cdot S_1|_{S_2}  &f_2 \cdot K_{S_2} 
    \end{pmatrix}
    = \begin{pmatrix}
      2 & -1 \\ -2 & 2
    \end{pmatrix} 
    \equiv  C_{C_2}
    \quad \Rightarrow \quad 
    \begin{cases}
    -J_\phi \cdot f_1 &= \langle \alpha_1 ,\phi \rangle \\
    -J_\phi \cdot f_2 &= \langle \alpha_2 ,\phi \rangle 
    \end{cases}
\end{align}
which identifies the $f_{1,2}$ as the simple roots $\alpha_{1,2}$ of the 5d $\sprmL(2)$ gauge algebra, see Appendix \ref{app:Sp_roots_weights}.
The gauge theory parameters can be incorporated into the K\"ahler form via
\begin{align}
    J= - \frac{1}{g^2}F+ \sum_{i=1}^2 \phi_i S_i +  m  N \,.
\end{align}
where the gluing curves of the non-compact surfaces $F$ and $N$  with the compact $S_i$ are determined as follows:
\begin{compactitem}
 \item Firstly, one identifies the $-1$ curves that give rise to the fundamental BPS particles. It is instructive  to evaluate the volumes of the blowups; in detail,
 \begin{align}
     -J_\phi \cdot x =  -J_\phi \cdot y =  2\phi_1 = - \langle -(2\alpha_1+\alpha_2) ,\phi \rangle
 \end{align}
 where $2\alpha_1+\alpha_2$ is the highest weight of $[2,0]_C$, see Appendix \ref{app:Sp_roots_weights}. Recalling \eqref{eq:roots_via_volumes_Sp2_Pi_1Adj}, the volumes of the relevant $-1$ curves are required to satisfy:
 \begin{align}
\raisebox{-.5\height}{
\begin{tikzpicture}
\node[draw,circle,inner sep=0.8pt,fill,black]  (a1) at (0,0) {};
 \node[draw,circle,inner sep=0.8pt,fill,black]  (a2) at (0,-1) {};
 \node[draw,circle,inner sep=0.8pt,fill,black]  (a3L) at (-1,-2) {};
 \node[draw,circle,inner sep=0.8pt,fill,black]  (a4L) at (-1,-3) {};
 \node[draw,circle,inner sep=0.8pt,fill,black]  (a5L) at (-1,-4) {};
 \node[draw,circle,inner sep=0.8pt,fill,black]  (a3R) at (1,-2) {};
 \node[draw,circle,inner sep=0.8pt,fill,black]  (a4R) at (1,-3) {};
 \node[draw,circle,inner sep=0.8pt,fill,black]  (a5R) at (1,-4) {};
 \node[draw,circle,inner sep=0.8pt,fill,black]  (a6) at (0,-5) {};
 \node[draw,circle,inner sep=0.8pt,fill,black]  (a7) at (0,-6) {};
 \draw (a1)--(a2)--(a3L)--(a4L)--(a5L)--(a6) (a2)--(a3R)--(a4R)--(a5R)--(a6)--(a7);
 \draw[red,dashed] (-0.2,-5.5)--(6,-5.5);
 \node [right=1ex of a1] {$\vol (4f_1 +f_2 - x)  = \langle (2\alpha_1 +\alpha_2) ,\phi \rangle+m$};
 \node [right=1ex of a2] {$\vol (3f_1 +2f_2 - x) = \langle (\alpha_1 +\alpha_2) ,\phi \rangle+m  $};
 \node [left=1ex of a3L] {$\vol (2f_1 +2f_2 - x) =\langle \alpha_2,\phi \rangle +m$};
 \node [right=1ex of a3R] {$\vol (3f_1 +f_2 - x)  = \langle \alpha_1,\phi \rangle +m$};
 \node [left=1ex of a4L] {$\vol (2f_1 +f_2 - x) = \langle 0 ,\phi \rangle +m$};
 \node [right=1ex of a4R] {$\vol (2f_1 +f_2 - x) = \langle 0 ,\phi \rangle +m$};
 \node [left=1ex of a5L] {$\vol (2f_1 - x) = \langle -\alpha_2,\phi \rangle +m$};
 \node [right=1ex of a5R] {$\vol (f_1 +f_2 - x) = \langle -\alpha_1,\phi \rangle +m$};
 \node [right=1ex of a6] {$\vol (f_1 - x) =\langle-( \alpha_1 +\alpha_2) ,\phi \rangle  +m$};
  \node [right=1ex of a7] {$\vol ( x) = -\left( \langle -(2\alpha_1 +\alpha_2) ,\phi \rangle  +m\right)$};
 \node at (-0.5,-0.5) {$\scriptstyle{\alpha_1 \; \downarrow}$};
 \node at (-0.75,-1.25) {$\scriptstyle{\alpha_1\; \swarrow}$};
 \node at (0.25,-1.75) {$\scriptstyle{\alpha_2\; \searrow}$};
 \node at (-0.6,-2.5) {$\scriptstyle{\downarrow \; \alpha_2  }$};
 \node at (-0.6,-3.5) {$\scriptstyle{\downarrow \;\alpha_2}$};
 \node at (0.6,-2.5) {$\scriptstyle{ \alpha_1 \; \downarrow }$};
 \node at (0.6,-3.5) {$\scriptstyle{\alpha_1 \; \downarrow}$};
 \node at (-0.75,-4.75) {$\scriptstyle{\alpha_1\; \searrow}$};
 \node at (0.25,-4.25) {$\scriptstyle{\alpha_2\; \swarrow}$};
 \node at (-0.5,-5.5) {$\scriptstyle{\alpha_1 \; \downarrow}$};
\end{tikzpicture}
}
\end{align}
 and the self-gluing curves in \eqref{eq:geom_su1_on_2_su1_on_2_loop_5d} impose 
 \begin{align}
     \vol (x) = \vol (y) \,.
 \end{align}
  One notes that these volumes motivate the choice of phase \eqref{eq:phase_choice_Sp2_Pi_1Adj}. 
\item Secondly, the geometric effective coupling \eqref{eq:effectice_coupling_geom}
needs to match the field theory result \eqref{eq:effective_coupling}. 
\end{compactitem}
Consequently, one finds 
\begin{align}
F|_{S_i} =
\begin{cases}
2 f_1  \;, & i=1\\
 f_2  \;, & i=2\\
\end{cases}
\,,\quad 
N |_{S_i} =
\begin{cases}
6f_1 -x-y  \;, & i=1 \;,\\
3f_2 \;, & i=2\,.
\end{cases}
\end{align}
One may note that $F|_{S_i}$ restricts to $2f_1$ and $f_2$ respectively, because the $f_i$ act as simple roots \eqref{eq:roots_via_volumes_Sp2_Pi_1Adj}. The prefactors equal $\frac{2}{\langle\alpha_i,\alpha_i \rangle}=2,1$ for the roots of $C_2$.
\paragraph{Fibre-base dual $\surm(3)$.}
Returning to the geometry \eqref{eq:geom_su1_on_2_su1_on_2_loop_5d}, one aims to utilise the same geometry for the $\surm(3)$ KK theory. As a first check, one computes
\begin{align}
\label{eq:roots_via_volume_SU3_CS=3/2_1Sym}
    -  
    \begin{pmatrix}
      f_1 \cdot  K_{S_1} & f_1 \cdot S_2|_{S_1} \\
      e_2 \cdot S_1|_{S_2} & e_2 \cdot  K_{S_2}
    \end{pmatrix}
    = 
    \begin{pmatrix}
      2 & -1 \\ -1 & 2 
    \end{pmatrix}
    = C_{A_2}
    \quad \Rightarrow \quad 
    \begin{cases}
    -J_\phi \cdot f_1 &= \langle\alpha_1 ,\phi \rangle \\
    -J_\phi \cdot e_2 &= \langle\alpha_2 ,\phi \rangle 
    \end{cases}
\end{align}
which indicates the 5d $\surmL(3)$ gauge algebra and identifies $f_1$, $e_2$ as simple roots $\alpha_i$, see Appendix \ref{app:SU_roots_weights}. Besides restricting to the Weyl-chamber of $A_2$, i.e.\ $\langle \phi,\alpha_i \rangle \geq 0$ for $i=1,2$, the following phase is chosen 
\begin{align}
    \raisebox{-.5\height}{
\begin{tikzpicture}
\node[draw,circle,inner sep=0.8pt,fill,black]  (v1) at (0,0) {};
 \node[draw,circle,inner sep=0.8pt,fill,black]  (v2) at (0,-1) {};
 \node[draw,circle,inner sep=0.8pt,fill,black]  (v3L) at (-1,-2) {};
 \node[draw,circle,inner sep=0.8pt,fill,black]  (v3R) at (1,-2) {};
 \node[draw,circle,inner sep=0.8pt,fill,black]  (v4) at (0,-3) {};
 \node[draw,circle,inner sep=0.8pt,fill,black]  (v5) at (0,-4) {};
 \draw (v1)--(v2)--(v3L)--(v4) (v2)--(v3R)--(v4)--(v5);
 \draw[red,dashed] (-0.2,-0.5)--(3,-0.5);
 \node[red] at (3.5,-0.5) {$\substack{\text{phase}\\ \text{choice}}$};
 \node [right=1ex of v1] {$\langle v_1 ,\phi \rangle +m \geq 0$};
 \node [right=1ex of v2] {$ \langle v_2 ,\phi \rangle +m \leq 0$};
 \node [left=1ex of v3L] {$ \langle v_3 ,\phi \rangle +m \leq 0$};
 \node [right=1ex of v3R] {$ \langle v_4 ,\phi \rangle +m \leq 0$};
 \node [right=1ex of v4] {$\langle v_5 ,\phi \rangle +m \leq 0$};
 \node [right=1ex of v5] {$\langle v_6 ,\phi \rangle +m \leq 0$};
 \node at (-0.5,-0.5) {$\scriptstyle{\alpha_2 \; \downarrow}$};
 \node at (-0.75,-1.25) {$\scriptstyle{\alpha_1\; \swarrow}$};
 \node at (0.25,-1.75) {$\scriptstyle{\alpha_2\; \searrow}$};
 \node at (-0.75,-2.75) {$\scriptstyle{\alpha_2\; \searrow}$};
 \node at (0.25,-2.25) {$\scriptstyle{\alpha_1\; \swarrow}$};
 \node at (-0.5,-3.5) {$\scriptstyle{\alpha_2 \; \downarrow}$};
\end{tikzpicture}
}
\label{eq:phase_choice_SU3_CS=3/2_1Sym}
\end{align}
with $v_i \in [2,0]_A$, see Appendix \ref{app:SU_roots_weights}. The prepotential becomes
\begin{align}
\begin{aligned}
  6\Fcal_{5d} &=
   \frac{6}{g^2} \left( \phi_1^2  - \phi_1 \phi_2 + \phi_2^2 \right)
 +12 \phi_1^2 \phi_2
 -18 \phi_1 \phi_2^2
 +8 \phi_2^3 
 \\
 &+3 m \phi_1^2
 -15 m \phi_1 \phi_2
 +15 m \phi_2^2
 -6 m^2 \phi_1
 \,,
\end{aligned}
\label{eq:prepot_field_theory_SU3_CS=1/2_1Sym}
 \end{align}
 with $g$ the gauge coupling and $m$ the mass parameter of the hypermultiplet in the 2nd rank symmetric representation.
The K\"ahler form is supplemented by non-compact surfaces as follows:
\begin{align}
    J = - \frac{1}{g^2}F +\sum_{i=1}^2 \phi_i S_i  +  \frac{m}{2}N \,. 
\end{align}
The gluing curves of $F$ and $N$ with the compact surfaces $S_i$ are determined via two constraints.
\begin{compactitem}
\item One identifies the $-1$ curves that give rise to the fundamental BPS particles. The blowups provide some insights
\begin{align}
    -J_\phi \cdot x = -J_\phi \cdot y = 2 \phi_1 = \langle v_1,\phi \rangle 
\end{align}
with the highest weight vector $v_1 \in [2,0]_A$. The identical volumes for $x$ and $y$ is a consequence of the self-gluing in \eqref{eq:geom_su1_on_2_su1_on_2_loop_5d}. Recalling \eqref{eq:roots_via_volume_SU3_CS=3/2_1Sym}, the set of $-1$ curves is given by
\begin{align}
    \raisebox{-.5\height}{
\begin{tikzpicture}
\node[draw,circle,inner sep=0.8pt,fill,black]  (v1) at (0,0) {};
 \node[draw,circle,inner sep=0.8pt,fill,black]  (v2) at (0,-1) {};
 \node[draw,circle,inner sep=0.8pt,fill,black]  (v3L) at (-1,-2) {};
 \node[draw,circle,inner sep=0.8pt,fill,black]  (v3R) at (1,-2) {};
 \node[draw,circle,inner sep=0.8pt,fill,black]  (v4) at (0,-3) {};
 \node[draw,circle,inner sep=0.8pt,fill,black]  (v5) at (0,-4) {};
 \draw (v1)--(v2)--(v3L)--(v4) (v2)--(v3R)--(v4)--(v5);
 \draw[red,dashed] (-0.2,-0.5)--(4,-0.5);
 \node [right=1ex of v1] {$\vol( x) \stackrel{!}{=} \langle v_1 ,\phi \rangle +m$};
 \node [right=1ex of v2] {$ \vol(f_1 - x) \stackrel{!}{=} - \left(\langle v_2 ,\phi \rangle +m\right)$};
 \node [left=1ex of v3L] {$ \vol(2f_1 -x) \stackrel{!}{=} - \left( \langle v_3 ,\phi \rangle +m \right)$};
 \node [right=1ex of v3R] {$\vol(f_1 +e_2 -x ) \stackrel{!}{=}- \left( \langle v_4 ,\phi \rangle +m \right)$};
 \node [right=1ex of v4] {$\vol(2f_1 +e_2 -x) \stackrel{!}{=} - \left( \langle v_5 ,\phi \rangle +m\right)$};
 \node [right=1ex of v5] {$\vol(2f_1 +2e_2 -x) \stackrel{!}{=} -\left( \langle v_6 ,\phi \rangle +m \right)$};
 \node at (-0.5,-0.5) {$\scriptstyle{\alpha_2 \; \downarrow}$};
 \node at (-0.75,-1.25) {$\scriptstyle{\alpha_1\; \swarrow}$};
 \node at (0.25,-1.75) {$\scriptstyle{\alpha_2\; \searrow}$};
 \node at (-0.75,-2.75) {$\scriptstyle{\alpha_2\; \searrow}$};
 \node at (0.25,-2.25) {$\scriptstyle{\alpha_1\; \swarrow}$};
 \node at (-0.5,-3.5) {$\scriptstyle{\alpha_2 \; \downarrow}$};
\end{tikzpicture}
}
\end{align}
and the self-gluing imposes
\begin{align}
    \vol(x) \stackrel{!}{=} \vol(y)  \;.
\end{align}
These conditions motivate the phase \eqref{eq:phase_choice_SU3_CS=3/2_1Sym}.
\item The effective coupling \eqref{eq:effectice_coupling_geom} derived from the geometry has to agree with field theory \eqref{eq:effective_coupling}.
\end{compactitem}
These constraints lead to 
\begin{align}
    F|_{S_i} =
    \begin{cases}
    f_1\;,  &i=1 \\
    e_2\;, &i=2
    \end{cases}
    \;, \quad 
    N|_{S_i}  = 
    \begin{cases}
    2 x +2 y - 5 f_1 \,,& i=1\\
    -5e_2 \,, &i=2
    \end{cases}
    \,.
\end{align}
As above, the $F|_{S_i}$ restrict to $f_1$ and $e_2$ respectively, because the $f_1$, $e_2$ act as simple roots \eqref{eq:roots_via_volume_SU3_CS=3/2_1Sym}. The prefactor of $1$ equals $\frac{2}{\langle\alpha_i,\alpha_i \rangle}$ for the roots of $A_2$.
Moreover, one verifies that $6\Fcal_{\geom} =  J^3$ matches the field theory result \eqref{eq:prepot_field_theory_SU3_CS=1/2_1Sym}.
\subsection{\texorpdfstring{5d $G_2$ +1 Adj -- 5d $\surm(3)_{\frac{15}{2}}$ + 1 F}{5d G2 +1Adj - 5d SU(3), CS=15/2+ 1F}}
The twisted circle reduction of 6d  $\Ncal=(2,0)$ $D_4$ \eqref{eq:geom_D4_twisted} is known to admit two different 5d descriptions:
$G_2$ with one adjoint hypermultiplet \cite{Tachikawa:2011ch} and $\surm(3)_{\frac{15}{2}}$ with one fundamental hypermultiplet \cite{Bhardwaj:2019jtr}.

The $G_2$ theory has already been discussed in Section \ref{sec:6d_D4_with_twist}, where the 5d frame has been obtained via fibre-base duality from the 6d frame.
In this section, it is demonstrated that a suitable 5d frame for $\surm(3)$ is related via a $\FF_0$ isomorphism $e\leftrightarrow f$ to another 5d frame for the $G_2$ description (in a different phase). In other words, this mimics a fibre-base like duality between the two dual 5d theories.
\paragraph{5d SU(3) description.}
The $\surm(3)$ gauge theory with Chern-Simons level $\kappa=\frac{15}{2}$ has one fundamental hypermultiplet. The prepotential is derived from \eqref{eq:F} via the data summarised in Appendix \ref{app:SU_roots_weights}. In addition to restricting to the $A_2$ Weyl chamber, a phase needs to be choose. A suitable choice is
\begin{align}
\raisebox{-.5\height}{
\begin{tikzpicture}
\node[draw,circle,inner sep=0.8pt,fill,black]  (w1) at (0,0) {};
 \node[draw,circle,inner sep=0.8pt,fill,black]  (w2) at (0,-1) {};
 \node[draw,circle,inner sep=0.8pt,fill,black]  (w3) at (0,-2) {};
 \draw (w1)--(w2)--(w3);
 \draw[red,dashed] (-0.2,-1.5)--(3,-1.5);
 \node[red] at (3.5,-1.5) {$\substack{\text{phase}\\ \text{choice}}$};
 \node at (1.5,0) {$ \langle \phi,w_1 \rangle +m_f \geq0$};
 \node at (1.5,-1) {$\langle \phi,w_2 \rangle +m_f \geq 0$};
 \node at (1.5,-2) {$ \langle \phi,w_3 \rangle +m_f \leq0$};
 \node at (-0.5,-0.5) {$\scriptstyle{\alpha_1 \; \downarrow}$};
 \node at (-0.5,-1.5) {$\scriptstyle{\alpha_2\; \downarrow}$};
\end{tikzpicture}
}
\label{eq:phase_choice_SU3_CS=15/2_1F}
\end{align}
with $w_i \in [1,0]_A$, see Appendix \ref{app:SU_roots_weights}.
In this phase, the prepotential becomes
\begin{align}
    \begin{aligned}
    6 \Fcal &=
    8 \phi_1^3 +18 \phi_1^2 \phi_2  -24 \phi_1 \phi_2^2  +7 \phi_2^3
    +\frac{6}{g^2} \left( \phi_1^2 - \phi_1 \phi_2 + \phi_2^2 \right) \\
    &\quad -3 m \phi_1^2 
    +3 m \phi_1 \phi_2
    -3 m^2 \phi_2
    \end{aligned} \,,
    \label{eq:prepot_field_theory_SU3_CS=15/2_1F}
\end{align}
with $g$ the gauge coupling and $m$ the mass of the fundamental hypermultiplet.
\paragraph{Geometry.}
For the $\surm(3)$ frame, one may employ the following geometry \cite{Bhardwaj:2019jtr}
\begin{align}
\raisebox{-.5\height}{
 \begin{tikzpicture}
  \node (v0) at (0,0) {$\mathbf{1}_{8}$};  
  \node (v1) at (3,0){$\mathbf{2}_{0}^{1}$};  
  \draw  (v0) edge (v1);
  \node at (0.5,0.2) {$\scriptstyle{e_1}$};
  \node at (2.25,0.2) {$\scriptstyle{e_2+3f_2}$};
 \end{tikzpicture}
 }
 \label{eq:geom_SU3_1F_CS=15/2}
\end{align} 
which is argued to be flop equivalent to \eqref{eq:geom_SU1_on_2_SU1_on_2_triple_edge_6d}.
One straightforwardly verifies that
\begin{align}
\label{eq:roots_via_volume_SU3_CS=15/2+1F}
    -\begin{pmatrix}
      f_1 \cdot K_{S_1} & f_1 \cdot S_2|_{S_1} \\
      f_2 \cdot S_1|_{S_2}  &f_2 \cdot K_{S_2} 
    \end{pmatrix}
    = \begin{pmatrix}
      2 & -1 \\ -1 & 2
    \end{pmatrix}
    \equiv  C_{A_2}
    \quad \Rightarrow \quad 
    \begin{cases}
    -J_\phi \cdot f_1 &= \langle \alpha_i ,\phi \rangle \\
    -J_\phi \cdot f_2 &= \langle \alpha_2 ,\phi \rangle
    \end{cases}
\end{align}
which identifies the Cartan matrix of the 5d $\surmL(3)$ gauge algebra. In addition, the fibres $f_i$ act as simple roots $\alpha_i$ of $A_2$, see Appendix \ref{app:SU_roots_weights}.

The gauge theory parameter can be included as follows:
\begin{align}
    J= - \frac{1}{g^2}F+ \sum_{i=1}^2 \phi_i S_i + \frac{m}{2}  N
\end{align}
and the gluing curves of the non-compact surfaces with the compact $S_i$ are determined via two conditions:
\begin{compactitem}
 \item Firstly, the volume of the $-1$ curves match BPS masses. Based on \eqref{eq:roots_via_volume_SU3_CS=15/2+1F} and
 \begin{align}
     -J_\phi \cdot x = \phi_2 = - \langle w_3 ,\phi \rangle\;,
 \end{align}
 with $w_2 \in [1,0]_A$,  one concludes that the BPS mass are given by
  \begin{align}
\raisebox{-.5\height}{
\begin{tikzpicture}
\node[draw,circle,inner sep=0.8pt,fill,black]  (w1) at (0,0) {};
 \node[draw,circle,inner sep=0.8pt,fill,black]  (w2) at (0,-1) {};
 \node[draw,circle,inner sep=0.8pt,fill,black]  (w3) at (0,-2) {};
 \draw (w1)--(w2)--(w3);
 \draw[red,dashed] (-0.2,-1.5)--(5,-1.5);
 \node [right=1ex of w1] {$\vol(f_1 + f_2 -x ) \stackrel{!}{=}  \langle w_1 ,\phi \rangle +m$};
 \node [right=1ex of w2] {$\vol( f_2 -x ) \stackrel{!}{=} \langle w_2 ,\phi \rangle + m$};
 \node [right=1ex of w3] {$  \vol(x )\stackrel{!}{=} -\left( \langle w_3 ,\phi \rangle +m \right)$};
 \node at (-0.5,-0.5) {$\scriptstyle{\alpha_1 \; \downarrow}$};
 \node at (-0.5,-1.5) {$\scriptstyle{\alpha_2\; \downarrow}$};
\end{tikzpicture}
}
\end{align}
  which motivates the phase \eqref{eq:phase_choice_SU3_CS=15/2_1F}.
\item The geometric effective gauge coupling \eqref{eq:effectice_coupling_geom}
needs to match the field theory result \eqref{eq:effective_coupling}.
\end{compactitem}
As a result, the non-compact surfaces are glued as follows: 
\begin{align}
F|_{S_i} = f_i \quad \text{for } i=1,2
\,,\quad 
N |_{S_i} =
\begin{cases}
 f_1& i=1 \\
 f_2 - 2x  & i=2
\end{cases}
\,.
\end{align}
The $F|_{S_i}$ restrict to the $f_i$ respectively, because the $f_i$ act as simple roots \eqref{eq:roots_via_volume_SU3_CS=15/2+1F}. The prefactor of $1$ equals $\frac{2}{\langle\alpha_i,\alpha_i \rangle}$ for the roots of $A_2$.
\paragraph{Fibre-base dual $G_2$.}
Inspecting \eqref{eq:geom_SU3_1F_CS=15/2} shows an interesting instance of fibre-base duality. In detail,
\begin{align}
    -\begin{pmatrix}
      f_1 \cdot K_{S_1} & f_1 \cdot S_2|_{S_1} \\
      e_2 \cdot S_1|_{S_2}  &e_2 \cdot K_{S_2} 
    \end{pmatrix}
    = \begin{pmatrix}
      2 & -1 \\ -3 & 2
    \end{pmatrix}
    \equiv  C_{G_2}
     \quad \Rightarrow \quad 
    \begin{cases}
    -J_\phi \cdot f_1 &= \langle \alpha_i ,\phi \rangle \\
    -J_\phi \cdot e_2 &= \langle \alpha_2 ,\phi \rangle
    \end{cases}
    \label{eq:roots_G2_1Adj_alternative}
\end{align}
such that $f_1$, $e_2$ act as simple roots $\alpha_i$ of $G_2$, see Appendix \ref{app:G2_roots_weights}. Therefore, one is lead to another description of $G_2$ with one adjoint hypermulitplet, but this time in a different phase
\begin{align}
    \raisebox{-.5\height}{
\begin{tikzpicture}
\node[draw,circle,inner sep=0.8pt,fill,black]  (a00) at (0,-1) {};
 \node (aux1) at (0,-2) {$\vdots$};
 \node[draw,circle,inner sep=0.8pt,fill,black]  (a2) at (0,-3) {};
 \node[draw,circle,inner sep=0.8pt,fill,black]  (a6) at (0,-5) {};
 \node[draw,circle,inner sep=0.8pt,fill,black]  (a8) at (0,-7) {};
\node (aux2) at (0,-6) {$\vdots$};
\node (aux3L) at (-1,-4) {$\vdots$};
\node (aux3R) at (1,-4) {$\vdots$};
 \node[draw,circle,inner sep=0.8pt,fill,black]  (a9) at (0,-8) {};
 \draw (a00)--(aux1)--(a2)--(aux3L)--(a6) (a2)--(aux3R)--(a6)--(aux2)--(a9);
 \draw[red,dashed] (-0.2,-7.5)--(5,-7.5);
 \node[red] at (5.55,-7.5) {$\substack{\text{phase}\\ \text{choice}}$};
 \node [right=1ex of a00] {$\langle \phi,3\alpha_1 +2\alpha_2 \rangle +m_f  \geq 0$};
 \node [right=1ex of a2] {$\langle \phi,\alpha_1 +\alpha_2 \rangle +m_f  \geq 0$};
 \node [right=1ex of a6] {$\langle \phi,-(\alpha_1 +\alpha_2) \rangle +m_f  \leq 0$};
  \node [right=1ex of a8] {$ \langle \phi,-(3\alpha_1 +\alpha_2) \rangle +m_f  \geq 0$};
  \node [right=1ex of a9] {$\langle \phi,-(3\alpha_1 +2\alpha_2) \rangle +m_f  \leq 0 $};
 \node at (-0.5,-1.5) {$\scriptstyle{\alpha_2 \; \downarrow}$};
 \node at (-0.75,-3.25) {$\scriptstyle{\alpha_1\; \swarrow}$};
 \node at (0.25,-3.75) {$\scriptstyle{\alpha_2\; \searrow}$};
 \node at (-0.75,-4.75) {$\scriptstyle{\alpha_1\; \searrow}$};
 \node at (0.25,-4.25) {$\scriptstyle{\alpha_2\; \swarrow}$};
 \node at (-0.5,-7.5) {$\scriptstyle{\alpha_2 \; \downarrow}$};
\end{tikzpicture}
}
\label{eq:phase_choice_G2_1Adj_alternative}
\end{align}
The prepotential in this phase equals 
\begin{align}
    \begin{aligned}
    6 \Fcal &=
    8 \phi_1^3 +18 \phi_1^2 \phi_2  -24 \phi_1 \phi_2^2  +7 \phi_2^3
    +\frac{6}{g^2} \left( 3\phi_1^2 -3 \phi_1 \phi_2 + \phi_2^2 \right) \\
    &\quad -72 m \phi_1^2 
    +72 m \phi_1 \phi_2
    -21 m \phi_2^2
    -3 m^2 \phi_2
    \end{aligned} \,.
\end{align}
The K\"ahler form can be expressed as 
\begin{align}
    J= - \frac{1}{g^2}F+ \sum_{i=1}^2 \phi_i S_i + m  N
\end{align}
and the non-compact surfaces $F$ and $N$ are glued to the $S_i$ via
\begin{align}
F|_{S_i} = 
\begin{cases}
3 f_1& i=1 \\
e_2  & i=2
\end{cases}
\,,\quad 
N |_{S_i} =
\begin{cases}
12 f_1& i=1 \\
4 f_2 - x  & i=2
\end{cases}
\,.
\end{align}
Analogously to the cases discussed above, the $F|_{S_i}$ restrict to $3f_1$ and $e_2$ respectively, because $f_1$ and $e_2$ act as simple roots \eqref{eq:roots_G2_1Adj_alternative}. The prefactors equal $\frac{2}{\langle\alpha_i,\alpha_i \rangle}=3,1$ for the roots of $G_2$.
Also, note that in this construction $-J_\phi \cdot x = - \langle -(3\alpha_1 +2\alpha_2),\phi \rangle $, which justifies the phase \eqref{eq:phase_choice_G2_1Adj_alternative}.
%
\section{Conclusions}
\label{sec:conclusions}
In this work, a geometric description of twisted circle compactifications of 6d $\Ncal=(1,0)$ SCFTs has been analysed with the aim to fully characterise the resulting 5d theory. For a number of such theories, we have shown how to recover the prepotential of the 5d gauge theory on the Coulomb branch from the geometric prepotential of M-theory on a Calabi-Yau threefold $X_S$. We explicitly described the non-compact divisors which must be included in the K\"ahler potential of $X_S$ to reflect the contribution of mass parameters and gauge couplings. The strategy to find these divisors is composed of three steps: firstly, determine a geometric description such that there exist curves of self-intersection zero which act as roots of the 5d gauge algebra. Secondly, identify the weights of the blowup modes under the gauge algebra and construct the $-1$ curves that correspond to the fundamental BPS particles. The volumes of these curves have to reproduce the BPS masses. Moreover, this step identifies the phase of the 5d theory on the Coulomb branch. Thirdly, the geometric expression for the matrix of effective gauge couplings has to reproduce the corresponding field theory expression. As a result, the non-compact surfaces in the K\"ahler form are determined by their intersection with the compact surfaces. This is sufficient information to reproduce the physical quantities and to compute the prepotential $\Fcal_{\geom}= \frac{1}{3!}J^3 $ up to constant terms. 

This description then allows us to identify the effective gauge coupling with the volume of the elliptic fibre of the F-Theory description of the parent 6d SCFTs. In particular, we made this identification for the 5d reduction of the E-string theory, the untwisted 5d reduction of $\surmL(n)$ on a $-2$, the $\Z_2$ twisted compactification of the 6d $\Ncal=(1,0)$ minimal $\surm(3)$ SCFT, and the $\Z_3$ twisted reduction of the minimal $\sorm(8)$ SCFT. 

In a number of examples of rank 1 and rank 2 KK theories, we observed a fibre-base duality between the 6d and 5d frame, which is realised via an exchange of fibre and base in $\FF_0$.  For non-gauge theoretic nodes and, in particular, non-geometric theories, the appearance of some form of fibre-base duality between the 6d and 5d frame is not guaranteed. In all cases considered in Section \ref{sec:fibre-base_6d-5d}, the geometry for the 6d frame, which manifests the elliptic fibre, is fibre-dual to one of the 5d descriptions, precisely by an $\FF_0$ isomorphism $e \leftrightarrow f$.

For M-theory on a given non-compact Calabi-Yau threefold, compact surfaces in the geometry may exhibit different rulings, which corresponds to different effective gauge theory descriptions. In the limit in which we shrink the compact surfaces, all of these theories flow to the same 5d SCFT, so that such a phenomenon encodes a UV duality. A particular instance of this phenomenon is when the fibre of one ruling becomes a section of another. Seen from 6d, the circle reduction admits more than one 5d description in this case, and not all 5d frames are related to the 6d frame via fibre-base duality. We have demonstrated that some of the dual 5d theories are related among each other by a fibre-base like duality relating different 5d frames. 

The geometric description of 5d theories by a collection of complex compact surfaces sitting inside a non-compact threefold is a resourceful approach to the understanding of quantum field theories. There are a number of open issues that would be interesting to address in future research. Starting from a specific twisted compactification, the resulting collection of surfaces and their gluing rules are far from unique. In this work, the geometric frame suitable for a gauge theory in one specific phase has been chosen. Hence, the transitions between different phases have not been addressed, although it is expected that these are straightforward. Moreover, one may also choose to follow the isomorphisms relating the different phases, and establish parameter maps between the different gauge theory descriptions. Fibre-base like dualities relating different 5d descriptions of the same SCFT have also been observed in \cite{Closset:2018bjz,Apruzzi:2019enx,Bhardwaj:2019jtr,Bhardwaj:2020gyu}. In the present work we have collected evidence for an elegant relation between the divisor $F$ and the fibres that establish the gauge algebra. One may wonder whether there is an analogous, systematic relation for the divisors $N_f$ associated to hypermultiplet mass parameters.

\paragraph{Acknowledgements.}
The work of J.C., B.H., M.S., and S.Y. is supported by the National Thousand-Young-Talents Program of China.
M.S. is further supported by the National Natural Science Foundation of China (grant no.\ 11950410497), and the China Postdoctoral Science Foundation (grant no.\ 2019M650616).
M.S. thanks Fudan University, Department of Physics for hospitality during the intermediate stage of this work.
\appendix
\section{Background material}
\label{app:background}
\subsection{Lie algebras and weight spaces}
\label{sec:liealg}
The evaluation of the prepotential \eqref{eq:F} relies on the gauge group as well as the representations of the matter content. This appendix provides a brief summary. For more details, the reader is, for instance, referred to \cite{Feger:2012bs,Feger:2019tvk} as well as \cite[App.\ A]{Jefferson:2017ahm}.
\subsubsection{Notation}
Give a Lie algebra, the simple roots are $\alpha_i$, and the simple coroots are $\alpha_i^\vee$. The Cartan matrix is defined via
 \begin{align}
  A_{ij} = \langle \alpha_i ,\alpha_j^\vee \rangle  = 2 \frac{\langle \alpha_i ,\alpha_j \rangle}{ \langle \alpha_i ,\alpha_i \rangle } \,.
 \end{align}
 A convenient basis is the Dynkin basis, or weight basis, defined by the fundamental weights being the basis vectors. Then, the Cartan matrix is the transformation matrix that provides the expansion of the simple roots $\alpha_i$ in the basis of the fundamental weights, i.e.
 \begin{align}
\alpha_i = \sum_k A_{ik} \omega_k
\qquad \Leftrightarrow \qquad 
\omega_i = \sum_k (A^{-1})_{ik} \alpha_k  
 \end{align}
Note that the $j$-th row of the Cartan matrix gives the simple root $\alpha_j$ in terms of the fundamental weights. 

The vector $\phi$ of Coulomb moduli is an expanded in the basis of simple coroot 
\begin{align}
 \phi = \sum_{i} \phi_i \alpha_i^\vee \,,
\end{align}
which defines the components $\phi_i$ used throughout this paper.
Next, consider the following expressions
\begin{align}
 \langle \phi,\alpha_i \rangle 
 = \sum_{j,k} \phi_j A_{ik} \langle \alpha_j^\vee ,\omega_k \rangle
 = \sum_{j,k} \phi_j A_{ik} \delta_{jk} = \sum_{k} A_{ik} \phi_k
\end{align}
likewise
\begin{align}
 \langle \phi,\phi \rangle = \sum_{i,j} \phi_i \phi_j \langle \alpha_i^\vee,\alpha_j^\vee \rangle
 = \sum_{i,j} \phi_j D_j^{-1} A_{ji} \phi_i \equiv \sum_{i,j} \phi_j h_{ji} \phi_i
 \label{eq:Cartan_vs_metric_tensor}
\end{align}
where $h_{ij}$ is the inverse of the metric tensor defined as follows
\begin{align}
 (h^{-1})_{ij} = (A^{-1})_{ij}D_j
\end{align}
where $D_j$ is defined via 
\begin{align}
 A_{ij} = \langle \alpha_i ,\alpha_j^\vee \rangle  = 2 \frac{\langle \alpha_i ,\alpha_j \rangle}{ \langle \alpha_i ,\alpha_i \rangle } = D_{j}^{-1} \langle \alpha_i ,\alpha_j \rangle 
 \quad \Leftrightarrow \quad 
  D^{-1} = \diag\left( 
  \frac{2}{ \langle \alpha_1 ,\alpha_1 \rangle }
  , \ldots, 
  \frac{2}{ \langle \alpha_r ,\alpha_r \rangle } \right)\,.
  \label{eq:D_factor}
\end{align}
Recall that the factors $ \frac{2}{ \langle \alpha_i ,\alpha_i \rangle }$ can only take three different values: $1$ for all roots on $A$, $D$, $E$ algebras and the long roots in $B$, $C$, $F_4$; $2$ for the short roots of $B$, $C$, and $F_4$; and $3$ for the short root of $G_2$.
The Chern-Simons term in \eqref{eq:F} can be expressed as follows:
\begin{align}
 6 \Fcal_{\text{tree-level}}^{\text{CS}} = \kappa \ d_{ijk} \ \phi^i \phi^j\phi^k 
 \qquad \text{with} \qquad 
 d_{ijk} = \frac{1}{2} \Tr_{\mathbf{F}} \left( t_i \left\{t_j,t_k \right\} \right) 
\\
\Tr_{\mathbf{R}_f} \left( t_i \left\{t_j,t_k \right\} \right) \phi^i \phi^j\phi^k 
= 2 c_{\mathbf{R}_f}^{(3)} d_{ijk} \phi^i \phi^j\phi^k 
=2 c_{\mathbf{R}_f}^{(3)} \sum_{w \in \mathbf{R}_f} \langle \phi, w \rangle^3 
\end{align}
with $c_{\mathbf{R}_f}^{(3)}$ the cubic Dynkin index. Since $c_{\mathbf{F}}^{(3)}=1$ for $\mathfrak{su}(n)$, one finds  
\begin{align}
 d_{ijk} \phi^i \phi^j\phi^k = \sum_{w \in \mathbf{F}} \langle \phi, w \rangle^3  \qquad \text{for }\mathfrak{su}(n) \,.
\end{align}

In the following subsection, the roots and weights for the Lie algebra relevant for this paper are summarised.

\subsubsection{\texorpdfstring{$\surm(2)$, $\surm(3)$, and $\surm(4)$}{SU(2), SU(3), and SU(4)} }
\label{app:SU_roots_weights}
\paragraph{SU(2).}
The simple root is $\alpha=2$ and the representations relevant for this work have weight systems given by
\begin{align}
    \raisebox{-.5\height}{
\begin{tikzpicture}
\node at (-1,0.5) {$[1]_A$};
\node[draw,circle,inner sep=0.8pt,fill,black]  (w1) at (0,0) {};
 \node[draw,circle,inner sep=0.8pt,fill,black]  (w2) at (0,-1) {};
 \draw (w1)--(w2);
 \node at (0.75,0) {$\scriptstyle{w_1=1}$};
 \node at (0.75,-1) {$\scriptstyle{w_2=-1}$};
 \node at (-0.5,-0.5) {$\scriptstyle{\alpha \; \downarrow}$};
\end{tikzpicture}
}
\qquad 
\raisebox{-.5\height}{
\begin{tikzpicture}
\node at (-1,0.5) {$[2]_A$};
\node[draw,circle,inner sep=0.8pt,fill,black]  (a1) at (0,0) {};
 \node[draw,circle,inner sep=0.8pt,fill,black]  (a2) at (0,-1) {};
 \node[draw,circle,inner sep=0.8pt,fill,black]  (a3) at (0,-2) {};
 \draw (a1)--(a2)--(a3);
 \node at (0.5,0) {$\scriptstyle{2}$};
 \node at (0.5,-1) {$\scriptstyle{0}$};
 \node at (0.5,-2) {$\scriptstyle{-2}$};
 \node at (-0.5,-0.5) {$\scriptstyle{\alpha \; \downarrow}$};
 \node at (-0.5,-1.5) {$\scriptstyle{\alpha\; \downarrow}$};
\end{tikzpicture}
}
\end{align}
where the arrows indicate the action of the simple roots.
The Cartan matrix, which equals the inverse metric tensor, reads
\begin{align} 
     C_{A_1} = 2= h \;.
 \end{align}
\paragraph{SU(3).}
The simple roots are  $\alpha_1=(2,-1)$ , $\alpha_2=(-1,2)$ and the weight systems for the relevant representation are as follows
\begin{align}
\raisebox{-.5\height}{
\begin{tikzpicture}
\node at (-1,0.5) {$[1,0]_A$};
\node[draw,circle,inner sep=0.8pt,fill,black]  (w1) at (0,0) {};
 \node[draw,circle,inner sep=0.8pt,fill,black]  (w2) at (0,-1) {};
 \node[draw,circle,inner sep=0.8pt,fill,black]  (w3) at (0,-2) {};
 \draw (w1)--(w2)--(w3);
 \node at (0.75,0) {$\scriptstyle{w_1=(1,0)}$};
 \node at (0.75,-1) {$\scriptstyle{w_2=(-1,1)}$};
 \node at (0.75,-2) {$\scriptstyle{w_3=(0,-1)}$};
 \node at (-0.5,-0.5) {$\scriptstyle{\alpha_1 \; \downarrow}$};
 \node at (-0.5,-1.5) {$\scriptstyle{\alpha_2\; \downarrow}$};
\end{tikzpicture}
}
\qquad 
\raisebox{-.5\height}{
\begin{tikzpicture}
\node at (-1,0.5) {$[2,0]_A$};
\node[draw,circle,inner sep=0.8pt,fill,black]  (v1) at (0,0) {};
 \node[draw,circle,inner sep=0.8pt,fill,black]  (v2) at (0,-1) {};
 \node[draw,circle,inner sep=0.8pt,fill,black]  (v3L) at (-1,-2) {};
 \node[draw,circle,inner sep=0.8pt,fill,black]  (v3R) at (1,-2) {};
 \node[draw,circle,inner sep=0.8pt,fill,black]  (v4) at (0,-3) {};
 \node[draw,circle,inner sep=0.8pt,fill,black]  (v5) at (0,-4) {};
 \draw (v1)--(v2)--(v3L)--(v4) (v2)--(v3R)--(v4)--(v5);
 \node at (0.75,0) {$\scriptstyle{v_1=(2,0)}$};
 \node at (0.75,-1) {$\scriptstyle{v_2=(0,1)}$};
 \node at (-1.75,-2) {$\scriptstyle{v_3=(-2,2)}$};
 \node at (1.75,-2) {$\scriptstyle{v_4=(1,-1)}$};
 \node at (0.75,-3) {$\scriptstyle{v_5=(-1,0)}$};
  \node at (0.75,-4) {$\scriptstyle{v_6=(0,-2)}$};
 \node at (-0.5,-0.5) {$\scriptstyle{\alpha_2 \; \downarrow}$};
 \node at (-0.75,-1.25) {$\scriptstyle{\alpha_1\; \swarrow}$};
 \node at (0.25,-1.75) {$\scriptstyle{\alpha_2\; \searrow}$};
 \node at (-0.75,-2.75) {$\scriptstyle{\alpha_2\; \searrow}$};
 \node at (0.25,-2.25) {$\scriptstyle{\alpha_1\; \swarrow}$};
 \node at (-0.5,-3.5) {$\scriptstyle{\alpha_2 \; \downarrow}$};
\end{tikzpicture}
}
\qquad 
\raisebox{-.5\height}{
\begin{tikzpicture}
\node at (-1,-0.5) {$[1,1]_A$};
 \node[draw,circle,inner sep=0.8pt,fill,black]  (a2) at (0,-1) {};
 \node[draw,circle,inner sep=0.8pt,fill,black]  (a3L) at (-1,-2) {};
 \node[draw,circle,inner sep=0.8pt,fill,black]  (a4L) at (-1,-3) {};
 \node[draw,circle,inner sep=0.8pt,fill,black]  (a5L) at (-1,-4) {};
 \node[draw,circle,inner sep=0.8pt,fill,black]  (a3R) at (1,-2) {};
 \node[draw,circle,inner sep=0.8pt,fill,black]  (a4R) at (1,-3) {};
 \node[draw,circle,inner sep=0.8pt,fill,black]  (a5R) at (1,-4) {};
 \node[draw,circle,inner sep=0.8pt,fill,black]  (a6) at (0,-5) {};
 \draw (a2)--(a3L)--(a4L)--(a5L)--(a6) (a2)--(a3R)--(a4R)--(a5R)--(a6);
 \node at (0.5,-1) {$\scriptstyle{(1,1)}$};
 \node at (-1.5,-2) {$\scriptstyle{(-1,2)}$};
 \node at (1.5,-2) {$\scriptstyle{(2,-1)}$};
 \node at (-1.5,-3) {$\scriptstyle{(0,0)}$};
 \node at (1.5,-3) {$\scriptstyle{(0,0)}$};
 \node at (-1.5,-4) {$\scriptstyle{(2,-1)}$};
 \node at (1.5,-4) {$\scriptstyle{(-2,1)}$};
 \node at (0.55,-5) {$\scriptstyle{(-1,-1)}$};
 \node at (-0.75,-1.25) {$\scriptstyle{\alpha_1\; \swarrow}$};
 \node at (0.25,-1.75) {$\scriptstyle{\alpha_2\; \searrow}$};
 \node at (-0.6,-2.5) {$\scriptstyle{\downarrow \; \alpha_2  }$};
 \node at (-0.6,-3.5) {$\scriptstyle{\downarrow \;\alpha_2}$};
 \node at (0.6,-2.5) {$\scriptstyle{ \alpha_1 \; \downarrow }$};
 \node at (0.6,-3.5) {$\scriptstyle{\alpha_1 \; \downarrow}$};
 \node at (-0.75,-4.75) {$\scriptstyle{\alpha_1\; \searrow}$};
 \node at (0.25,-4.25) {$\scriptstyle{\alpha_2\; \swarrow}$};
\end{tikzpicture}
}
\end{align}
where the arrows indicate the action of the simple roots. In addition, the Cartan matrix and the inverse metric tensor are given by
 \begin{align}
     C_{A_2} = \begin{pmatrix}
       2 & -1  \\ -1 & 2 
     \end{pmatrix}
     = h_{ij} \,.
 \end{align}
\paragraph{SU(4).}
The simple roots are $\alpha_1=(2,-1,0)$ , $\alpha_2=(-1,2,-1)$, $\alpha_3=(0,-1,2)$ and the Cartan matrix, which equals the inverse metric tensor, is given by
\begin{align}
     C_{A_3} = \begin{pmatrix}
       2 & -1 & 0 \\ -1 & 2 & -1 \\ 0 & -1 & 2
     \end{pmatrix}
     = h_{ij} \,.
 \end{align}
\subsubsection{\texorpdfstring{$\sprm(2)$}{Sp(2)} }
\label{app:Sp_roots_weights}
The simple roots are $\alpha_1=(2,-1)$ , $\alpha_2=(-2,2)$ and $a_1$ is the short root. The representations relevant for this work have the weights systems given by
\begin{align}
\raisebox{-.5\height}{
\begin{tikzpicture}
\node at (-1,0.5) {$[1,0]_C$};
\node[draw,circle,inner sep=0.8pt,fill,black]  (w1) at (0,0) {};
 \node[draw,circle,inner sep=0.8pt,fill,black]  (w2) at (0,-1) {};
 \node[draw,circle,inner sep=0.8pt,fill,black]  (w3) at (0,-2) {};
 \node[draw,circle,inner sep=0.8pt,fill,black]  (w4) at (0,-3) {};
 \draw (w1)--(w2)--(w3)--(w4);
 \node at (0.75,0) {$\scriptstyle{w_1=(1,0)}$};
 \node at (0.75,-1) {$\scriptstyle{w_2=(-1,1)}$};
 \node at (0.75,-2) {$\scriptstyle{w_3=(1,-1)}$};
 \node at (0.75,-3) {$\scriptstyle{w_4=(-1,0)}$};
 \node at (-0.5,-0.5) {$\scriptstyle{\alpha_1 \; \downarrow}$};
 \node at (-0.5,-1.5) {$\scriptstyle{\alpha_2\; \downarrow}$};
 \node at (-0.5,-2.5) {$\scriptstyle{\alpha_1 \; \downarrow}$};
\end{tikzpicture}
}
\qquad 
\raisebox{-.5\height}{
\begin{tikzpicture}
\node at (-1,0.5) {$[0,1]_C$};
\node[draw,circle,inner sep=0.8pt,fill,black]  (v1) at (0,0) {};
 \node[draw,circle,inner sep=0.8pt,fill,black]  (v2) at (0,-1) {};
 \node[draw,circle,inner sep=0.8pt,fill,black]  (v3) at (0,-2) {};
 \node[draw,circle,inner sep=0.8pt,fill,black]  (v4) at (0,-3) {};
 \node[draw,circle,inner sep=0.8pt,fill,black]  (v5) at (0,-4) {};
 \draw (v1)--(v2)--(v3)--(v4)--(v5);
 \node at (0.75,0) {$\scriptstyle{v_1=(0,1)}$};
 \node at (0.75,-1) {$\scriptstyle{v_2=(2,-1)}$};
 \node at (0.75,-2) {$\scriptstyle{v_3=(0,0)}$};
 \node at (0.75,-3) {$\scriptstyle{v_4=(-2,1)}$};
  \node at (0.75,-4) {$\scriptstyle{v_5=(0,-1)}$};
 \node at (-0.5,-0.5) {$\scriptstyle{\alpha_2 \; \downarrow}$};
 \node at (-0.5,-1.5) {$\scriptstyle{\alpha_1\; \downarrow}$};
 \node at (-0.5,-2.5) {$\scriptstyle{\alpha_1 \; \downarrow}$};
 \node at (-0.5,-3.5) {$\scriptstyle{\alpha_2 \; \downarrow}$};
\end{tikzpicture}
}
\qquad 
\raisebox{-.5\height}{
\begin{tikzpicture}
\node at (-1,0.5) {$[2,0]_C$};
\node[draw,circle,inner sep=0.8pt,fill,black]  (a1) at (0,0) {};
 \node[draw,circle,inner sep=0.8pt,fill,black]  (a2) at (0,-1) {};
 \node[draw,circle,inner sep=0.8pt,fill,black]  (a3L) at (-1,-2) {};
 \node[draw,circle,inner sep=0.8pt,fill,black]  (a4L) at (-1,-3) {};
 \node[draw,circle,inner sep=0.8pt,fill,black]  (a5L) at (-1,-4) {};
 \node[draw,circle,inner sep=0.8pt,fill,black]  (a3R) at (1,-2) {};
 \node[draw,circle,inner sep=0.8pt,fill,black]  (a4R) at (1,-3) {};
 \node[draw,circle,inner sep=0.8pt,fill,black]  (a5R) at (1,-4) {};
 \node[draw,circle,inner sep=0.8pt,fill,black]  (a6) at (0,-5) {};
 \node[draw,circle,inner sep=0.8pt,fill,black]  (a7) at (0,-6) {};
 \draw (a1)--(a2)--(a3L)--(a4L)--(a5L)--(a6) (a2)--(a3R)--(a4R)--(a5R)--(a6)--(a7);
 \node at (0.5,0) {$\scriptstyle{(2,0)}$};
 \node at (0.5,-1) {$\scriptstyle{(0,1)}$};
 \node at (-1.5,-2) {$\scriptstyle{(-2,2)}$};
 \node at (1.5,-2) {$\scriptstyle{(2,-1)}$};
 \node at (-1.5,-3) {$\scriptstyle{(0,0)}$};
 \node at (1.5,-3) {$\scriptstyle{(0,0)}$};
 \node at (-1.5,-4) {$\scriptstyle{(2,-2)}$};
 \node at (1.5,-4) {$\scriptstyle{(-2,1)}$};
 \node at (0.5,-5) {$\scriptstyle{(0,-1)}$};
  \node at (0.5,-6) {$\scriptstyle{(-2,0)}$};
 \node at (-0.5,-0.5) {$\scriptstyle{\alpha_1 \; \downarrow}$};
 \node at (-0.75,-1.25) {$\scriptstyle{\alpha_1\; \swarrow}$};
 \node at (0.25,-1.75) {$\scriptstyle{\alpha_2\; \searrow}$};
 \node at (-0.6,-2.5) {$\scriptstyle{\downarrow \; \alpha_2  }$};
 \node at (-0.6,-3.5) {$\scriptstyle{\downarrow \;\alpha_2}$};
 \node at (0.6,-2.5) {$\scriptstyle{ \alpha_1 \; \downarrow }$};
 \node at (0.6,-3.5) {$\scriptstyle{\alpha_1 \; \downarrow}$};
 \node at (-0.75,-4.75) {$\scriptstyle{\alpha_1\; \searrow}$};
 \node at (0.25,-4.25) {$\scriptstyle{\alpha_2\; \swarrow}$};
 \node at (-0.5,-5.5) {$\scriptstyle{\alpha_1 \; \downarrow}$};
\end{tikzpicture}
}
\end{align}
where the arrows indicate the action of the simple roots. In addition, the
Cartan matrix and the inverse metric tensor $h_{ij}$ are given by
 \begin{align}
     C_{C_2} = \begin{pmatrix}
      2 & -1  \\ -2 & 2 
     \end{pmatrix}
     \; ,\qquad 
      (h_{ij}) = \begin{pmatrix}
      4 & -2  \\ -2 & 2 
     \end{pmatrix} \,.
     \end{align}
\subsubsection{\texorpdfstring{$G_2$}{G2}}
\label{app:G2_roots_weights}
The simple roots are given by $\alpha_1=(2,-1)$ , $\alpha_2=(-3,2)$ and $\alpha_1$ is the short root. The representations relevant for this paper have the following weight systems
\begin{align}
\raisebox{-.5\height}{
\begin{tikzpicture}
\node at (-1,0.5) {$[1,0]_G$};
\node[draw,circle,inner sep=0.8pt,fill,black]  (v1) at (0,0) {};
 \node[draw,circle,inner sep=0.8pt,fill,black]  (v2) at (0,-1) {};
 \node[draw,circle,inner sep=0.8pt,fill,black]  (v3) at (0,-2) {};
 \node[draw,circle,inner sep=0.8pt,fill,black]  (v4) at (0,-3) {};
 \node[draw,circle,inner sep=0.8pt,fill,black]  (v5) at (0,-4) {};
 \node[draw,circle,inner sep=0.8pt,fill,black]  (v6) at (0,-5) {};
 \node[draw,circle,inner sep=0.8pt,fill,black]  (v7) at (0,-6) {};
 \draw (v1)--(v2)--(v3)--(v4)--(v5)--(v6)--(v7);
 \node at (0.75,0) {$\scriptstyle{w_1=(1,0)}$};
 \node at (0.75,-1) {$\scriptstyle{w_2=(-1,1)}$};
 \node at (0.75,-2) {$\scriptstyle{w_3=(2,-1)}$};
 \node at (0.75,-3) {$\scriptstyle{w_4=(0,0)}$};
  \node at (0.75,-4) {$\scriptstyle{w_5=(-2,1)}$};
  \node at (0.75,-5) {$\scriptstyle{w_6=(1,-1)}$};
  \node at (0.75,-6) {$\scriptstyle{w_7=(-1,0)}$};
 \node at (-0.5,-0.5) {$\scriptstyle{\alpha_1 \; \downarrow}$};
 \node at (-0.5,-1.5) {$\scriptstyle{\alpha_2\; \downarrow}$};
 \node at (-0.5,-2.5) {$\scriptstyle{\alpha_1 \; \downarrow}$};
 \node at (-0.5,-3.5) {$\scriptstyle{\alpha_1 \; \downarrow}$};
 \node at (-0.5,-4.5) {$\scriptstyle{\alpha_2\; \downarrow}$};
 \node at (-0.5,-5.5) {$\scriptstyle{\alpha_1 \; \downarrow}$};
\end{tikzpicture}
}
\qquad 
\raisebox{-.5\height}{
\begin{tikzpicture}
\node at (-1,2.5) {$[0,1]_G$};
\node[draw,circle,inner sep=0.8pt,fill,black]  (a00) at (0,2) {};
 \node[draw,circle,inner sep=0.8pt,fill,black]  (a0) at (0,1) {};
\node[draw,circle,inner sep=0.8pt,fill,black]  (a1) at (0,0) {};
 \node[draw,circle,inner sep=0.8pt,fill,black]  (a2) at (0,-1) {};
 \node[draw,circle,inner sep=0.8pt,fill,black]  (a3L) at (-1,-2) {};
 \node[draw,circle,inner sep=0.8pt,fill,black]  (a4L) at (-1,-3) {};
 \node[draw,circle,inner sep=0.8pt,fill,black]  (a5L) at (-1,-4) {};
 \node[draw,circle,inner sep=0.8pt,fill,black]  (a3R) at (1,-2) {};
 \node[draw,circle,inner sep=0.8pt,fill,black]  (a4R) at (1,-3) {};
 \node[draw,circle,inner sep=0.8pt,fill,black]  (a5R) at (1,-4) {};
 \node[draw,circle,inner sep=0.8pt,fill,black]  (a6) at (0,-5) {};
 \node[draw,circle,inner sep=0.8pt,fill,black]  (a7) at (0,-6) {};
 \node[draw,circle,inner sep=0.8pt,fill,black]  (a8) at (0,-7) {};
 \node[draw,circle,inner sep=0.8pt,fill,black]  (a9) at (0,-8) {};
 \draw (a00)--(a0)--(a1)--(a2)--(a3L)--(a4L)--(a5L)--(a6) (a2)--(a3R)--(a4R)--(a5R)--(a6)--(a7)--(a8)--(a9);
 \node at (0.5,2) {$\scriptstyle{(0,1)}$};
 \node at (0.5,1) {$\scriptstyle{(3,-1)}$};
 \node at (0.5,0) {$\scriptstyle{(1,0)}$};
 \node at (0.5,-1) {$\scriptstyle{(-1,1)}$};
 \node at (-1.5,-2) {$\scriptstyle{(-3,2)}$};
 \node at (1.5,-2) {$\scriptstyle{(2,-1)}$};
 \node at (-1.5,-3) {$\scriptstyle{(0,0)}$};
 \node at (1.5,-3) {$\scriptstyle{(0,0)}$};
 \node at (-1.5,-4) {$\scriptstyle{(3,-2)}$};
 \node at (1.5,-4) {$\scriptstyle{(-2,1)}$};
 \node at (0.5,-5) {$\scriptstyle{(1,-1)}$};
  \node at (0.5,-6) {$\scriptstyle{(-1,0)}$};
  \node at (0.5,-7) {$\scriptstyle{(-3,1)}$};
  \node at (0.5,-8) {$\scriptstyle{(0,-1)}$};
 \node at (-0.5,1.5) {$\scriptstyle{\alpha_2 \; \downarrow}$};
 \node at (-0.5,0.5) {$\scriptstyle{\alpha_1 \; \downarrow}$};
 \node at (-0.5,-0.5) {$\scriptstyle{\alpha_1 \; \downarrow}$};
 \node at (-0.75,-1.25) {$\scriptstyle{\alpha_1\; \swarrow}$};
 \node at (0.25,-1.75) {$\scriptstyle{\alpha_2\; \searrow}$};
 \node at (-0.6,-2.5) {$\scriptstyle{\downarrow \; \alpha_2  }$};
 \node at (-0.6,-3.5) {$\scriptstyle{\downarrow \;\alpha_2}$};
 \node at (0.6,-2.5) {$\scriptstyle{ \alpha_1 \; \downarrow }$};
 \node at (0.6,-3.5) {$\scriptstyle{\alpha_1 \; \downarrow}$};
 \node at (-0.75,-4.75) {$\scriptstyle{\alpha_1\; \searrow}$};
 \node at (0.25,-4.25) {$\scriptstyle{\alpha_2\; \swarrow}$};
 \node at (-0.5,-5.5) {$\scriptstyle{\alpha_1 \; \downarrow}$};
 \node at (-0.5,-6.5) {$\scriptstyle{\alpha_1 \; \downarrow}$};
 \node at (-0.5,-7.5) {$\scriptstyle{\alpha_2 \; \downarrow}$};
\end{tikzpicture}
}
\end{align}
where the arrows indicate the action of the simple roots. In addition, the
Cartan matrix and the inverse metric tensor $h_{ij}$ are given by
\begin{align}
     C_{G_2} = \begin{pmatrix}
       2 & -1  \\ -3 & 2 
     \end{pmatrix}
     \; ,\qquad 
      (h_{ij}) = \begin{pmatrix}
       6 & -3  \\ -3 & 2 
     \end{pmatrix}
     \;.
 \end{align}
%
\subsection{Geometry of Hirzebruch surfaces}
\label{sec:hirzebruch}
A Hirzebruch surface is a $\mathbb{P}^1$ fibration over $\mathbb{P}^1$, and one denotes by $\FF_n$ a Hirzebruch surface with a degree $-n$ fibration. The fibre $\mathbb{P}^1$ is denoted by $f$, while the base $\mathbb{P}^1$ is $e$. The intersections numbers are 
\begin{align}
    e^2 = -n
    \;,\quad 
    f^2 =0
    \;,\quad
    e\cdot f =1 \,.
\end{align}
Another curve inside $\FF_n$ is defined by $h\coloneqq e +nf$, which for $\FF_0$ becomes equal to $e$. For $n\geq 0$, the set of holomorphic curves, the \emph{Mori cone}, is generated by $e$ and $f$.

One may also consider surfaces that arise from blowing up $\FF_n$ is a number of points. Let $b$ be the number of blowups, such that the blowup of $\FF_n$ at $b$ points is denoted by $\FF_n^b$. The exceptional divisors created by the blowups are the curves $x_i$ for $i=1,\ldots,b$. The intersection number of $e$, $f$, $h$ with the $x_i$ are given by
\begin{align}
    x_i \cdot x_j = - \delta_{i,j}
    \;, \quad 
    e\cdot x_i =0
    \;, \quad 
    f\cdot x_i =0
    \;, \quad 
    h\cdot x_i =0 \,.
\end{align}
Following the conventions of \cite{Bhardwaj:2019fzv}, the total transforms of the curves $e$, $f$, $h$ are denoted by the same names $e$, $f$, $h$ in $\FF_n^b$.

A single theory may enjoy many (isomorphic) geometric descriptions. To transition between them, it is useful to recall the  $\FF_n^b \to \FF_{n+1}^b$ isomorphism, see for instance \cite[eqs.\ (2.19)-(2.22)]{Bhardwaj:2019ngx}, 
\begin{align}
\begin{aligned}
 e-x_i &\to e \\
 f-x_i &\to x_i \\
 x_i &\to f-x_i \\
 x_j &\to x_j \qquad j\neq i
 \end{aligned}
 \label{eq:iso_Fn_Fn+1}
\end{align}
which comes into play in Section \ref{sec:affine_quiver}.
\section{Rulings on rational elliptic surfaces}
\label{app:rulingsofdp9}
In this appendix, we give an explicit realisation of a rational elliptic surface $dP_9$ \cite{miranda1989basic} that shows that we can also 
view it as a blowup $\mathbb{F}_0^8$ of the Hirzebruch surface $\mathbb{F}_0 = \P^1 \times \P^1$ at eight points. 
Seen in this way, $dP_9$ inherits the two rulings on $\mathbb{F}_0$, and we can derive the inner form between divisors.   

Consider a hypersurface $S$ of degree $(1,2,2)$ inside $\P^1_w \times \P^1_y \times \P^1_z$. Denoting the 
homogeneous coordinates of the three $\P^1$s by $[w_1:w_2]$, $[y_1:y_2]$ and $[z_1:z_2]$, such a hypersurface can be written as 
\begin{equation}
\label{eq:dp9hypers}
P(y,z) w_1 =  Q(y,z) w_2   
\end{equation}
for two homogeneous polynomials $P$ and $Q$ that have both degrees $2$ in $ [y_1:y_2]$ and $[z_1:z_2]$. 

Projecting to $\P^1_w$, this surface has the structure of an elliptic fibration, with the fibre embedded in $\P_y 
\times \P_z$. Denoting the hyperplane divisors of the three $\P^1$s by $H_w, H_y, H_z$, we find that $c_1(S) = H_w$, 
which is represented by the generic fibre of this elliptic fibration. We hence identify $S$ as a rational elliptic surface. 

Let us now discuss the ruling. For a generic point on $\P^1_y \times \P^1_z$, \eqref{eq:dp9hypers} fixes a unique 
point on $\P^1_w$. However, over points $(q_i,p_i)$ on $\P^1_y \times \P^1_z$ where $Q = P = 0$, there is no constraint on $\P^1_w$ and 
we find another $\P^1$ in $S$. As $Q$ and $P$ are in the class $2H_y + 2 H_z$, there are
\begin{equation}
\int_{\P^1_y \times \P^1_z}  (2 H_y + 2 H_z)^2 = 8 
\end{equation}
such points. We hence see that $dP_9$ is $\mathbb{F}_0$ blown up at 8 points $(q_i,p_i)$. We can now describe the inner form among divisors 
on $dP_9$ from this perspective. Let us denote $H_y$ restricted to $S$ by $e$ and $H_z$ restricted to $S$ by $f$. We then have 
that $e^2 = f^2 = 0$ (as $H_y \cdot H_y = H_z \cdot H_z=0$) and $e \cdot f = 1$ (as $H_y \cdot H_z = 1$ on S). Let us denote the 
$\P^1$s sitting at the 8 points $Q=P=0$ by $x_i$. As these are $\P^1$s and 
$\int_S c_1(S) \cdot x_i = \int_{\P^1_w \times \P^1_y \times \P^1_z} H_w \cdot x_i = 1$ we find using adjunction that 
$x_i \cdot x_i = -1$. All other intersections must vanish, $x_i \cdot x_j = - \delta_{ij}$ as these sit over 
different points of $\P^1_y \times \P^1_z$, and similarly $x_i \cdot e = x_i \cdot f = 0 $ as a generic hyperplane section 
misses any of the points $Q=P=0$. In summary, the only non-vanishing intersections of the inner form are
\begin{equation}
\begin{aligned}
e \cdot f = 1 \hspace{1cm} x_i^2 = -1 \\
\end{aligned}
\end{equation}
which in particular shows that this lattice is unimodular, i.e.\ $e,f,x_i$ generate $H^2(S,\mathbb{Z})$. Note that the curves $x_i$ all become 
sections of the elliptic fibration on $S$: for every $x_i$, we may assign every point on a copy of the base $\P^1_w$ to the point $(p_i,q_i)$
in the fibre. 

We can also discuss the details of the (blown-up) ruling. As we want $f$ to be class represented by the fibre and $f= H_z$,
i.e. the fibre class is given by e.g.\ $z_1=0$, the projection of the ruling acts on the level of the ambient space as 
$\pi: \P^1_x \times \P^1_y \times \P^1_z \rightarrow \P^1_z$, i.e.\ it projects $S$ down to $\P^1_z$. Over a generic point $z$, the fibre of the 
ruling is the $\P^1$ described by 
\begin{equation}
Q(z,y) w_1 = P(z,y) w_2 \, ,
\end{equation}
which is smooth as $Q(z,y)$ and $P(z,y)$ do not vanish simultaneously for any $y$ for such a fixed $z$. The class of the fibre $f$. 
If we now take $[z_1:z_2]$ to correspond to the image of one of the $(q_i,p_i)$, the fibre is given by 
\begin{equation}
Q(q_i,y) w_1 = P(q_i,y) w_2 \, ,
\end{equation}
where now $P=Q=0$ for $y = p_i$. This implies that this set is the union of the $\P^1_y$ times the unique point in $\P^1_w$ given by the above equation, together 
with the $\P^1$ at $p_i \times \P^1_w = x_i$. If we denote the first curve by $\hat{x}_i$, linear equivalence gives
\begin{equation}
f = \hat{x}_i + x_i 
\end{equation}
for all $i$. In other words, the fibre of the ruling splits into pairs of $\P^1$s of classes $x_i$ and $\hat{x}_i = f- x_i$ over 8 points on $\P^1_z$. 

As we may as well project down to $\P^1_y$, there is another ruling with fibre $e$ and base $f$. The fibre components of this ruling are seen to be 
$x_i$ and $\check{x}_i = e-x_i$ by repeating the same arguments as above. 

As $h^{1,1}(S) = 10$, the ten classes $e,f,x_i$ span all of $H^{1,1}(S,\mathbb{Q})$ and we can work out the class of 
$c_1(S)$ in terms of them by computing intersections. As
\begin{equation}
\begin{aligned}
c_1(S) \cdot e  &=  H_w \cdot H_y =  2 \\
c_1(S) \cdot f  &=  H_w \cdot H_z =  2 \\
c_1(S) \cdot x_i&=  H_w \cdot x_i = 1 
\end{aligned}
\end{equation}
we find that 
\begin{equation}
c_1(S) = H_w = 2 e + 2 f - \sum_{i=1}^8 x_i \, .
\end{equation}
This is of course precisely the result we are expecting for a surface that is reached from $\mathbb{F}_0$ by blowing up at 
8 points.

By construction, the cone of effective curves contains $e,f$ and the $x_i$, as well as $c_1(S)$. Furthermore, we have seen that there are the additional 
effective curves $\hat{x}_i = f - x_i$ and $\check{x}_i= e -x_i$. All of these must have positive volume and due their negative self-intersection they 
are among the extremal generators of the Mori cone. Note that we can write $f = x_i + \hat{x}_i$, $e = x_i + \check{x}_i$, but that $c_1(S)$ is not 
a positive linear combination of $e,f,x_i,\hat{x}_i,\check{x}_i$, so that is also among the extremal rays of the cone of effective curves. Note that in the explicit realisation of $S$ we have given, the volumes of the $x_i$ are locked to all be identical. 
%
%
 \bibliographystyle{JHEP}     
 {\footnotesize{\bibliography{references}}}

\providecommand{\href}[2]{#2}\begingroup\raggedright\begin{thebibliography}{100}

\bibitem{Seiberg:1996bd}
N.~Seiberg, \emph{{Five-dimensional SUSY field theories, nontrivial fixed
  points and string dynamics}},
  \href{https://doi.org/10.1016/S0370-2693(96)01215-4}{\emph{Phys. Lett. B}
  {\bfseries 388} (1996) 753}
  [\href{https://arxiv.org/abs/hep-th/9608111}{{\ttfamily hep-th/9608111}}].

\bibitem{Morrison:1996xf}
D.~R. Morrison and N.~Seiberg, \emph{{Extremal transitions and five-dimensional
  supersymmetric field theories}},
  \href{https://doi.org/10.1016/S0550-3213(96)00592-5}{\emph{Nucl. Phys. B}
  {\bfseries 483} (1997) 229}
  [\href{https://arxiv.org/abs/hep-th/9609070}{{\ttfamily hep-th/9609070}}].

\bibitem{Intriligator:1997pq}
K.~A. Intriligator, D.~R. Morrison and N.~Seiberg, \emph{{Five-dimensional
  supersymmetric gauge theories and degenerations of Calabi-Yau spaces}},
  \href{https://doi.org/10.1016/S0550-3213(97)00279-4}{\emph{Nucl. Phys. B}
  {\bfseries 497} (1997) 56}
  [\href{https://arxiv.org/abs/hep-th/9702198}{{\ttfamily hep-th/9702198}}].

\bibitem{Aharony:1997ju}
O.~Aharony and A.~Hanany, \emph{{Branes, superpotentials and superconformal
  fixed points}},
  \href{https://doi.org/10.1016/S0550-3213(97)00472-0}{\emph{Nucl. Phys. B}
  {\bfseries 504} (1997) 239}
  [\href{https://arxiv.org/abs/hep-th/9704170}{{\ttfamily hep-th/9704170}}].

\bibitem{Aharony:1997bh}
O.~Aharony, A.~Hanany and B.~Kol, \emph{{Webs of (p,q) five-branes,
  five-dimensional field theories and grid diagrams}},
  \href{https://doi.org/10.1088/1126-6708/1998/01/002}{\emph{JHEP} {\bfseries
  01} (1998) 002} [\href{https://arxiv.org/abs/hep-th/9710116}{{\ttfamily
  hep-th/9710116}}].

\bibitem{DeWolfe:1999hj}
O.~DeWolfe, A.~Hanany, A.~Iqbal and E.~Katz, \emph{{Five-branes, seven-branes
  and five-dimensional E(n) field theories}},
  \href{https://doi.org/10.1088/1126-6708/1999/03/006}{\emph{JHEP} {\bfseries
  03} (1999) 006} [\href{https://arxiv.org/abs/hep-th/9902179}{{\ttfamily
  hep-th/9902179}}].

\bibitem{Douglas:1996xp}
M.~R. Douglas, S.~H. Katz and C.~Vafa, \emph{{Small instantons, Del Pezzo
  surfaces and type I-prime theory}},
  \href{https://doi.org/10.1016/S0550-3213(97)00281-2}{\emph{Nucl. Phys. B}
  {\bfseries 497} (1997) 155}
  [\href{https://arxiv.org/abs/hep-th/9609071}{{\ttfamily hep-th/9609071}}].

\bibitem{Kim:2012gu}
H.-C. Kim, S.-S. Kim and K.~Lee, \emph{{5-dim Superconformal Index with
  Enhanced En Global Symmetry}},
  \href{https://doi.org/10.1007/JHEP10(2012)142}{\emph{JHEP} {\bfseries 10}
  (2012) 142} [\href{https://arxiv.org/abs/1206.6781}{{\ttfamily 1206.6781}}].

\bibitem{Bergman:2013ala}
O.~Bergman, D.~Rodr\'\i{}guez-G\'omez and G.~Zafrir, \emph{{Discrete $\theta$
  and the 5d superconformal index}},
  \href{https://doi.org/10.1007/JHEP01(2014)079}{\emph{JHEP} {\bfseries 01}
  (2014) 079} [\href{https://arxiv.org/abs/1310.2150}{{\ttfamily 1310.2150}}].

\bibitem{Bergman:2013aca}
O.~Bergman, D.~Rodr\'\i{}guez-G\'omez and G.~Zafrir, \emph{{5-Brane Webs,
  Symmetry Enhancement, and Duality in 5d Supersymmetric Gauge Theory}},
  \href{https://doi.org/10.1007/JHEP03(2014)112}{\emph{JHEP} {\bfseries 03}
  (2014) 112} [\href{https://arxiv.org/abs/1311.4199}{{\ttfamily 1311.4199}}].

\bibitem{Zafrir:2014ywa}
G.~Zafrir, \emph{{Duality and enhancement of symmetry in 5d gauge theories}},
  \href{https://doi.org/10.1007/JHEP12(2014)116}{\emph{JHEP} {\bfseries 12}
  (2014) 116} [\href{https://arxiv.org/abs/1408.4040}{{\ttfamily 1408.4040}}].

\bibitem{Tachikawa:2015mha}
Y.~Tachikawa, \emph{{Instanton operators and symmetry enhancement in 5d
  supersymmetric gauge theories}},
  \href{https://doi.org/10.1093/ptep/ptv040}{\emph{PTEP} {\bfseries 2015}
  (2015) 043B06} [\href{https://arxiv.org/abs/1501.01031}{{\ttfamily
  1501.01031}}].

\bibitem{Zafrir:2015uaa}
G.~Zafrir, \emph{{Instanton operators and symmetry enhancement in 5d
  supersymmetric USp, SO and exceptional gauge theories}},
  \href{https://doi.org/10.1007/JHEP07(2015)087}{\emph{JHEP} {\bfseries 07}
  (2015) 087} [\href{https://arxiv.org/abs/1503.08136}{{\ttfamily
  1503.08136}}].

\bibitem{Yonekura:2015ksa}
K.~Yonekura, \emph{{Instanton operators and symmetry enhancement in 5d
  supersymmetric quiver gauge theories}},
  \href{https://doi.org/10.1007/JHEP07(2015)167}{\emph{JHEP} {\bfseries 07}
  (2015) 167} [\href{https://arxiv.org/abs/1505.04743}{{\ttfamily
  1505.04743}}].

\bibitem{Cremonesi:2015lsa}
S.~Cremonesi, G.~Ferlito, A.~Hanany and N.~Mekareeya, \emph{{Instanton
  Operators and the Higgs Branch at Infinite Coupling}},
  \href{https://doi.org/10.1007/JHEP04(2017)042}{\emph{JHEP} {\bfseries 04}
  (2017) 042} [\href{https://arxiv.org/abs/1505.06302}{{\ttfamily
  1505.06302}}].

\bibitem{Jefferson:2017ahm}
P.~Jefferson, H.-C. Kim, C.~Vafa and G.~Zafrir, \emph{{Towards Classification
  of 5d SCFTs: Single Gauge Node}},
  \href{https://arxiv.org/abs/1705.05836}{{\ttfamily 1705.05836}}.

\bibitem{Ferlito:2017xdq}
G.~Ferlito, A.~Hanany, N.~Mekareeya and G.~Zafrir, \emph{{3d Coulomb branch and
  5d Higgs branch at infinite coupling}},
  \href{https://doi.org/10.1007/JHEP07(2018)061}{\emph{JHEP} {\bfseries 07}
  (2018) 061} [\href{https://arxiv.org/abs/1712.06604}{{\ttfamily
  1712.06604}}].

\bibitem{Bao:2011rc}
L.~Bao, E.~Pomoni, M.~Taki and F.~Yagi, \emph{{M5-Branes, Toric Diagrams and
  Gauge Theory Duality}},
  \href{https://doi.org/10.1007/JHEP04(2012)105}{\emph{JHEP} {\bfseries 04}
  (2012) 105} [\href{https://arxiv.org/abs/1112.5228}{{\ttfamily 1112.5228}}].

\bibitem{Bergman:2014kza}
O.~Bergman and G.~Zafrir, \emph{{Lifting 4d dualities to 5d}},
  \href{https://doi.org/10.1007/JHEP04(2015)141}{\emph{JHEP} {\bfseries 04}
  (2015) 141} [\href{https://arxiv.org/abs/1410.2806}{{\ttfamily 1410.2806}}].

\bibitem{Kim:2015jba}
S.-S. Kim, M.~Taki and F.~Yagi, \emph{{Tao Probing the End of the World}},
  \href{https://doi.org/10.1093/ptep/ptv108}{\emph{PTEP} {\bfseries 2015}
  (2015) 083B02} [\href{https://arxiv.org/abs/1504.03672}{{\ttfamily
  1504.03672}}].

\bibitem{Hayashi:2015fsa}
H.~Hayashi, S.-S. Kim, K.~Lee, M.~Taki and F.~Yagi, \emph{{A new 5d description
  of 6d D-type minimal conformal matter}},
  \href{https://doi.org/10.1007/JHEP08(2015)097}{\emph{JHEP} {\bfseries 08}
  (2015) 097} [\href{https://arxiv.org/abs/1505.04439}{{\ttfamily
  1505.04439}}].

\bibitem{Gaiotto:2015una}
D.~Gaiotto and H.-C. Kim, \emph{{Duality walls and defects in 5d $
  \mathcal{N}=1 $ theories}},
  \href{https://doi.org/10.1007/JHEP01(2017)019}{\emph{JHEP} {\bfseries 01}
  (2017) 019} [\href{https://arxiv.org/abs/1506.03871}{{\ttfamily
  1506.03871}}].

\bibitem{Bergman:2015dpa}
O.~Bergman and G.~Zafrir, \emph{{5d fixed points from brane webs and
  O7-planes}}, \href{https://doi.org/10.1007/JHEP12(2015)163}{\emph{JHEP}
  {\bfseries 12} (2015) 163}
  [\href{https://arxiv.org/abs/1507.03860}{{\ttfamily 1507.03860}}].

\bibitem{Zafrir:2015rga}
G.~Zafrir, \emph{{Brane webs, $5d$ gauge theories and $6d$ $\mathcal{N}=(1,0)$
  SCFT's}}, \href{https://doi.org/10.1007/JHEP12(2015)157}{\emph{JHEP}
  {\bfseries 12} (2015) 157}
  [\href{https://arxiv.org/abs/1509.02016}{{\ttfamily 1509.02016}}].

\bibitem{Hayashi:2015zka}
H.~Hayashi, S.-S. Kim, K.~Lee and F.~Yagi, \emph{{6d SCFTs, 5d Dualities and
  Tao Web Diagrams}},
  \href{https://doi.org/10.1007/JHEP05(2019)203}{\emph{JHEP} {\bfseries 05}
  (2019) 203} [\href{https://arxiv.org/abs/1509.03300}{{\ttfamily
  1509.03300}}].

\bibitem{Ohmori:2015tka}
K.~Ohmori and H.~Shimizu, \emph{{$S^1/T^2$ compactifications of 6d $
  \mathcal{N}=\left(1,\;0\right) $ theories and brane webs}},
  \href{https://doi.org/10.1007/JHEP03(2016)024}{\emph{JHEP} {\bfseries 03}
  (2016) 024} [\href{https://arxiv.org/abs/1509.03195}{{\ttfamily
  1509.03195}}].

\bibitem{Zafrir:2015ftn}
G.~Zafrir, \emph{{Brane webs and $O5$-planes}},
  \href{https://doi.org/10.1007/JHEP03(2016)109}{\emph{JHEP} {\bfseries 03}
  (2016) 109} [\href{https://arxiv.org/abs/1512.08114}{{\ttfamily
  1512.08114}}].

\bibitem{Hayashi:2015vhy}
H.~Hayashi, S.-S. Kim, K.~Lee, M.~Taki and F.~Yagi, \emph{{More on 5d
  descriptions of 6d SCFTs}},
  \href{https://doi.org/10.1007/JHEP10(2016)126}{\emph{JHEP} {\bfseries 10}
  (2016) 126} [\href{https://arxiv.org/abs/1512.08239}{{\ttfamily
  1512.08239}}].

\bibitem{Zafrir:2016jpu}
G.~Zafrir, \emph{{Brane webs in the presence of an O5$^{−}$-plane and 4d
  class S theories of type D}},
  \href{https://doi.org/10.1007/JHEP07(2016)035}{\emph{JHEP} {\bfseries 07}
  (2016) 035} [\href{https://arxiv.org/abs/1602.00130}{{\ttfamily
  1602.00130}}].

\bibitem{Hayashi:2016abm}
H.~Hayashi, S.-S. Kim, K.~Lee and F.~Yagi, \emph{{Equivalence of several
  descriptions for 6d SCFT}},
  \href{https://doi.org/10.1007/JHEP01(2017)093}{\emph{JHEP} {\bfseries 01}
  (2017) 093} [\href{https://arxiv.org/abs/1607.07786}{{\ttfamily
  1607.07786}}].

\bibitem{Hayashi:2017btw}
H.~Hayashi, S.-S. Kim, K.~Lee and F.~Yagi, \emph{{Discrete theta angle from an
  O5-plane}}, \href{https://doi.org/10.1007/JHEP11(2017)041}{\emph{JHEP}
  {\bfseries 11} (2017) 041}
  [\href{https://arxiv.org/abs/1707.07181}{{\ttfamily 1707.07181}}].

\bibitem{Hayashi:2018bkd}
H.~Hayashi, S.-S. Kim, K.~Lee and F.~Yagi, \emph{{5-brane webs for 5d $
  \mathcal{N} $ = 1 G$_{2}$ gauge theories}},
  \href{https://doi.org/10.1007/JHEP03(2018)125}{\emph{JHEP} {\bfseries 03}
  (2018) 125} [\href{https://arxiv.org/abs/1801.03916}{{\ttfamily
  1801.03916}}].

\bibitem{Hayashi:2018lyv}
H.~Hayashi, S.-S. Kim, K.~Lee and F.~Yagi, \emph{{Dualities and 5-brane webs
  for 5d rank 2 SCFTs}},
  \href{https://doi.org/10.1007/JHEP12(2018)016}{\emph{JHEP} {\bfseries 12}
  (2018) 016} [\href{https://arxiv.org/abs/1806.10569}{{\ttfamily
  1806.10569}}].

\bibitem{Cabrera:2018jxt}
S.~Cabrera, A.~Hanany and F.~Yagi, \emph{{Tropical Geometry and Five
  Dimensional Higgs Branches at Infinite Coupling}},
  \href{https://doi.org/10.1007/JHEP01(2019)068}{\emph{JHEP} {\bfseries 01}
  (2019) 068} [\href{https://arxiv.org/abs/1810.01379}{{\ttfamily
  1810.01379}}].

\bibitem{Hayashi:2019yxj}
H.~Hayashi, S.-S. Kim, K.~Lee and F.~Yagi, \emph{{Rank-3 antisymmetric matter
  on 5-brane webs}}, \href{https://doi.org/10.1007/JHEP05(2019)133}{\emph{JHEP}
  {\bfseries 05} (2019) 133}
  [\href{https://arxiv.org/abs/1902.04754}{{\ttfamily 1902.04754}}].

\bibitem{Hayashi:2019jvx}
H.~Hayashi, S.-S. Kim, K.~Lee and F.~Yagi, \emph{{Complete prepotential for 5d
  $ \mathcal{N} $ = 1 superconformal field theories}},
  \href{https://doi.org/10.1007/JHEP02(2020)074}{\emph{JHEP} {\bfseries 02}
  (2020) 074} [\href{https://arxiv.org/abs/1912.10301}{{\ttfamily
  1912.10301}}].

\bibitem{Bourget:2020gzi}
A.~Bourget, J.~F. Grimminger, A.~Hanany, M.~Sperling and Z.~Zhong,
  \emph{{Magnetic Quivers from Brane Webs with O5 Planes}},
  \href{https://doi.org/10.1007/JHEP07(2020)204}{\emph{JHEP} {\bfseries 07}
  (2020) 204} [\href{https://arxiv.org/abs/2004.04082}{{\ttfamily
  2004.04082}}].

\bibitem{vanBeest:2020civ}
M.~van Beest, A.~Bourget, J.~Eckhard and S.~Schafer-Nameki, \emph{{(5d RG-flow)
  Trees in the Tropical Rain Forest}},
  \href{https://arxiv.org/abs/2011.07033}{{\ttfamily 2011.07033}}.

\bibitem{vanBeest:2020kou}
M.~van Beest, A.~Bourget, J.~Eckhard and S.~Schafer-Nameki, \emph{{(Symplectic)
  Leaves and (5d Higgs) Branches in the Poly(go)nesian Tropical Rain Forest}},
  \href{https://doi.org/10.1007/JHEP11(2020)124}{\emph{JHEP} {\bfseries 11}
  (2020) 124} [\href{https://arxiv.org/abs/2008.05577}{{\ttfamily
  2008.05577}}].

\bibitem{Akhond:2020vhc}
M.~Akhond, F.~Carta, S.~Dwivedi, H.~Hayashi, S.-S. Kim and F.~Yagi,
  \emph{{Five-brane webs, Higgs branches and unitary/orthosymplectic magnetic
  quivers}}, \href{https://doi.org/10.1007/JHEP12(2020)164}{\emph{JHEP}
  {\bfseries 12} (2020) 164}
  [\href{https://arxiv.org/abs/2008.01027}{{\ttfamily 2008.01027}}].

\bibitem{Hayashi:2021pcj}
H.~Hayashi, H.-C. Kim and K.~Ohmori, \emph{{6d/5d exceptional gauge theories
  from web diagrams}},  \href{https://arxiv.org/abs/2103.02799}{{\ttfamily
  2103.02799}}.

\bibitem{DelZotto:2017pti}
M.~Del~Zotto, J.~J. Heckman and D.~R. Morrison, \emph{{6D SCFTs and Phases of
  5D Theories}}, \href{https://doi.org/10.1007/JHEP09(2017)147}{\emph{JHEP}
  {\bfseries 09} (2017) 147}
  [\href{https://arxiv.org/abs/1703.02981}{{\ttfamily 1703.02981}}].

\bibitem{Xie:2017pfl}
D.~Xie and S.-T. Yau, \emph{{Three dimensional canonical singularity and five
  dimensional $ \mathcal{N} $ = 1 SCFT}},
  \href{https://doi.org/10.1007/JHEP06(2017)134}{\emph{JHEP} {\bfseries 06}
  (2017) 134} [\href{https://arxiv.org/abs/1704.00799}{{\ttfamily
  1704.00799}}].

\bibitem{Esole:2017rgz}
M.~Esole, P.~Jefferson and M.~J. Kang, \emph{{The Geometry of F$_4$-Models}},
  \href{https://arxiv.org/abs/1704.08251}{{\ttfamily 1704.08251}}.

\bibitem{Esole:2017qeh}
M.~Esole, R.~Jagadeesan and M.~J. Kang, \emph{{The Geometry of G$_2$, Spin(7),
  and Spin(8)-models}},  \href{https://arxiv.org/abs/1709.04913}{{\ttfamily
  1709.04913}}.

\bibitem{Esole:2017hlw}
M.~Esole, M.~J. Kang and S.-T. Yau, \emph{{Mordell-Weil Torsion, Anomalies, and
  Phase Transitions}},  \href{https://arxiv.org/abs/1712.02337}{{\ttfamily
  1712.02337}}.

\bibitem{Jefferson:2018irk}
P.~Jefferson, S.~Katz, H.-C. Kim and C.~Vafa, \emph{{On Geometric
  Classification of 5d SCFTs}},
  \href{https://doi.org/10.1007/JHEP04(2018)103}{\emph{JHEP} {\bfseries 04}
  (2018) 103} [\href{https://arxiv.org/abs/1801.04036}{{\ttfamily
  1801.04036}}].

\bibitem{Esole:2018csl}
M.~Esole and M.~J. Kang, \emph{{Flopping and slicing: SO(4) and
  Spin(4)-models}},
  \href{https://doi.org/10.4310/ATMP.2019.v23.n4.a2}{\emph{Adv. Theor. Math.
  Phys.} {\bfseries 23} (2019) 1003}
  [\href{https://arxiv.org/abs/1802.04802}{{\ttfamily 1802.04802}}].

\bibitem{Esole:2018mqb}
M.~Esole and M.~J. Kang, \emph{{The Geometry of the SU(2)$\times$
  G$_2$-model}}, \href{https://doi.org/10.1007/JHEP02(2019)091}{\emph{JHEP}
  {\bfseries 02} (2019) 091}
  [\href{https://arxiv.org/abs/1805.03214}{{\ttfamily 1805.03214}}].

\bibitem{Bhardwaj:2018yhy}
L.~Bhardwaj and P.~Jefferson, \emph{{Classifying 5d SCFTs via 6d SCFTs: Rank
  one}}, \href{https://doi.org/10.1007/JHEP07(2019)178}{\emph{JHEP} {\bfseries
  07} (2019) 178} [\href{https://arxiv.org/abs/1809.01650}{{\ttfamily
  1809.01650}}].

\bibitem{Bhardwaj:2018vuu}
L.~Bhardwaj and P.~Jefferson, \emph{{Classifying 5d SCFTs via 6d SCFTs:
  Arbitrary rank}}, \href{https://doi.org/10.1007/JHEP10(2019)282}{\emph{JHEP}
  {\bfseries 10} (2019) 282}
  [\href{https://arxiv.org/abs/1811.10616}{{\ttfamily 1811.10616}}].

\bibitem{Apruzzi:2018nre}
F.~Apruzzi, L.~Lin and C.~Mayrhofer, \emph{{Phases of 5d SCFTs from M-/F-theory
  on Non-Flat Fibrations}},
  \href{https://doi.org/10.1007/JHEP05(2019)187}{\emph{JHEP} {\bfseries 05}
  (2019) 187} [\href{https://arxiv.org/abs/1811.12400}{{\ttfamily
  1811.12400}}].

\bibitem{Banerjee:2018syt}
S.~Banerjee, P.~Longhi and M.~Romo, \emph{{Exploring 5d BPS Spectra with
  Exponential Networks}},
  \href{https://doi.org/10.1007/s00023-019-00851-x}{\emph{Annales Henri
  Poincare} {\bfseries 20} (2019) 4055}
  [\href{https://arxiv.org/abs/1811.02875}{{\ttfamily 1811.02875}}].

\bibitem{Closset:2018bjz}
C.~Closset, M.~Del~Zotto and V.~Saxena, \emph{{Five-dimensional SCFTs and gauge
  theory phases: an M-theory/type IIA perspective}},
  \href{https://doi.org/10.21468/SciPostPhys.6.5.052}{\emph{SciPost Phys.}
  {\bfseries 6} (2019) 052} [\href{https://arxiv.org/abs/1812.10451}{{\ttfamily
  1812.10451}}].

\bibitem{Esole:2019hgr}
M.~Esole and P.~Jefferson, \emph{{The Geometry of SO(3), SO(5), and SO(6)
  models}},  \href{https://arxiv.org/abs/1905.12620}{{\ttfamily 1905.12620}}.

\bibitem{Esole:2019asj}
M.~Esole, R.~Jagadeesan and M.~J. Kang, \emph{{48 Crepant Paths to
  $\text{SU}(2)\!\times\!\text{SU}(3)$}},
  \href{https://arxiv.org/abs/1905.05174}{{\ttfamily 1905.05174}}.

\bibitem{Apruzzi:2019vpe}
F.~Apruzzi, C.~Lawrie, L.~Lin, S.~Sch\"afer-Nameki and Y.-N. Wang, \emph{{5d
  Superconformal Field Theories and Graphs}},
  \href{https://doi.org/10.1016/j.physletb.2019.135077}{\emph{Phys. Lett. B}
  {\bfseries 800} (2020) 135077}
  [\href{https://arxiv.org/abs/1906.11820}{{\ttfamily 1906.11820}}].

\bibitem{Apruzzi:2019opn}
F.~Apruzzi, C.~Lawrie, L.~Lin, S.~Sch\"afer-Nameki and Y.-N. Wang,
  \emph{{Fibers add Flavor, Part I: Classification of 5d SCFTs, Flavor
  Symmetries and BPS States}},
  \href{https://doi.org/10.1007/JHEP11(2019)068}{\emph{JHEP} {\bfseries 11}
  (2019) 068} [\href{https://arxiv.org/abs/1907.05404}{{\ttfamily
  1907.05404}}].

\bibitem{Apruzzi:2019enx}
F.~Apruzzi, C.~Lawrie, L.~Lin, S.~Sch\"afer-Nameki and Y.-N. Wang,
  \emph{{Fibers add Flavor, Part II: 5d SCFTs, Gauge Theories, and Dualities}},
  \href{https://doi.org/10.1007/JHEP03(2020)052}{\emph{JHEP} {\bfseries 03}
  (2020) 052} [\href{https://arxiv.org/abs/1909.09128}{{\ttfamily
  1909.09128}}].

\bibitem{Bhardwaj:2019jtr}
L.~Bhardwaj, \emph{{On the classification of 5d SCFTs}},
  \href{https://doi.org/10.1007/JHEP09(2020)007}{\emph{JHEP} {\bfseries 09}
  (2020) 007} [\href{https://arxiv.org/abs/1909.09635}{{\ttfamily
  1909.09635}}].

\bibitem{Bhardwaj:2019fzv}
L.~Bhardwaj, P.~Jefferson, H.-C. Kim, H.-C. Tarazi and C.~Vafa, \emph{{Twisted
  Circle Compactifications of 6d SCFTs}},
  \href{https://arxiv.org/abs/1909.11666}{{\ttfamily 1909.11666}}.

\bibitem{Bhardwaj:2019ngx}
L.~Bhardwaj, \emph{{Dualities of 5d gauge theories from S-duality}},
  \href{https://doi.org/10.1007/JHEP07(2020)012}{\emph{JHEP} {\bfseries 07}
  (2020) 012} [\href{https://arxiv.org/abs/1909.05250}{{\ttfamily
  1909.05250}}].

\bibitem{Saxena:2019wuy}
V.~Saxena, \emph{{Rank-two 5d SCFTs from M-theory at isolated toric
  singularities: a systematic study}},
  \href{https://doi.org/10.1007/JHEP04(2020)198}{\emph{JHEP} {\bfseries 04}
  (2020) 198} [\href{https://arxiv.org/abs/1911.09574}{{\ttfamily
  1911.09574}}].

\bibitem{Bhardwaj:2019xeg}
L.~Bhardwaj, \emph{{Do all 5d SCFTs descend from 6d SCFTs?}},
  \href{https://arxiv.org/abs/1912.00025}{{\ttfamily 1912.00025}}.

\bibitem{Apruzzi:2019kgb}
F.~Apruzzi, S.~Schafer-Nameki and Y.-N. Wang, \emph{{5d SCFTs from Decoupling
  and Gluing}}, \href{https://doi.org/10.1007/JHEP08(2020)153}{\emph{JHEP}
  {\bfseries 08} (2020) 153}
  [\href{https://arxiv.org/abs/1912.04264}{{\ttfamily 1912.04264}}].

\bibitem{Closset:2019juk}
C.~Closset and M.~Del~Zotto, \emph{{On 5d SCFTs and their BPS quivers. Part I:
  B-branes and brane tilings}},
  \href{https://arxiv.org/abs/1912.13502}{{\ttfamily 1912.13502}}.

\bibitem{Kashani-Poor:2019jyo}
A.-K. Kashani-Poor, \emph{{Determining F-theory matter via Gromov-Witten
  invariants}},  \href{https://arxiv.org/abs/1912.10009}{{\ttfamily
  1912.10009}}.

\bibitem{Closset:2020scj}
C.~Closset, S.~Schafer-Nameki and Y.-N. Wang, \emph{{Coulomb and Higgs Branches
  from Canonical Singularities: Part 0}},
  \href{https://doi.org/10.1007/JHEP02(2021)003}{\emph{JHEP} {\bfseries 02}
  (2021) 003} [\href{https://arxiv.org/abs/2007.15600}{{\ttfamily
  2007.15600}}].

\bibitem{Closset:2020afy}
C.~Closset, S.~Giacomelli, S.~Sch\"afer-Nameki and Y.-N. Wang, \emph{{5d and 4d
  SCFTs: Canonical Singularities, Trinions and S-Dualities}},
  \href{https://arxiv.org/abs/2012.12827}{{\ttfamily 2012.12827}}.

\bibitem{Duan:2020imo}
Z.~Duan, D.~J. Duque and A.-K. Kashani-Poor, \emph{{Weyl invariant Jacobi forms
  along Higgsing trees}},  \href{https://arxiv.org/abs/2012.10427}{{\ttfamily
  2012.10427}}.

\bibitem{Morrison:2020ool}
D.~R. Morrison, S.~Schafer-Nameki and B.~Willett, \emph{{Higher-Form Symmetries
  in 5d}}, \href{https://doi.org/10.1007/JHEP09(2020)024}{\emph{JHEP}
  {\bfseries 09} (2020) 024}
  [\href{https://arxiv.org/abs/2005.12296}{{\ttfamily 2005.12296}}].

\bibitem{Bhardwaj:2020phs}
L.~Bhardwaj and S.~Sch\"afer-Nameki, \emph{{Higher-form symmetries of 6d and 5d
  theories}}, \href{https://doi.org/10.1007/JHEP02(2021)159}{\emph{JHEP}
  {\bfseries 02} (2021) 159}
  [\href{https://arxiv.org/abs/2008.09600}{{\ttfamily 2008.09600}}].

\bibitem{Albertini:2020mdx}
F.~Albertini, M.~Del~Zotto, I.~n. Garc\'\i{}a~Etxebarria and S.~S. Hosseini,
  \emph{{Higher Form Symmetries and M-theory}},
  \href{https://doi.org/10.1007/JHEP12(2020)203}{\emph{JHEP} {\bfseries 12}
  (2020) 203} [\href{https://arxiv.org/abs/2005.12831}{{\ttfamily
  2005.12831}}].

\bibitem{Razamat:2016dpl}
S.~S. Razamat, C.~Vafa and G.~Zafrir, \emph{{4d $ \mathcal{N}=1 $ from 6d (1,
  0)}}, \href{https://doi.org/10.1007/JHEP04(2017)064}{\emph{JHEP} {\bfseries
  04} (2017) 064} [\href{https://arxiv.org/abs/1610.09178}{{\ttfamily
  1610.09178}}].

\bibitem{Bah:2017gph}
I.~Bah, A.~Hanany, K.~Maruyoshi, S.~S. Razamat, Y.~Tachikawa and G.~Zafrir,
  \emph{{4d $ \mathcal{N}=1 $ from 6d $ \mathcal{N}=\left(1,0\right) $ on a
  torus with fluxes}},
  \href{https://doi.org/10.1007/JHEP06(2017)022}{\emph{JHEP} {\bfseries 06}
  (2017) 022} [\href{https://arxiv.org/abs/1702.04740}{{\ttfamily
  1702.04740}}].

\bibitem{Kim:2017toz}
H.-C. Kim, S.~S. Razamat, C.~Vafa and G.~Zafrir, \emph{{E-String Theory on
  Riemann Surfaces}},
  \href{https://doi.org/10.1002/prop.201700074}{\emph{Fortsch. Phys.}
  {\bfseries 66} (2018) 1700074}
  [\href{https://arxiv.org/abs/1709.02496}{{\ttfamily 1709.02496}}].

\bibitem{Kim:2018bpg}
H.-C. Kim, S.~S. Razamat, C.~Vafa and G.~Zafrir, \emph{{D-type Conformal Matter
  and SU/USp Quivers}},
  \href{https://doi.org/10.1007/JHEP06(2018)058}{\emph{JHEP} {\bfseries 06}
  (2018) 058} [\href{https://arxiv.org/abs/1802.00620}{{\ttfamily
  1802.00620}}].

\bibitem{Kim:2018lfo}
H.-C. Kim, S.~S. Razamat, C.~Vafa and G.~Zafrir, \emph{{Compactifications of
  ADE conformal matter on a torus}},
  \href{https://doi.org/10.1007/JHEP09(2018)110}{\emph{JHEP} {\bfseries 09}
  (2018) 110} [\href{https://arxiv.org/abs/1806.07620}{{\ttfamily
  1806.07620}}].

\bibitem{Razamat:2018gro}
S.~S. Razamat and G.~Zafrir, \emph{{Compactification of 6d minimal SCFTs on
  Riemann surfaces}},
  \href{https://doi.org/10.1103/PhysRevD.98.066006}{\emph{Phys. Rev. D}
  {\bfseries 98} (2018) 066006}
  [\href{https://arxiv.org/abs/1806.09196}{{\ttfamily 1806.09196}}].

\bibitem{Chen:2019njf}
J.~Chen, B.~Haghighat, S.~Liu and M.~Sperling, \emph{{4d $N$=1 from 6d D-type
  $N$=(1,0)}}, \href{https://doi.org/10.1007/JHEP01(2020)152}{\emph{JHEP}
  {\bfseries 01} (2020) 152}
  [\href{https://arxiv.org/abs/1907.00536}{{\ttfamily 1907.00536}}].

\bibitem{Razamat:2019mdt}
S.~S. Razamat, E.~Sabag and G.~Zafrir, \emph{{From 6d flows to 4d flows}},
  \href{https://doi.org/10.1007/JHEP12(2019)108}{\emph{JHEP} {\bfseries 12}
  (2019) 108} [\href{https://arxiv.org/abs/1907.04870}{{\ttfamily
  1907.04870}}].

\bibitem{Pasquetti:2019hxf}
S.~Pasquetti, S.~S. Razamat, M.~Sacchi and G.~Zafrir, \emph{{Rank $Q$ E-string
  on a torus with flux}},
  \href{https://doi.org/10.21468/SciPostPhys.8.1.014}{\emph{SciPost Phys.}
  {\bfseries 8} (2020) 014} [\href{https://arxiv.org/abs/1908.03278}{{\ttfamily
  1908.03278}}].

\bibitem{Haghighat:2011xx}
B.~Haghighat and S.~Vandoren, \emph{{Five-dimensional gauge theory and
  compactification on a torus}},
  \href{https://doi.org/10.1007/JHEP09(2011)060}{\emph{JHEP} {\bfseries 09}
  (2011) 060} [\href{https://arxiv.org/abs/1107.2847}{{\ttfamily 1107.2847}}].

\bibitem{Haghighat:2012bm}
B.~Haghighat, J.~Manschot and S.~Vandoren, \emph{{A 5d/2d/4d correspondence}},
  \href{https://doi.org/10.1007/JHEP03(2013)157}{\emph{JHEP} {\bfseries 03}
  (2013) 157} [\href{https://arxiv.org/abs/1211.0513}{{\ttfamily 1211.0513}}].

\bibitem{Beaujard:2020sgs}
G.~Beaujard, J.~Manschot and B.~Pioline, \emph{{Vafa-Witten invariants from
  exceptional collections}},
  \href{https://arxiv.org/abs/2004.14466}{{\ttfamily 2004.14466}}.

\bibitem{Katz:1996fh}
S.~H. Katz, A.~Klemm and C.~Vafa, \emph{{Geometric engineering of quantum field
  theories}}, \href{https://doi.org/10.1016/S0550-3213(97)00282-4}{\emph{Nucl.
  Phys. B} {\bfseries 497} (1997) 173}
  [\href{https://arxiv.org/abs/hep-th/9609239}{{\ttfamily hep-th/9609239}}].

\bibitem{Katz:1997eq}
S.~Katz, P.~Mayr and C.~Vafa, \emph{{Mirror symmetry and exact solution of 4-D
  N=2 gauge theories: 1.}},
  \href{https://doi.org/10.4310/ATMP.1997.v1.n1.a2}{\emph{Adv. Theor. Math.
  Phys.} {\bfseries 1} (1998) 53}
  [\href{https://arxiv.org/abs/hep-th/9706110}{{\ttfamily hep-th/9706110}}].

\bibitem{Gu:2018gmy}
J.~Gu, B.~Haghighat, K.~Sun and X.~Wang, \emph{{Blowup Equations for 6d SCFTs.
  I}}, \href{https://doi.org/10.1007/JHEP03(2019)002}{\emph{JHEP} {\bfseries
  03} (2019) 002} [\href{https://arxiv.org/abs/1811.02577}{{\ttfamily
  1811.02577}}].

\bibitem{Gu:2019dan}
J.~Gu, A.~Klemm, K.~Sun and X.~Wang, \emph{{Elliptic blowup equations for 6d
  SCFTs. Part II. Exceptional cases}},
  \href{https://doi.org/10.1007/JHEP12(2019)039}{\emph{JHEP} {\bfseries 12}
  (2019) 039} [\href{https://arxiv.org/abs/1905.00864}{{\ttfamily
  1905.00864}}].

\bibitem{Gu:2019pqj}
J.~Gu, B.~Haghighat, A.~Klemm, K.~Sun and X.~Wang, \emph{{Elliptic blowup
  equations for 6d SCFTs. Part III. E-strings, M-strings and chains}},
  \href{https://doi.org/10.1007/JHEP07(2020)135}{\emph{JHEP} {\bfseries 07}
  (2020) 135} [\href{https://arxiv.org/abs/1911.11724}{{\ttfamily
  1911.11724}}].

\bibitem{Gu:2020fem}
J.~Gu, B.~Haghighat, A.~Klemm, K.~Sun and X.~Wang, \emph{{Elliptic Blowup
  Equations for 6d SCFTs. IV: Matters}},
  \href{https://arxiv.org/abs/2006.03030}{{\ttfamily 2006.03030}}.

\bibitem{Kim:2020hhh}
H.-C. Kim, M.~Kim, S.-S. Kim and K.-H. Lee, \emph{{Bootstrapping BPS spectra of
  5d/6d field theories}},  \href{https://arxiv.org/abs/2101.00023}{{\ttfamily
  2101.00023}}.

\bibitem{Heckman:2013pva}
J.~J. Heckman, D.~R. Morrison and C.~Vafa, \emph{{On the Classification of 6D
  SCFTs and Generalized ADE Orbifolds}},
  \href{https://doi.org/10.1007/JHEP05(2014)028}{\emph{JHEP} {\bfseries 05}
  (2014) 028} [\href{https://arxiv.org/abs/1312.5746}{{\ttfamily 1312.5746}}].

\bibitem{Douglas:2010iu}
M.~R. Douglas, \emph{{On D=5 super Yang-Mills theory and (2,0) theory}},
  \href{https://doi.org/10.1007/JHEP02(2011)011}{\emph{JHEP} {\bfseries 02}
  (2011) 011} [\href{https://arxiv.org/abs/1012.2880}{{\ttfamily 1012.2880}}].

\bibitem{Lambert:2010iw}
N.~Lambert, C.~Papageorgakis and M.~Schmidt-Sommerfeld, \emph{{M5-Branes,
  D4-Branes and Quantum 5D super-Yang-Mills}},
  \href{https://doi.org/10.1007/JHEP01(2011)083}{\emph{JHEP} {\bfseries 01}
  (2011) 083} [\href{https://arxiv.org/abs/1012.2882}{{\ttfamily 1012.2882}}].

\bibitem{Bhardwaj:2020gyu}
L.~Bhardwaj and G.~Zafrir, \emph{{Classification of 5d N=1 gauge theories}},
  \href{https://doi.org/10.1007/JHEP12(2020)099}{\emph{JHEP} {\bfseries 12}
  (2020) 099} [\href{https://arxiv.org/abs/2003.04333}{{\ttfamily
  2003.04333}}].

\bibitem{Tachikawa:2011ch}
Y.~Tachikawa, \emph{{On S-duality of 5d super Yang-Mills on $S^1$}},
  \href{https://doi.org/10.1007/JHEP11(2011)123}{\emph{JHEP} {\bfseries 11}
  (2011) 123} [\href{https://arxiv.org/abs/1110.0531}{{\ttfamily 1110.0531}}].

\bibitem{Feger:2012bs}
R.~Feger and T.~W. Kephart, \emph{{LieART\textemdash{}A Mathematica application
  for Lie algebras and representation theory}},
  \href{https://doi.org/10.1016/j.cpc.2014.12.023}{\emph{Comput. Phys. Commun.}
  {\bfseries 192} (2015) 166}
  [\href{https://arxiv.org/abs/1206.6379}{{\ttfamily 1206.6379}}].

\bibitem{Feger:2019tvk}
R.~Feger, T.~W. Kephart and R.~J. Saskowski, \emph{{LieART 2.0 \textendash{} A
  Mathematica application for Lie Algebras and Representation Theory}},
  \href{https://doi.org/10.1016/j.cpc.2020.107490}{\emph{Comput. Phys. Commun.}
  {\bfseries 257} (2020) 107490}
  [\href{https://arxiv.org/abs/1912.10969}{{\ttfamily 1912.10969}}].

\bibitem{miranda1989basic}
R.~Miranda and U.~di~Pisa. Dipartimento~di matematica, \emph{The Basic Theory
  of Elliptic Surfaces: Notes of Lectures}, Dottorato di ricerca in matematica
  / Universit{\`a} di Pisa, Dipartimento di Matematica. ETS Editrice, 1989.

\end{thebibliography}\endgroup
\end{document}